\documentclass[times,final]{elsarticle}
\usepackage{jcomp}
\usepackage{amsmath,amssymb,bm}
\usepackage{booktabs}
\usepackage{multirow}
\usepackage{hyperref}
\usepackage{float}
\hypersetup{
    colorlinks = true,
    urlcolor   = blue,
    citecolor  = red,
}

\usepackage{algorithm}
\usepackage{algpseudocode}

\journal{Journal of Computational Physics}

\begin{document}


\verso{C. Zhang et~al.}

\begin{frontmatter}

  \title{Coupled Rigid-Body Motion and Binary Phase Change: A Sharp Level-Set/Embedded-Boundary Method}%

  \author[1]{Chao-Fan Zhang}
  \author[1]{Zhong-Han Xue}
  \author[1]{Jie Zhang\corref{cor1}}
  \cortext[cor1]{Corresponding author:}
  \ead{j\_zhang@xjtu.edu.cn}

  \address[1]{ State Key Laboratory for Strength and Vibration of Mechanical Structures, School of Aerospace, Xi'an Jiaotong University, Xi'an 710049, China}

\begin{abstract}
We develop a sharp-interface method for freely moving solid particles undergoing pure or binary melting. A Level-Set representation preserves the particle geometry during long-distance translation and rotation, while an embedded-boundary formulation imposes hydrodynamic, thermal and solutal conditions sharply on a Eulerian Cartesian mesh. Liquid temperature and solutal concentration are advanced with a conservative moving-cut-cell discretization, whereas the solid temperature is transported through a geometry-consistent conservative tracer. Hydrodynamic loads, rigid-body motion and phase change are coupled iteratively within each physical time step.
The method is validated progressively for rigid-body motion, prescribed-motion pure and binary melting, and fully coupled melting with free particle motion. It reproduces established reference solutions and captures wake-induced rotation of a melting sphere and the simultaneous translation, rotation and thermosolutal melting of a three-dimensional oblate ellipsoid. As a final application, simulations of a floating ice disc reveal how thermoconvective instability, geometric confinement and melting jointly determine its spontaneous motion, with phase change feeding back on the instability through the evolving particle geometry. Additional tests establish the importance of long-time geometrical fidelity, local scalar conservation and geometry-consistent solid-temperature transport. The resulting framework provides a sharp and conservative approach for strongly coupled moving-body phase-change problems.
\end{abstract}


\end{frontmatter}

\section{Introduction}
\label{sec:introduction}

The melting or solidification of freely moving bodies arises in a broad range of natural and engineering processes, including drifting and capsizing ice bodies \citep{cenedese2023icebergs,bellincioni2025melting,johnsonShapeEvolutionCapsize2025}, as well as the growth, sedimentation and interaction of particles or crystals in solidifying melts \citep{Appolaire1998FreeGrowthSettling,Badillo2007DendriteTipKinetics,karagadde_coupled_2012}. Unlike phase change at a fixed boundary, these problems involve an intrinsic two-way coupling between body motion and interfacial thermodynamics: translation and rotation continuously modify the relative flow, the thermal and solutal boundary layers, and hence the local phase-change rate, while melting or solidification alters the body geometry and consequently its hydrodynamic force, torque and subsequent motion. Recent reviews \citep{du2024physics} have emphasized that such flow--phase-change interactions are central to many geophysical systems and that the melting of finite-size bodies transported by the surrounding flow remains comparatively unexplored, particularly when compositional effects are involved. High-fidelity numerical simulation is therefore especially valuable, since simultaneously resolving the evolving geometry, near-interface transport and free-body dynamics remains extremely challenging experimentally.

From a numerical standpoint, the difficulty is not merely the coexistence of fluid--solid interaction and phase change, but the requirement that the same evolving solid--liquid interface simultaneously serves as the carrier of the solid geometry, the hydrodynamic boundary for the surrounding flow, and the thermodynamic boundary for heat and mass transfer. Its motion contains two distinct contributions: rigid-body translation and rotation determined by the hydrodynamic force and torque, and local normal displacement generated by phase change. Consequently, an error in the transported geometry directly contaminates the surface stress and hence the subsequent body motion, whereas a loss of local conservation in the near-interface temperature or concentration fields alters the one-sided fluxes that determine the phase-change rate. The coupling becomes even tighter for binary phase change, where the interfacial concentration, equilibrium temperature and phase-change velocity must satisfy simultaneously the liquidus relation, solute-rejection condition and Stefan balance \citep{davis2001theory,karma_phase-field_2001,ramirez_phase-field_2004}. A successful numerical framework must therefore preserve the geometry of a freely translating and rotating body over long distances, impose the hydrodynamic and thermosolutal conditions sharply and conservatively on the same instantaneous interface, and close the mutual dependence among flow, rigid-body motion and phase change within each physical time step.

Existing numerical approaches address these requirements in different ways. Diffuse-interface methods, particularly phase-field formulations, have been widely used for solidification because anisotropy, solute partition and complex topological changes can be incorporated naturally within a continuous interfacial description \citep{beckermann1999modeling,boettinger_phase-field_2002,echebarria2004quantitative}. Coupling phase-field models with lattice-Boltzmann, immersed-boundary or fictitious-domain solvers has further enabled simulations of growing and freely moving dendrites undergoing translation, rotation and collision \citep{Rojas2015PhaseFieldLBM,Takaki2018PhaseFieldLBMGrainGrowth,Sakane2019PhaseFieldLBM,meng2020phase,mao2024anisotropic}. Their defining feature, however, is that the solid--liquid interface is represented by a finite-thickness transition layer, so that the interfacial thermal and solutal conditions are recovered through regularized or thin-interface formulations rather than imposed directly at a geometrically sharp boundary \citep{karma1998quantitative,hester_improved_2020}. For the present problem, where the local melting or solidification rate is determined directly by one-sided heat and solute fluxes and where the same surface must also support the hydrodynamic boundary condition and load evaluation, a sharp-interface formulation is particularly attractive.

Sharp-interface methods instead retain a geometrically localized solid--liquid boundary and impose the associated interfacial conditions directly \citep{udaykumar_computation_1999,yang2005sharp,theillard_sharp_2015,limare2023hybrid,bochkov2021sharp,xue2023three}, while their extension to freely moving phase-changing bodies has only recently received substantial attention. Early body-fitted formulations, such as that of Gan \textit{et al}.~\citep{gan2003simulation}, naturally accommodate rigid-body motion but require mesh deformation or regeneration as the particle translates, rotates and changes shape. More recently, Zhong \textit{et al}.~\citep{zhongFronttrackingImmersedboundaryFramework2025} developed one of the first non-body-conformal Lagrangian frameworks specifically for freely moving melting bodies, combining front tracking with an immersed-boundary formulation. The explicit triangulated surface provides direct access to the particle geometry and hydrodynamic loads, but continuous phase change requires dynamic maintenance of the Lagrangian mesh as its area and spacing evolve. Almost in parallel with the development of the present framework, Xu \textit{et al}.~\citep{xu2025EulerianLSIBM} proposed one of the first fully Eulerian fixed-grid approaches for freely moving phase-changing solids, coupling an advected level-set representation with an Eulerian immersed-boundary method. Their formulation avoids Lagrangian remeshing and accounts for both rigid-body motion and phase-change propagation in the interface transport. Nevertheless, several difficulties central to the present problem remain open. In particular, binary phase change requires the interfacial concentration, equilibrium temperature and phase-change velocity to be solved as a mutually coupled sharp closure; meanwhile, local scalar conservation must be maintained as the moving boundary continuously changes the physical volume of cut cells. Moreover, the hydrodynamic boundary condition, surface stress and thermosolutal fluxes should ideally be evaluated on the same reconstructed sharp geometry rather than through distinct regularized or volumetric representations. These requirements motivate a sharp, conservative moving-boundary formulation in which geometry transport, interfacial scalar fluxes and hydrodynamic loading are treated consistently within a common fixed-grid framework.


Our previous work provides a natural starting point for addressing these issues. Xue \textit{et al}.~\citep{xue2023three,xue2026wake} developed a three-dimensional (3D) sharp and conservative VOF framework for binary solidification and melting, in which the solid--liquid interface was reconstructed geometrically and the temperature and concentration fields were solved separately in the two phases, with their interfacial conditions imposed sharply through an embedded-boundary formulation. This approach accurately resolved the thermosolutal coupling associated with a phase-changing interface, but the solid itself was not allowed to translate or rotate in response to hydrodynamic loads. Once free rigid-body motion is introduced, a new incompatibility emerges between long-time geometry transport and local conservation. In particular, our tests (see \S~\ref{sec:numerical_tests}) show that repeated VOF reconstruction and advection of a non-spherical body may gradually accumulate shape errors during prolonged translation and rotation even when its enclosed volume remains well preserved; such errors directly contaminate the surface stress, torque and subsequent body motion. This observation motivates a deliberate division of labor in the present method: the Level-Set representation is used as the primary carrier of the moving geometry, whereas the same zero-level surface is reconstructed as a sharp embedded boundary for the conservative finite-volume treatment of momentum, heat and solute transport. In this way, geometrical fidelity during long-distance rigid-body motion and local conservation near the phase boundary are treated by the numerical representations best suited to each task, without introducing independent interface descriptions.

The present study develops a 3D sharp and conservative framework for freely moving bodies undergoing pure or binary melting/solidification. The method is constructed around the three numerical difficulties identified above:
\begin{itemize}
\item First, the solid geometry is transported by a Level-Set formulation in which rigid-body translation and rotation are combined with the local phase-change velocity. This treatment avoids the cumulative shape distortion observed with repeated geometrical VOF transport during long-distance body motion, while providing a common sharp geometry for the subsequent flow, scalar and load calculations.
    
\item Second, heat and solute transport are formulated directly on the time-dependent cut-cell geometry. Sharp one-sided diffusive fluxes are imposed at the reconstructed interface, while the change of physical phase volume caused by boundary motion is incorporated into the conservative scalar advection. In contrast to momentum, for which a locally relaxed conservative treatment is employed for small-cell stability, temperature and concentration retain strict local conservation because their near-interface errors directly contaminate the Stefan and solutal balances. The solid temperature is further transported together with the reconstructed solid volume, keeping its numerical support geometrically consistent with the moving body.
    
\item Third, the flow solution, sharp hydrodynamic loads, rigid-body dynamics, interfacial thermodynamics and phase-change motion are closed through a fixed-time iterative procedure. The force and torque are evaluated directly on the same reconstructed surface used for the velocity and thermosolutal boundary conditions, and the coupled variables are iterated without repeatedly advancing the transient equations over the same physical interval. The resulting converged state therefore provides mutually consistent geometry, flow, heat and mass transfer, phase-change rate and rigid-body motion at each time level.
\end{itemize}

Together, these developments extend the sharp binary phase-change framework of Xue \textit{et al}.~\citep{xue2023three} from thermodynamically evolving interfaces without hydrodynamically induced rigid-body motion to freely translating and rotating bodies for which geometry transport, local conservation and multiphysics coupling must be satisfied simultaneously. The remainder of the paper is organized as follows. \S~\ref{sec:problem} formulates the freely moving binary phase-change problem, including the governing equations, sharp interfacial conditions and the numerical challenges arising from the simultaneous rigid-body and phase-change motions. \S~\ref{sec:numerical_methods} presents the numerical framework and the three techniques developed to address these challenges, namely long-time transport of the sharp moving geometry, sharp and conservative transport on moving cut cells, and the strongly coupled hydrodynamic-load and rigid-body update. \S~\ref{sec:numerical_tests} assesses the method progressively through benchmark problems that isolate interface transport, rigid-body motion and scalar transport before considering fully coupled pure and binary melting. Finally, the main conclusions are summarized in \S~\ref{sec:conclusion}.

\section{Problem formulation and numerical challenges}
\label{sec:problem}

\subsection{Sharp binary phase change on a freely moving solid boundary}
\label{sec:configuration}

\begin{figure}
	\centering
	\includegraphics[width=0.5\textwidth]{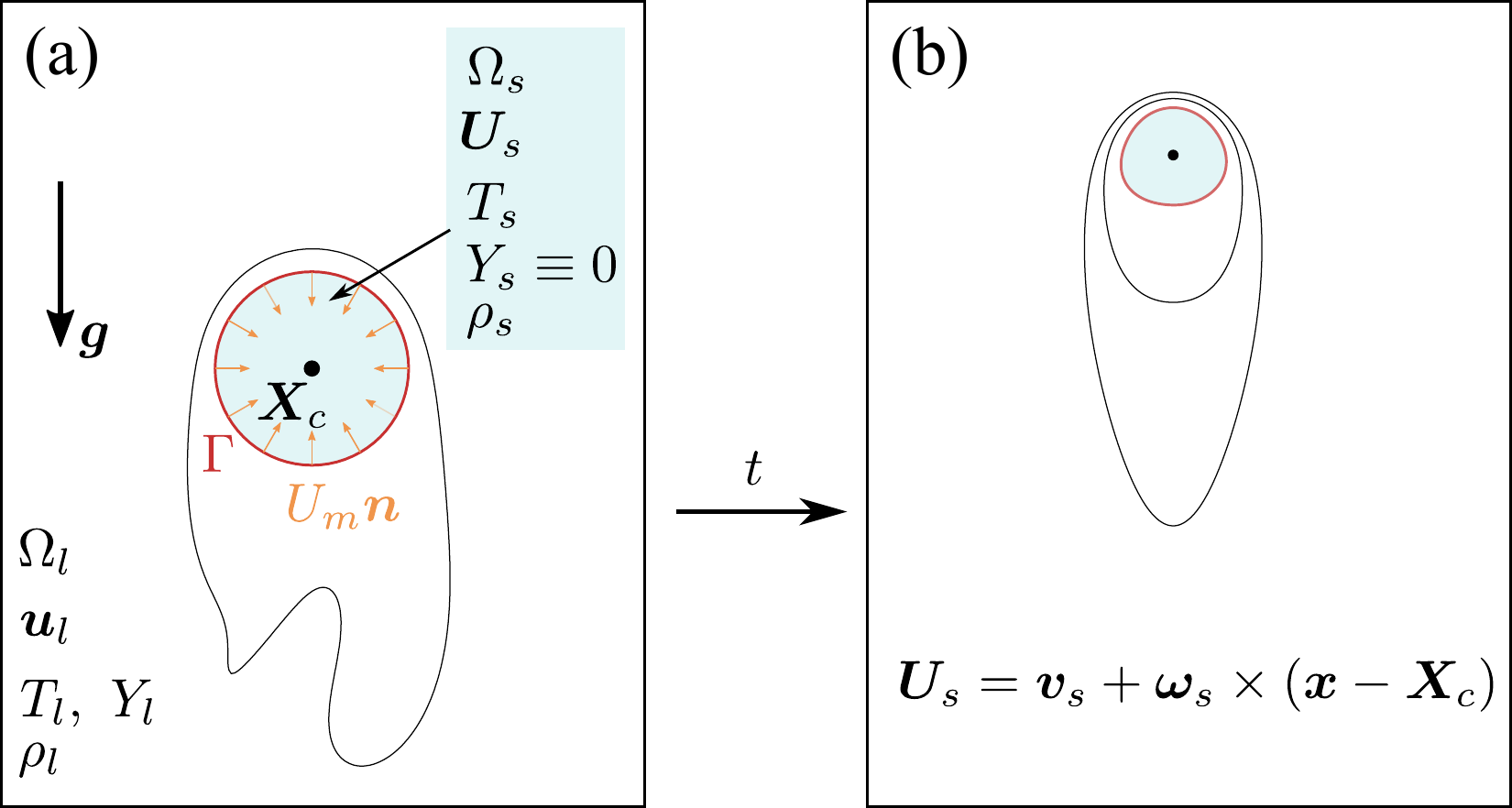}
	\caption{
Schematic of a freely moving solid body undergoing binary melting.
The sharp interface $\Gamma$ separates the solid domain $\Omega_s$ from the liquid domain $\Omega_l$, where the velocity, temperature and solute fields are defined separately and coupled through the interfacial conditions; for the NaCl--water system, $Y_s\equiv0$.
The solid motion is described by its centroid $\bm{X}_c$, translational velocity $\bm{v}_s$ and angular velocity $\bm{\omega}_s$, with $\bm{U}_s=\bm{v}_s+\bm{\omega}_s\times(\bm{x}-\bm{X}_c)$.
The black solid lines in both panels denote isotherms.
Panels (a) and (b) illustrate the coupled evolution of the body geometry, motion and surrounding thermal field during melting.}
	\label{fig:schematic}
\end{figure}

We consider a rigid solid body immersed in an incompressible liquid and undergoing melting or solidification while being free to translate and rotate under the hydrodynamic loads exerted by the surrounding fluid, as illustrated schematically in Fig.~\ref{fig:schematic}. At any instant, the computational domain $\Omega$ is partitioned into the time-dependent liquid and solid subdomains, $\Omega_l(t)$ and $\Omega_s(t)$, separated by a sharp solid--liquid interface $\Gamma(t)$,
\begin{equation}
    \Omega
    =
    \Omega_l(t)
    \cup
    \Gamma(t)
    \cup
    \Omega_s(t).
    \label{eq:domain_decomposition}
\end{equation}
The subscripts $l$ and $s$ denote quantities in the liquid and solid phases, respectively. For compactness, the explicit time dependence of $\Omega_l$, $\Omega_s$, $\Gamma$ and other time-dependent quantities is omitted below unless it is required for clarity. For later numerical representation, the sharp interface is associated with a signed-distance function $\phi(\bm x,t)$, defined by $\phi>0$ in $\Omega_l$, $\phi<0$ in $\Omega_s$, and $\Gamma=\{\bm x:\phi(\bm x,t)=0\}$.

Physically, $\Gamma$ is regarded as a zero-thickness internal boundary rather than a numerically smeared transition region. Accordingly, the temperature and concentration fields are solved separately in $\Omega_l$ and $\Omega_s$, while the thermodynamic equilibrium, heat-flux balance and solute-rejection conditions are imposed directly on $\Gamma$. In our previous formulation \citep{xue2023three}, the interface evolved solely through melting or solidification, whereas the solid phase itself did not undergo hydrodynamically induced rigid-body motion. The essential extension considered here is that the entire sharp boundary is now transported by the translation and rotation of the solid while simultaneously being displaced locally by phase change. Hence, the zero-thickness boundary carrying the thermal and solutal interfacial conditions not only deforms because of phase change, but also continuously moves across the underlying Cartesian mesh as a consequence of FSI.

Let $\bm X_c$ denote the instantaneous centroid of the solid body, and let $\bm v_s$ and $\bm\omega_s$ denote its translational and angular velocities, respectively. The rigid-body velocity field within the instantaneous solid domain is
\begin{equation}
    \bm U_s(\bm x)
    =
    \bm v_s
    +
    \bm\omega_s\times
    \left(\bm x-\bm X_c\right),
    \qquad
    \bm x\in\Omega_s\cup\Gamma .
    \label{eq:rigid_body_velocity}
\end{equation}
Equation~\eqref{eq:rigid_body_velocity} describes the transport of the instantaneous solid geometry due solely to rigid-body motion and therefore does not, by itself, change the volume of solid. Phase change introduces an additional displacement of $\Gamma$ relative to this moving solid. We define $\bm n$ as the unit normal pointing from the liquid toward the solid and $U_m$ as the local phase-change velocity, taken to be positive for melting. The geometrical velocity of the solid--liquid interface is then
\begin{equation}
    \bm V_\Gamma
    =
    \bm U_s
    +
    U_m\bm n ,
    \qquad
    \bm x\in\Gamma .
    \label{eq:interface_velocity}
\end{equation}
Thus, rigid-body motion transports the instantaneous solid geometry globally, whereas phase change modifies the interface locally through $U_m$.

This decomposition has an important numerical consequence. The same sharp interface $\Gamma$ simultaneously serves as the moving hydrodynamic boundary for the flow problem and as the internal boundary on which the thermal and solutal conditions are imposed. Its geometrical representation must therefore remain mutually consistent among the flow, heat-transfer, solute-transport and rigid-body calculations. This requirement becomes particularly restrictive during long-distance translation and rotation, since small errors accumulated through repeated interface transport may gradually distort the body geometry and thereby contaminate the hydrodynamic force, torque and subsequent trajectory. Meanwhile, as $\Gamma$ sweeps across the Cartesian mesh, the liquid and solid portions of interfacial control volumes vary continuously, imposing an additional requirement of conservative scalar transport on the moving geometry.

The present problem is therefore more demanding than either conventional rigid-body FSI with a prescribed geometry or binary phase change involving a kinematically fixed solid. An appropriate numerical framework must preserve the sharp solid geometry during long-time rigid-body transport, while retaining a geometrically consistent representation of the same interface for hydrodynamics and interfacial heat and mass transfer. These requirements motivate the interface treatment developed in \S~\ref{sec:numerical_methods}; the additional difficulties associated with scalar conservation and the coupled binary interfacial conditions are identified in the following subsections.

\subsection{Governing equations for fluid transport and rigid-body dynamics}
\label{sec:governing_equations}

With the sharp-domain decomposition introduced in \S~\ref{sec:configuration}, the flow and scalar fields are solved only in their respective physical domains. In the liquid region $\Omega_l$, the motion of the incompressible Newtonian fluid is governed by the Navier--Stokes equations under the Boussinesq approximation,
\begin{equation}
    \rho_l
    \left(
    \frac{\partial \bm u_l}{\partial t}
    +
    \bm u_l\cdot\nabla\bm u_l
    \right)
    =
    -\nabla p
    +
    \nabla\cdot\mu_l[\nabla\bm u_l+\nabla\bm u_l^\mathrm{T}]
    +
    \rho'\bm g,
    \qquad
    \bm x\in\Omega_l ,
    \label{eq:NS}
\end{equation}
together with
\begin{equation}
    \nabla\cdot\bm u_l=0,
    \qquad
    \bm x\in\Omega_l .
    \label{eq:continuity}
\end{equation}
Here, $\bm u_l$ and $p$ denote the liquid velocity and dynamic pressure with the reference hydrostatic contribution removed, respectively; $\rho_l$ and $\mu_l$ are the reference density and dynamic viscosity of the liquid; and $\bm g$ is the gravitational acceleration. Under the Boussinesq approximation, the density fluctuation entering the buoyancy term is $\rho'=\rho-\rho_l=\rho_l[-\beta_T(T_l-T_{\rm ref})+\beta_Y(Y_l-Y_{\rm ref})]$, where $T$ and $Y$ are the temperature and solutal concentration, with $\beta_T$ and $\beta_Y$ denoting the thermal and solutal expansion coefficients, respectively, while all other fluid properties are treated as constants. For the saline solutions considered here, increasing temperature decreases the liquid density whereas increasing solute concentration increases it; thermal and solutal stratifications may therefore compete in driving the surrounding flow and, consequently, modify the interfacial heat and mass transfer \citep{xue2023three,guo2025effects}. It is also worth noting that Eq.~\eqref{eq:continuity} remains strictly divergence-free in the liquid domain, distinguished from some other diffusive approaches. Since the two phases are treated as incompressible and phase change is localized at the zero-thickness interface, the volume change associated with $\rho_l\neq\rho_s$ is represented through the interfacial mass-balance condition rather than through a volumetric source term in Eq.~\eqref{eq:continuity}, as adopted in some one-domain formulations \citep{xu2025EulerianLSIBM}.

The temperature is defined on both sides of the sharp solid--liquid interface. Introducing $\zeta=l$ or $s$ for the liquid and solid phases, respectively, its governing equation can be written compactly as
\begin{equation}
    \frac{\partial T_\zeta}{\partial t}
    +
    \bm u_\zeta\cdot\nabla T_\zeta
    =
    \alpha_\zeta\nabla^2T_\zeta,
    \qquad
    \bm x\in\Omega_\zeta,
    \qquad
    \zeta=l,s ,
    \label{eq:temperature}
\end{equation}
where $\alpha_\zeta=k_\zeta/(\rho_\zeta c_{p,\zeta})$ is the thermal diffusivity, with $k_\zeta$ the heat conductivity and $c_{p, \zeta}$ the heat capacity. Although Eq.~\eqref{eq:temperature} has the same differential form in both phases, the advecting velocities are physically distinct: $\bm u_\zeta=\bm u_l$ in $\Omega_l$, whereas $\bm u_\zeta=\bm U_s$ in $\Omega_s$, with $\bm U_s$ defined by Eq.~\eqref{eq:rigid_body_velocity}. Thus, $T_l$ is transported by the local fluid motion, while $T_s$ is carried together with the translating and rotating solid. The latter introduces an additional numerical requirement on a fixed Cartesian mesh, since the support of $T_s$ moves together with $\Omega_s$ and changes continuously through phase change. Its transport must therefore remain geometrically consistent with the instantaneous solid domain while preserving thermal conservation.

For the NaCl--water system considered in the present work, salt is assumed to be completely rejected by the ice phase below the eutectic composition, so that $Y_s\equiv0$ and no solute equation is solved in $\Omega_s$. The liquid-phase solute concentration satisfies
\begin{equation}
    \frac{\partial Y_l}{\partial t}
    +
    \bm u_l\cdot\nabla Y_l
    =
    D_l\nabla^2Y_l,
    \qquad
    \bm x\in\Omega_l ,
    \label{eq:concentration}
\end{equation}
where $D_l$ is the solutal diffusivity. In the present sharp formulation, phase change is not represented by volumetrically distributed source terms in Eqs.~\eqref{eq:temperature} and \eqref{eq:concentration}. Instead, the heat and solute fluxes are coupled directly through the conditions imposed on the zero-thickness interface $\Gamma$, as will be introduced in \S~\ref{sec:interface_conditions}. This distinction is particularly important for binary melting, since the local phase-change velocity is determined directly by the one-sided thermal and solutal fluxes at the interface.

The surrounding flow, in turn, drives the translation and rotation of the solid through the hydrodynamic force and torque acting on its instantaneous boundary. Denoting the solid volume and mass by $V_s$ and $M_s=\rho_sV_s$, respectively, and its inertia tensor about the centroid $\bm X_c$ by $\bm I_s=\int_{\Omega_s}\rho_s[|\bm r|^2\bm I-\bm r\bm r]\,{\rm d}V$, with $\bm r=\bm x-\bm X_c$, the rigid-body motion is governed by
\begin{equation}
    M_s\frac{\mathrm d\bm v_s}{\mathrm dt}
    =
    \bm F_H
    +
    (\rho_s-\rho_l)V_s\bm g ,
    \label{eq:translation}
\end{equation}
and
\begin{equation}
    \bm I_s\frac{\mathrm d\bm\omega_s}{\mathrm dt}
    +
    \bm\omega_s\times
    \left(
    \bm I_s\bm\omega_s
    \right)
    =
    \bm T_H ,
    \label{eq:rotation}
\end{equation}
where $\bm F_H$ and $\bm T_H$ denote the hydrodynamic force and torque exerted by the liquid on the solid, respectively. With $\bm n$ defined in \S~\ref{sec:configuration} as pointing from the liquid toward the solid, they are obtained from the interfacial stress as $\bm F_H=\int_\Gamma (p\bm I-\bm\tau_l)\cdot\bm n\,{\rm d}A$ and $\bm T_H=\int_\Gamma \bm r\times[(p\bm I-\bm\tau_l)\cdot\bm n]\,{\rm d}A$, where $\bm\tau_l=\mu_l[\nabla\bm u_l+(\nabla\bm u_l)^T]$ is the viscous stress tensor. Since $p$ excludes the reference hydrostatic contribution, the corresponding buoyancy contribution appears explicitly as $(\rho_s-\rho_l)V_s\bm g$ in Eq.~\eqref{eq:translation}.

Because phase change continuously modifies the solid geometry, $V_s$, $\bm X_c$ and $\bm I_s$ generally vary with time. In the present formulation, however, the instantaneous mass and inertia properties are regarded as fixed during each rigid-body update and are renewed after the phase-change-induced geometry evolution. This treatment assumes that the characteristic phase-change time scale is sufficiently longer than that associated with rigid-body motion. A closely related approximation was adopted by Zhong \textit{et al}.~\citep{zhongFronttrackingImmersedboundaryFramework2025}, who retained $M_s^{n+1}\approx M_s^n$ during the translational update when melting-induced mass variation is slow compared with particle dynamics. The conventional Newton--Euler formulation has also been employed in recent Eulerian simulations of freely moving phase-changing solids by Xu \textit{et al}.~\citep{xu2025EulerianLSIBM}. More generally, Suzuki \textit{et al}.~\citep{suzuki2026equations} recently derived phase-change-corrected equations of motion that explicitly account for momentum and angular-momentum fluxes associated with melting and solidification, including those arising from non-uniform interface deformation and density change. These additional mechanical corrections are beyond the scope of the present work.

Equations~\eqref{eq:NS}--\eqref{eq:rotation} govern the bulk fluid motion, thermal and solutal transport, and rigid-body dynamics, but they do not yet constitute a closed system. The liquid velocity at the moving boundary, the interfacial temperature and concentration, and the local phase-change velocity $U_m$ must all be determined consistently on $\Gamma$. In a binary system, these quantities are mutually coupled through local mass conservation, thermodynamic equilibrium, the Stefan condition and solute rejection relation. The corresponding sharp interfacial conditions are introduced next.

\subsection{Sharp interfacial conditions for binary phase change}
\label{sec:interface_conditions}

The bulk equations introduced in \S~\ref{sec:governing_equations} are coupled through the sharp solid--liquid interface $\Gamma$. In the present problem, $\Gamma$ plays three roles simultaneously: it serves as a moving hydrodynamic boundary for the liquid flow, a thermodynamic-equilibrium boundary coupling temperature and solute concentration, and a phase-change boundary across which heat and solute fluxes determine the local melting rate. The corresponding interfacial conditions are specified below.

\textit{\textbf{Kinematic condition.}} As defined in Eq.~\eqref{eq:interface_velocity}, the geometrical interface velocity is $\bm V_\Gamma=\bm U_s+U_m\bm n$, where $\bm U_s$ represents rigid-body translation and rotation and $U_m>0$ denotes melting. Local mass conservation across the moving interface requires $\rho_l(\bm u_{l,\Gamma}-\bm V_\Gamma)\cdot\bm n=\rho_s(\bm U_s-\bm V_\Gamma)\cdot\bm n$. Together with continuity of the tangential velocity, this gives
\begin{equation}
    \bm u_{l,\Gamma}
    =
    \bm U_s
    +
    \left(
    1-\frac{\rho_s}{\rho_l}
    \right)U_m\bm n,
    \qquad
    \bm x\in\Gamma ,
    \label{eq:velocity_interface}
\end{equation}
where $\bm u_{l,\Gamma}$ denotes the liquid velocity evaluated at $\Gamma$. Hence, the velocity imposed on the liquid flow is generally different from the geometrical velocity $\bm V_\Gamma$ used to advance the interface. When $\rho_s\neq\rho_l$, phase change induces a normal Stefan flow associated with local volumetric expansion or contraction; in the equal-density limit, Eq.~\eqref{eq:velocity_interface} reduces to the conventional no-slip condition $\bm u_{l,\Gamma}=\bm U_s$.

\textit{\textbf{Thermal conditions.}} The temperature is continuous across the solid--liquid interface, i.e. $T_{l,\Gamma}=T_{s,\Gamma}=T_\Gamma$. For a binary system, the equilibrium interfacial temperature further depends on the liquid-side solute concentration through the Gibbs--Thomson relation. In its general form, $T_\Gamma=T_m+m_LY_{l,\Gamma}-\epsilon_\kappa\kappa-\epsilon_VU_m$, where $T_m$ is the melting temperature of the pure solvent, $m_L$ is the liquidus slope, and the last two terms represent capillary and kinetic undercooling, respectively. Since these microscopic corrections are negligible for the macroscopic melting problems considered here, as also highlighted by other studies~\citep{yang2023morphology,xue2026wake}, the equilibrium condition reduces to
\begin{equation}
    T_\Gamma
    =
    T_{l,\Gamma}
    =
    T_{s,\Gamma}
    =
    T_m+m_LY_{l,\Gamma},
    \qquad
    \bm x\in\Gamma .
    \label{eq:gibbs_thomson}
\end{equation}
Unlike pure-substance melting, for which the interfacial temperature is prescribed by $T_m$, $T_\Gamma$ is therefore an unknown in the binary system and varies locally with $Y_{l,\Gamma}$. Energy conservation across $\Gamma$ provides the complementary flux condition determining the local phase-change velocity,
\begin{equation}
    \rho_s L U_m
    =
    \left[
        (k\nabla T)_s
        -
        (k\nabla T)_l
    \right]\cdot\bm n,
    \qquad
    \bm x\in\Gamma ,
    \label{eq:stefan}
\end{equation}
where $L$ is the latent heat of fusion. Equation~\eqref{eq:stefan} directly highlights the need for a sharp treatment of the thermal field: $U_m$ is determined by the difference between the one-sided heat fluxes evaluated at the zero-thickness interface.

\textit{\textbf{Solutal condition.}} For a general binary system, local solute conservation requires $(Y_l-Y_s)U_m=[(D\nabla Y)_s-(D\nabla Y)_l]\cdot\bm n$, while thermodynamic partitioning gives $Y_{s,\Gamma}=k_pY_{l,\Gamma}$, with $k_p$ denoting the equilibrium partition coefficient. For the NaCl--water system considered here, $k_p=0$ and $Y_s\equiv0$, as discussed in \S~\ref{sec:governing_equations}. The interfacial solute balance consequently reduces to
\begin{equation}
    U_mY_{l,\Gamma}
    +
    D_l
    \left.
    \frac{\partial Y_l}{\partial n}
    \right|_{\Gamma}
    =
    0,
    \qquad
    \bm x\in\Gamma .
    \label{eq:solute_interface}
\end{equation}
Equation~\eqref{eq:solute_interface} represents solute rejection at the phase-change front and constitutes a moving-boundary Robin condition, in which the unknown interfacial concentration is coupled simultaneously to the local phase-change velocity and the liquid-side concentration gradient.

Equations~\eqref{eq:gibbs_thomson}--\eqref{eq:solute_interface} expose the essential numerical difficulty of binary phase change: $T_\Gamma$, $Y_{l,\Gamma}$ and $U_m$ cannot be determined independently. The local concentration determines $T_\Gamma$ through Eq.~\eqref{eq:gibbs_thomson}; $T_\Gamma$ then provides the boundary condition for the thermal problem, whose one-sided fluxes determine $U_m$ through Eq.~\eqref{eq:stefan}; the resulting $U_m$, in turn, enters Eq.~\eqref{eq:solute_interface} and modifies $Y_{l,\Gamma}$. The interfacial closure therefore forms the closed thermosolutal loop $Y_{l,\Gamma}\rightarrow T_\Gamma\rightarrow\left.\nabla T\right|_\Gamma\rightarrow U_m\rightarrow\left.\nabla Y_l\right|_\Gamma\rightarrow Y_{l,\Gamma}$, which must be satisfied consistently on the moving interface.

Free rigid-body motion further embeds this local thermosolutal loop into a larger FSI--phase-change feedback. The rigid-body velocity $\bm U_s$ enters the liquid boundary condition \eqref{eq:velocity_interface} and advects the solid temperature through Eq.~\eqref{eq:temperature}; the surrounding flow determines the thermal and solutal boundary layers and hence the local phase-change velocity $U_m$; meanwhile, temperature and concentration modify the liquid motion through the buoyancy term in Eq.~\eqref{eq:NS}, and the resulting phase-change-induced evolution of $\Gamma$ alters the hydrodynamic force and torque governing $\bm v_s$ and $\bm\omega_s$. The global coupling may therefore be summarized as $\Gamma\rightarrow(\bm u_l,p,T_l,Y_l)\rightarrow(\bm F_H,\bm T_H)\rightarrow(\bm v_s,\bm\omega_s)\rightarrow\Gamma$, with the local thermosolutal closure embedded at the interface. Consequently, the present problem cannot be advanced accurately by treating flow, rigid-body motion, scalar transport and phase change as independent sequential subproblems. Instead, the interfacial quantities, bulk fields and rigid-body state must remain mutually consistent within each physical time step, which motivates the coupled iterative strategy developed in \S~\ref{sec:numerical_methods}.

\subsection{Numerical challenges}
\label{sec:numerical_challenges}

The formulation in \S\S~\ref{sec:configuration}--\ref{sec:interface_conditions} shows that free rigid-body motion introduces three additional numerical challenges beyond the sharp binary phase-change problem.

\textit{Long-time geometry fidelity.}
The interface undergoes rigid-body translation/rotation together with local phase-change deformation. Accumulated transport errors may distort the body shape and thereby contaminate the hydrodynamic force, torque and subsequent motion. The same sharp geometry must therefore remain consistent throughout the coupled calculation.

\textit{Sharp and conservative transport on moving cut cells.}
As the interface crosses the Cartesian mesh, the physical phase volumes of cut cells vary continuously. Temperature and concentration must therefore be transported conservatively on the evolving geometry, since errors in the near-interface fields directly affect the thermal and solutal fluxes, and hence $U_m$. The solid temperature must additionally remain confined to the moving solid domain.

\textit{Strong multiphysics closure.}
The interface geometry, hydrodynamic loads, rigid-body motion and thermosolutal phase-change rate form a closed nonlinear feedback. These quantities must therefore be made mutually consistent within each physical time step.

These three challenges motivate Techniques~I--III developed in the following section.

\section{Numerical method}
\label{sec:numerical_methods}

The numerical method is developed on a fixed adaptive Cartesian mesh, with the solid--liquid interface $\Gamma$ represented by the zero level set of the signed-distance function $\phi$ introduced in \S~\ref{sec:configuration}. Throughout the computation, the same instantaneous interface geometry is used consistently for interface advancement, embedded-boundary discretization, interfacial flux evaluation and hydrodynamic-load calculation. Three ingredients are central to the present framework. First, the Level-Set representation is used as the primary carrier of the moving geometry, allowing long-distance rigid-body translation and rotation to be combined with local deformation induced by phase change. Second, sharp finite-volume operators are constructed on the corresponding moving embedded geometry, with locally conservative treatments for thermal and solutal transport. Third, the flow, interfacial thermosolutal closure, hydrodynamic loads and rigid-body dynamics are coupled within a common nonlinear iteration so that the geometry, transport fields and body motion remain consistent at the same physical time level. The implementation builds on our previous sharp binary phase-change solver \citep{xue2023three} and is developed within the adaptive quadtree/octree framework of \textit{Basilisk} \citep{popinet2009accurate,popinet2015quadtree}, with the additional formulations required for Level-Set geometry transport, free rigid-body motion and their strongly coupled moving-boundary-phase-change treatment developed in the present work.

\subsection{Overall discrete framework and coupled time advancement}
\label{sec:overall_algorithm}

Let the superscript $n$ denote the converged solution at time $t^n$, and let $r$ denote the Picard iteration used to determine the solution at $t^{n+1}=t^n+\Delta t$. The quantities entering the nonlinear coupling are collected as
$\mathcal{Q}=(T_\Gamma,Y_{l,\Gamma},U_m,\bm v_s,\bm\omega_s)$.
Their $r$th approximation at the new time level is denoted by $\mathcal{Q}^{n+1,r}$, with $\mathcal{Q}^{n+1,0}=\mathcal{Q}^{n}$ used to initialize each physical time step.
A key feature of the present coupling is that all Picard iterations advance the solution over the same physical interval $[t^n,t^{n+1}]$. At the beginning of each iteration, the transient fields $(\bm u_l,p,T_l,T_s,Y_l,\phi)$ are restored to their converged states at $t^n$ and advanced again over $\Delta t$ using the latest estimate $\mathcal{Q}^{n+1,r-1}$. Hence, the iteration corrects only the nonlinear coupling, rather than repeatedly advancing the transient equations in pseudo-time. Upon convergence, the interface geometry, Eulerian fields and rigid-body variables therefore all correspond to the same physical time $t^{n+1}$.

The resulting advancement is summarized in Algorithm~\ref{alg:overall}. For a given estimate $\mathcal{Q}^{n+1,r-1}$, the sharp interface $\Gamma^{n+1}$ is first advanced from $\Gamma^n$ using
$\bm V_\Gamma=\bm U_s+U_m\bm n$, after which the corresponding embedded geometry is reconstructed.
On this updated geometry, the liquid concentration is advanced first and the interfacial concentration
$Y_{l,\Gamma}^{n+1,r}$ is obtained from the sharp solute-rejection condition. The liquidus relation then gives
$T_\Gamma^{n+1,r}$, which is imposed as the sharp interfacial condition for advancing the liquid and solid
temperature fields. Their one-sided heat fluxes determine the updated melting velocity
$U_m^{n+1,r}$ through the Stefan condition. The incompressible flow equations are subsequently solved
using the updated thermal and solutal fields together with the moving-boundary condition on the same
reconstructed interface. Finally, the resulting hydrodynamic force and torque are evaluated and used to
update $\bm v_s^{n+1,r}$ and $\bm\omega_s^{n+1,r}$. The procedure is repeated until the interfacial and
rigid-body quantities have converged.

The overall advancement may thus be written schematically as
\[
\begin{array}{rcl}
\mathcal{Q}^{n+1,r-1}
\;\longrightarrow\;
\Gamma^{n+1,r}
\;\longrightarrow\;
Y_l^{n+1,r}
\;\longrightarrow\;
Y_{l,\Gamma}^{n+1,r}
& \longrightarrow &
T_\Gamma^{n+1,r}
\\[-0.1em]
&& \Big\downarrow
\\[-0.1em]
(\bm v_s,\bm\omega_s)^{n+1,r}
\;\longleftarrow\;
(\bm F_H,\bm T_H)^{n+1,r}
\;\longleftarrow\;
(\bm u_l,p)^{n+1,r}
\;\longleftarrow\;
U_m^{n+1,r}
& \longleftarrow &
(T_l,T_s)^{n+1,r}.
\end{array}
\]
Convergence is reached when the componentwise relative variations of
$\mathcal{Q}$ between two successive Picard iterations fall below a
prescribed tolerance $\varepsilon$, which is set to $10^{-3}$ in the present simulations. The converged interface, Eulerian
fields and rigid-body variables are then assigned to time level
$t^{n+1}$.

\begin{algorithm}[t]
\caption{Coupled advancement from $t^n$ to $t^{n+1}$.}
\label{alg:overall}
\begin{algorithmic}[1]
\State Given the converged state
$(\bm u_l,p,T_l,T_s,Y_l,\phi,\mathcal Q)^n$.
\State Set $\mathcal Q^{n+1,0}=\mathcal Q^n$ and $r=0$.
\Repeat
    \State $r\leftarrow r+1$.
    \State Restore
    $(\bm u_l,p,T_l,T_s,Y_l,\phi)
    \leftarrow
    (\bm u_l,p,T_l,T_s,Y_l,\phi)^n$.
    \State Advance $\phi^n$ using
    $\bm V_\Gamma^{n+1,r-1}
    =\bm U_s^{n+1,r-1}
    +U_m^{n+1,r-1}\bm n$
    and reconstruct $\Gamma^{n+1,r}$ and the associated embedded geometry.
    \State Advance $Y_l$ on $\Omega_l^{n+1,r}$ and determine
    $Y_{l,\Gamma}^{n+1,r}$ from Eq.~\eqref{eq:solute_interface}.
    \State Set
    $T_\Gamma^{n+1,r}
    =T_m+m_LY_{l,\Gamma}^{n+1,r}$
    from Eq.~\eqref{eq:gibbs_thomson}, and advance
    $T_l$ and $T_s$.
    \State Evaluate $U_m^{n+1,r}$ from Eq.~\eqref{eq:stefan}.
    \State Solve the incompressible flow equations on
    $\Omega_l^{n+1,r}$ using the updated thermal and solutal fields
    and the moving-boundary condition of Eq.~\eqref{eq:velocity_interface}.
    \State Evaluate
    $(\bm F_H,\bm T_H)^{n+1,r}$
    and update
    $(\bm v_s,\bm\omega_s)^{n+1,r}$.
\Until{$\|\mathcal Q^{n+1,r}-\mathcal Q^{n+1,r-1}\| / \| \mathcal Q^{n+1,r-1}\|<\epsilon$}
\State Assign the converged solution to time level $n+1$.
\end{algorithmic}
\end{algorithm}

The following subsections detail the three numerical treatments required
by this coupled advancement: long-time transport of the moving sharp
geometry, sharp and conservative transport on the evolving embedded
geometry, and sharp hydrodynamic-load evaluation coupled to the
rigid-body update. Note that these techniques describe the individual numerical components invoked within each Picard sweep of Algorithm~\ref{alg:overall}. To avoid cumbersome notation, the Picard index $r$ is suppressed unless it is required explicitly to describe the nonlinear update. Thus, quantities at the new or intermediate time levels should be understood as the current Picard estimate, e.g. $\chi^{n+1}\equiv\chi^{n+1,r}$ and $\chi^{n+\frac12}\equiv\chi^{n+\frac12,r}$, whereas the superscript $n$ always denotes the converged state restored at the beginning of each sweep.

\subsection{Technique I: Long-time transport of the freely moving sharp interface}
\label{sec:interface_transport}

The first numerical task is to transport the solid geometry accurately under the combined action of rigid-body motion and phase change. The signed-distance function $\phi$ introduced in \S~\ref{sec:configuration} is used as the primary geometrical representation, and its zero level set defines the solid--liquid interface. During each Picard iteration, $\phi$ is advanced from its converged state at $t^n$ according to
\begin{equation}
    \frac{\partial \phi}{\partial t}
    +
    \bm V_\Gamma^{\rm ext}\cdot\nabla\phi
    =
    0,
    \label{eq:levelset_advection}
\end{equation}
where $\bm V_\Gamma^{\rm ext}$ is a narrow-band extension of the geometrical interface velocity $\bm V_\Gamma$ defined in Eq.~\eqref{eq:interface_velocity}. The rigid-body contribution translates and rotates the existing geometry, whereas $U_m\bm n$ produces the local deformation associated with phase change. Both are therefore incorporated into the same geometrical transport of the zero level set.

The Level-Set advection and reinitialization follow the formulations of Peng \textit{et al}.~\citep{peng1999pde} and Limare \textit{et al}.~\citep{limare2023hybrid}. Since $U_m$, and hence $\bm V_\Gamma$, is known initially only at the reconstructed interface, its velocity is extended to a narrow band before solving Eq.~\eqref{eq:levelset_advection}. With the sign convention $\phi>0$ in $\Omega_l$, $\phi<0$ in $\Omega_s$ and $\bm n$ pointing from liquid to solid, $\bm n=-\nabla\phi/|\nabla\phi|$. The normal extension is obtained from
\begin{equation}
    \frac{\partial \bm V_\Gamma^{\rm ext}}{\partial \tau}
    -
    \delta_e S(\phi)\bm n\cdot\nabla\bm V_\Gamma^{\rm ext}
    =
    0,
    \label{eq:velocity_extension}
\end{equation}
where $\tau$ is a fictitious time, $S(\phi)$ is the sign function, and $\delta_e$ is zero in interfacial cells and unity elsewhere. The interfacial values are therefore kept fixed while $\bm V_\Gamma^{\rm ext}$ is propagated approximately constantly along the Level-Set normal. After several pseudo-time sweeps, the values in the interfacial cells are corrected for the interpolation error between the cell centers and the interface centroids using quadratic Lagrange interpolation on a $3^{\mathcal D}$ stencil. The extended velocity $\bm V_\Gamma^{\rm ext}$ is used exclusively for geometrical transport; the liquid velocity imposed on the physical moving boundary remains $\bm u_{l,\Gamma}$ given by Eq.~\eqref{eq:velocity_interface}.

\begin{figure}
	\centering
	\includegraphics[width=0.6\textwidth]{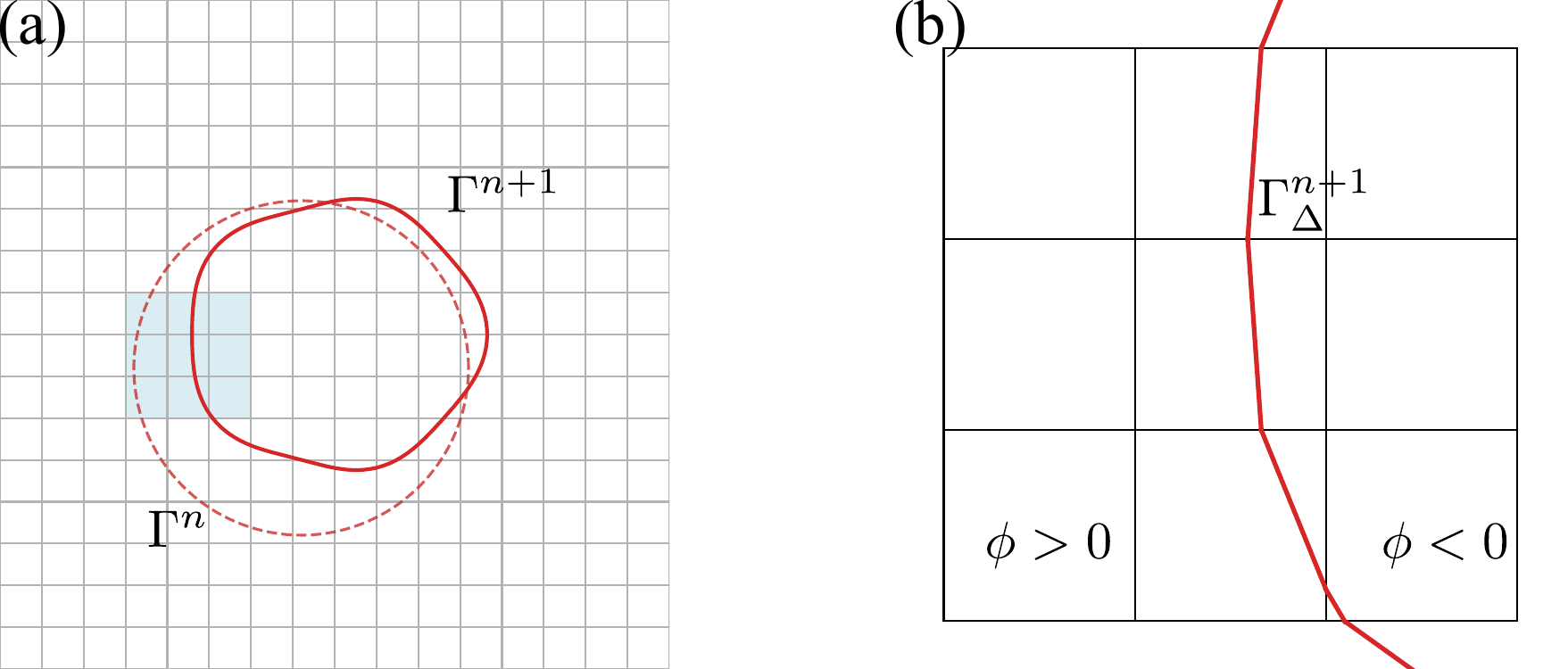}
	\caption{Schematic of the Level-Set method used for (a) interface transport, and (b) reconstruction of the melting interface remarked by the colored region in panel (a).}	
    \label{fig:level}
\end{figure}

After the interface-advection step, $\phi$ generally no longer satisfies the signed-distance condition $|\nabla\phi|=1$, although its zero level set has been transported to the new interface location. The distance property is therefore recovered by solving the Hamilton--Jacobi equation \citep{sussman1994level}
\begin{equation}
    \frac{\partial \phi}{\partial \tau}
    +
    S(\phi_0)
    \left(
        |\nabla\phi|-1
    \right)
    =
    0,
    \qquad
    \phi(\bm x,0)=\phi_0(\bm x),
    \label{eq:levelset_reinit}
\end{equation}
where $\phi_0$ denotes the advected Level-Set field. The redistancing follows the subcell-accurate procedure of Min and Gibou~\citep{min2007second}, in which the original zero-level-set position is explicitly incorporated into the near-interface stencil to minimize artificial interface displacement. Away from the interface, standard ENO-type differences are employed. The resulting signed-distance field provides regular geometric information for evaluating interface normals and reconstructing the sharp solid--liquid boundary. The advection of the interface from $\Gamma^{n}$ to $\Gamma^{n+1}$, followed by redistancing while preserving the zero-level-set location, is illustrated schematically in Fig.~\ref{fig:level}(a).

The Level-Set representation is adopted here specifically to address the long-distance rigid-body transport encountered for freely moving phase-changing particles. The VOF framework of Xue \textit{et al}.~\citep{xue2023three} provides conservative and accurate tracking of interfaces whose evolution is dominated by local phase change. For prolonged translation and rotation, however, repeated geometrical VOF reconstruction and advection in our implementation gradually accumulate shape errors even when the enclosed volume remains well preserved. As shown in \ref{sec:appA}, these errors eventually affect the reconstructed body geometry and hence the hydrodynamic loads and trajectory. The Level-Set representation is therefore adopted as the primary geometry carrier for long-duration rigid-body motion. As expected for a non-conservative interface representation, this improvement in geometrical fidelity is accompanied by somewhat poorer global mass conservation in the same tests. This trade-off is acceptable here because local conservation required by the thermal and solutal transport is recovered separately through the reconstructed embedded finite-volume formulation. This choice addresses the specific requirement considered here and does not imply a general superiority of Level Set over VOF.

After advection and redistancing, the zero level set of $\phi$ is reconstructed on the adaptive Cartesian mesh as a piecewise-planar sharp interface, denoted by $\Gamma_\Delta$, as illustrated in Fig.~\ref{fig:level}(b). From this geometry, the liquid volume fraction $c$, open-face fractions, interface area, centroid and normal are determined and supplied to the embedded-boundary finite-volume discretization. Importantly, these reconstructed volume fractions are geometric measures derived from the instantaneous Level-Set field rather than an independent interface tracker. Thus, the Level-Set field provides long-time geometrical fidelity, while the reconstructed embedded geometry supplies the local geometric information required for conservative and sharp discretization of flow, heat and solute transport \citep{johansen1998cartesian,xue2023three,limare2023hybrid}. The corresponding finite-volume operators are developed next.

\subsection{Technique II: Sharp and conservative transport on the moving cut-cell geometry}
\label{sec:transport}

Once the sharp interface $\Gamma_\Delta$ has been reconstructed, the velocity and scalar fields are solved in their respective physical subdomains. Despite their different physical meanings, the corresponding transport equations share a common convection--diffusion structure. Denoting by $\chi$ one component of $\bm u_l$, or one of the scalar fields $T_l$, $Y_l$ and $T_s$, the transport core may be written generically as
\begin{equation}
    \frac{\partial \chi}{\partial t}
    +
    \underbrace{
    \nabla\cdot(\bm u_\chi\chi)
    }_{\mathcal A_{\mathrm{adv}}}
    =
    \underbrace{
    \nabla\cdot(\lambda_\chi\nabla\chi)
    }_{\mathcal L_{\mathrm{dif}}}
    +
    \mathcal S_\chi ,
    \label{eq:generic_transport}
\end{equation}
where $\bm u_\chi=\bm u_l$ for the liquid velocity components, $T_l$ and $Y_l$, while $\bm u_\chi=\bm U_s$ for $T_s$. Accordingly, $\lambda_\chi$ denotes $\mu_l/\rho_l$, $\alpha_l$, $D_l$ or $\alpha_s$. For a velocity component, $\mathcal S_\chi$ contains the pressure-gradient and buoyancy terms, which are treated within the projection procedure described below, whereas $\mathcal S_\chi=0$ for the scalar equations.

The moving sharp boundary introduces three distinct discretization requirements. First, $\mathcal L_{\mathrm{dif}}$ requires accurate one-sided gradients at $\Gamma_\Delta$ while satisfying the different velocity, thermal and solutal boundary conditions. Second, $\mathcal A_{\mathrm{adv}}$ acts on cut-cell control volumes whose physical phase volumes vary as $\Gamma_\Delta$ moves across the fixed mesh; this geometric variation must enter the discrete balance to preserve local conservation. Third, the numerical support of $T_s$ must remain locked to the moving solid geometry rather than leak across $\Gamma_\Delta$. These requirements are addressed by Techniques~II-A, II-B and II-C, respectively. 

\subsubsection{Technique II-A: Sharp discretization of interfacial diffusion}
\label{sec:sharp_diffusion}

We first consider the diffusion operator $\mathcal L_{\mathrm{dif}}$ in Eq.~\eqref{eq:generic_transport}. The zero level set $\phi=0$ is reconstructed as a piecewise-planar interface $\Gamma_\Delta$, which partitions each interfacial Cartesian cell into liquid and solid portions. The key idea is to treat $\Gamma_\Delta$ as an internal boundary of the corresponding phase and impose the interfacial condition directly at its reconstructed location. The diffusion operator can therefore be evaluated separately on either side of the sharp interface without introducing a regularized volumetric representation.

\begin{figure}
    \centering
    \includegraphics[width=0.8\textwidth]{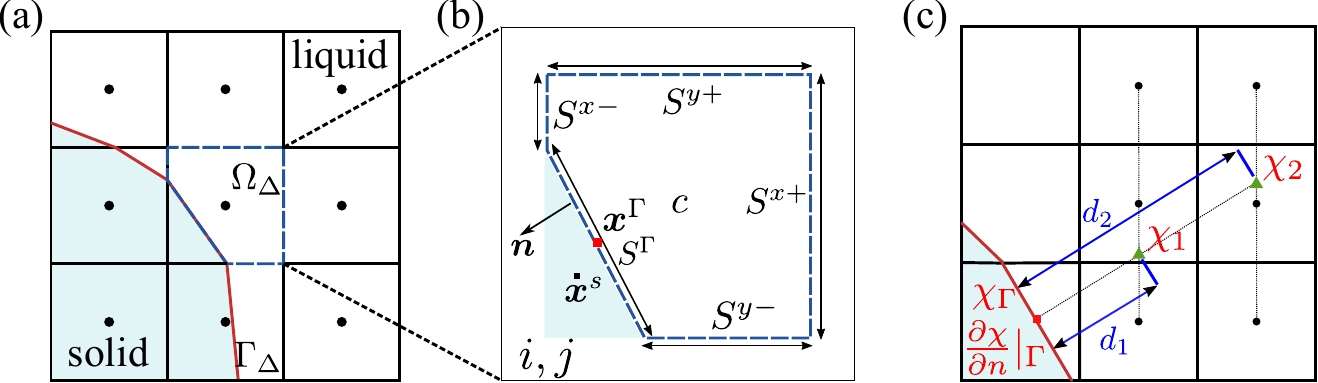}
    \caption{
    Embedded-boundary geometry and interfacial reconstruction in 2D.
    (a) A Cartesian cell $\Omega_\Delta$ cut by $\Gamma_\Delta$.
    (b) Geometric quantities of the resulting cut cell.
    (c) Reconstruction of $\chi_\Gamma$ and $\left.\partial\chi/\partial n\right|_\Gamma$ using auxiliary values $\chi_1$ and $\chi_2$ located at normal distances $d_1$ and $d_2$ from the interface centroid.
    }
    \label{fig:EBM}
\end{figure}

Consider a Cartesian cell $(i,j)$ of size $\Delta$ intersected by $\Gamma_\Delta$, as illustrated in Fig.~\ref{fig:EBM}(a,b). The liquid volume fraction is
$c_{i,j}=V_{i,j}^l/\Delta^{\mathcal D}$, where $\mathcal D=2$ or $3$ denotes the spatial dimension. For each Cartesian face $f$, $S_{i,j}^f=A_{i,j}^f/\Delta^{\mathcal D-1}$ denotes the open liquid-face fraction, while $S_{i,j}^\Gamma=A_{i,j}^\Gamma/\Delta^{\mathcal D-1}$ denotes the normalized area of the embedded facet. Its centroid is denoted by $\bm x_{i,j}^\Gamma$, and $\bm n_{i,j}$ points from liquid to solid, consistently with \S~\ref{sec:configuration}. The piecewise-planar interface normal is obtained using the Mixed-Youngs-Centred (MYC) reconstruction of Aulisa \textit{et al}.~\citep{aulisaGeometricalAreapreservingVolumeofFluid2003}. The corresponding solid-side quantities are obtained from the complementary portion of the same cut cell. These geometric measures are constructed within the embedded-boundary framework of \textit{Basilisk}~\citep{Popinet2026Basilisk}, following the Cartesian cut-cell formulation of Johansen and Colella~\citep{johansen1998cartesian}.

\textit{\textbf{Spatial discretization of diffusion.}}
For the liquid phase, integration of the diffusion operator over the physical portion of a cut cell gives
\begin{equation}
    \mathcal L_{l,\mathrm{dif}}(\chi)
    =
    \frac{1}{V^l}
    \int_{\Omega_\Delta^l}
    \nabla\cdot(\lambda_\chi\nabla\chi)\,\mathrm dV
    \simeq
    \frac{1}{c\Delta}
    \left[
        \sum_f S^f F_D^f
        +
        S^\Gamma F_D^\Gamma
    \right],
    \label{eq:cutcell_divergence}
\end{equation}
where $F_D^f=(\lambda_\chi\partial\chi/\partial n)_f$ and
$F_D^\Gamma=(\lambda_\chi\partial\chi/\partial n)_\Gamma$
are the outward normal diffusive-flux densities through the Cartesian faces and the embedded interface, respectively. Here and below, the cell indices $(i,j)$ are omitted for clarity. Equation~\eqref{eq:cutcell_divergence} forms the conservative finite-volume diffusion operator used throughout the present method. The fluxes through fully and partially open Cartesian faces are evaluated using the second-order embedded-interface reconstruction described in our previous studies~\citep{zhao2022boiling,xue2023three} and are not repeated here. The remaining key ingredient is the embedded-interface flux $F_D^\Gamma$, which depends explicitly on the physical boundary condition imposed on $\Gamma_\Delta$. The solid-side operator has the same form, with $c$ replaced by $(1-c)$ and the corresponding solid-side geometric quantities and outward normal.

As illustrated in Fig.~\ref{fig:EBM}(c), a normal line is constructed from the interface centroid $\bm x^\Gamma$ into the phase under consideration. Two auxiliary values, $\chi_1$ and $\chi_2$, are reconstructed at normal distances $d_1$ and $d_2$ from $\Gamma_\Delta$ using neighboring cell-centered values and the transverse interpolation employed in the embedded-boundary formulation of Johansen and Colella~\citep{johansen1998cartesian} and implemented in \textit{Basilisk}~\citep{popinet2015quadtree}. Once $\chi_\Gamma$ is specified, the triplet $(\chi_\Gamma,\chi_1,\chi_2)$ defines a quadratic reconstruction along the interface-normal direction, from which the one-sided derivative at $\Gamma_\Delta$ is obtained. The construction extends directly to three dimensions \citep{schwartz2006cartesian}. In the present problem, however, $\chi_\Gamma$ is not always prescribed: depending on the interfacial condition, it may either be known directly or determined together with the normal derivative. These two cases are treated below.

For a prescribed Dirichlet condition $\chi=\chi_\Gamma$ on $\Gamma_\Delta$, the one-sided normal derivative can be reconstructed as
\begin{equation}
    \left.
    \frac{\partial\chi}{\partial n}
    \right|_\Gamma
    =
    \frac{1}{d_2-d_1}
    \left[
        \frac{d_2}{d_1}
        \left(\chi_\Gamma-\chi_1\right)
        -
        \frac{d_1}{d_2}
        \left(\chi_\Gamma-\chi_2\right)
    \right]
    +
    \mathcal{O}(\Delta^2).
    \label{eq:dirichlet_gradient}
\end{equation}
Here Eq.~\eqref{eq:dirichlet_gradient} is written for the liquid-side reconstruction, for which the auxiliary points lie inside $\Omega_l$, i.e. opposite to the common normal $\bm n$ pointing from liquid to solid. The solid-side counterpart is obtained analogously by reversing the reconstruction direction while retaining the same normal convention when the gradient is used in the interfacial conditions. The corresponding embedded-boundary flux is
$F_D^\Gamma=(\lambda_\chi\partial\chi/\partial n)_\Gamma$.
Equation~\eqref{eq:dirichlet_gradient} is used to reconstruct independently the liquid- and solid-side temperature gradients entering the Stefan condition, as well as the velocity gradients required for evaluating the viscous stress. In these cases, the interfacial Dirichlet values are already determined by the coupled interfacial relations: $T_\Gamma$ follows from $Y_{l,\Gamma}$ through the equilibrium relation~\eqref{eq:gibbs_thomson}, while $\bm u_{l,\Gamma}$ is prescribed by the kinematic condition~\eqref{eq:velocity_interface}. The remaining task is therefore to recover the corresponding one-sided gradients from the phase-wise field values.

For solute transport, $Y_{l,\Gamma}$ is not prescribed independently but must be determined together with its normal gradient. To treat this condition within the same sharp reconstruction, consider the general Robin condition
\begin{equation}
    \left.
    \frac{\partial\chi}{\partial n}
    \right|_\Gamma
    +
    g_1\chi_\Gamma
    =
    g_2 .
    \label{eq:general_robin}
\end{equation}
Combining Eq.~\eqref{eq:general_robin} with the same quadratic normal reconstruction Eq.~\eqref{eq:dirichlet_gradient} gives
\begin{equation}
    \chi_\Gamma
    =
    \frac{
        d_2^2\chi_1-d_1^2\chi_2
        +
        d_1d_2(d_2-d_1)g_2
    }{
        d_2^2-d_1^2
        +
        d_1d_2(d_2-d_1)g_1
    }
    +
    \mathcal{O}(\Delta^2),
    \label{eq:robin_value}
\end{equation}
and
\begin{equation}
    \left.
    \frac{\partial\chi}{\partial n}
    \right|_\Gamma
    =
    \frac{
        (d_2^2-d_1^2)g_2
        -
        (d_2^2\chi_1-d_1^2\chi_2)g_1
    }{
        d_2^2-d_1^2
        +
        d_1d_2(d_2-d_1)g_1
    }
    +
    \mathcal{O}(\Delta^2).
    \label{eq:robin_gradient}
\end{equation}
For the NaCl--water system, the solute-rejection condition~\eqref{eq:solute_interface} corresponds to $g_1=U_m/D_l$ and $g_2=0$. Equations~\eqref{eq:robin_value} and \eqref{eq:robin_gradient} therefore provide simultaneously $Y_{l,\Gamma}$ and its liquid-side normal gradient $(\left.\partial Y_l/\partial n\right)|_\Gamma$.

The significance of the present reconstruction extends beyond the construction of individual boundary stencils. All quantities entering the binary phase-change closure are evaluated consistently on the same instantaneous interface $\Gamma_\Delta$. Specifically, $T_\Gamma$ provides the phase-wise thermal boundary value, the independently reconstructed gradients $(\left.\partial T_l/\partial n\right)|_\Gamma$ and $(\left.\partial T_s/\partial n\right)|_\Gamma$ determine $U_m$ through Eq.~\eqref{eq:stefan}, while the Robin reconstruction yields simultaneously $Y_{l,\Gamma}$ and $(\left.\partial Y_l/\partial n\right)|_\Gamma$. The thermosolutal closure defined in \S~\ref{sec:interface_conditions} is thus enforced directly and consistently on a single geometrically sharp interface, without spreading the latent-heat or solute-rejection conditions over a finite numerical thickness.

\textit{\textbf{Temporal integration of diffusion.}}
The sharp spatial discretization above is combined with a temporally robust treatment of diffusion on the reconstructed cut-cell geometry. A first-order implicit-Euler scheme provides strong damping but sacrifices temporal accuracy, whereas conventional second-order schemes such as Crank--Nicolson may exhibit non-physical oscillations for stiff diffusion operators and rapidly varying near-interface fields. We therefore employ the second-order $L_0$-stable Twizell--Gumel--Arigu (TGA) scheme \citep{twizellSecondorderL0stableMethods1996,mccorquodaleCartesianGridEmbedded2001}. Denoting by $\chi^{\rm adv}$ the state obtained after advection, the diffusion update is performed in two implicit stages,
\begin{equation}
    \frac{\chi^\star-\chi^{\rm adv}}{\Delta t}
    =
    \mu_2\mathcal L_{\mathrm{dif}}(\chi^\star)
    +
    \mu_3\mathcal L_{\mathrm{dif}}(\chi^{\rm adv}),
    \label{eq:TGA_first}
\end{equation}
followed by
\begin{equation}
    \frac{\chi^{n+1}-\chi^\star}{\Delta t}
    =
    \mu_1\mathcal L_{\mathrm{dif}}(\chi^{n+1}),
    \label{eq:TGA_second}
\end{equation}
where
\begin{equation}
    \mu_1=
    \frac{a-\sqrt{a^2-4a+2}}{2},
    \qquad
    \mu_2=
    \frac{a+\sqrt{a^2-4a+2}}{2},
    \qquad
    \mu_3=1-a,
    \label{eq:TGA_coefficients}
\end{equation}
with $a=2-\sqrt{2}-\epsilon$, where $\epsilon$ denotes machine precision. These coefficients satisfy $\mu_1+\mu_2+\mu_3=1$.
At each Picard iteration, the same sharp embedded-boundary diffusion operator constructed on the current interface geometry is used in both implicit stages, with the corresponding Dirichlet or Robin condition imposed directly on $\Gamma_\Delta$. The resulting Poisson-type linear systems are solved using a multigrid method. Hence, the TGA treatment provides a second-order, strongly damped temporal integration of the sharp diffusion operator without altering its embedded-boundary treatment.

The remaining quantity, $\chi^{\rm adv}$, must be obtained conservatively while $\Gamma_\Delta$ sweeps across the fixed Cartesian mesh. The resulting moving-control-volume advection is addressed by Technique~II-B in \S~\ref{sec:conservative_advection}.

\subsubsection{Technique II-B: Consistent advection with a moving embedded-boundary}
\label{sec:conservative_advection}

We next consider the advection operator $\mathcal A_{\mathrm{adv}}$ in Eq.~\eqref{eq:generic_transport}. For the liquid-phase fields $\bm u_l$, $T_l$ and $Y_l$, the physical control volume changes as $\Gamma_\Delta$ sweeps across the fixed Cartesian mesh. A conservative update must therefore account for both the convective fluxes through the instantaneous cell boundary and the amount of transported quantity associated with the changing liquid volume. Omitting the latter produces a transport imbalance even when all face fluxes are evaluated conservatively.

Consider a liquid cut cell whose volume fraction changes from $c^n$ to $c^{n+\frac12}$ during the advection update. For a generic liquid-phase quantity $\chi$, the corresponding operator is written as
\begin{equation}
    \mathcal A_{\mathrm{adv}}^{\,n+\frac12}(\chi)
    =
    \frac{
        c^{n+\frac12}-c^n
    }{
        c^{n+\frac12}
    }
    \frac{
        \chi^n-\chi_\Gamma^{\,n+\frac12}
    }{
        \Delta t
    }
    +
    \frac{1}{
        c^{n+\frac12}\Delta
    }
    \left[
        \sum_f
        S^f F_C^f
        +
        S^\Gamma F_C^\Gamma
    \right]^{n+\frac12}.
    \label{eq:moving_conservative_advection}
\end{equation}
Here, $F_C^f$ and $F_C^\Gamma$ denote the convective-flux densities through the Cartesian faces and the embedded boundary, respectively. For a scalar field,
$F_C^f=(\bm u_l^f\cdot\bm n_f)\chi^f$ and
$F_C^\Gamma=(\bm u_{l,\Gamma}\cdot\bm n)\chi_\Gamma$,
where $\bm n_f$ is the outward normal of Cartesian face $f$. The face states $\chi^f$ are reconstructed using the second-order upwind Bell--Colella--Glaz scheme \citep{bell1989second}, with the open-face fraction $S^f$ accounting for partial faces. At $\Gamma_\Delta$, $\chi_\Gamma$ is supplied by the corresponding sharp interfacial condition developed in Technique~II-A; for momentum, the boundary state is given by $\bm u_{l,\Gamma}$ from Eq.~\eqref{eq:velocity_interface}.

The two contributions in Eq.~\eqref{eq:moving_conservative_advection} have distinct roles. The second is the conventional finite-volume divergence of the convective fluxes over the instantaneous liquid control volume, whereas the first is a swept-volume correction associated solely with the motion of $\Gamma_\Delta$. It is the discrete counterpart of the geometric conservation requirement for the time-dependent phase volume. Because $c$, $S^f$, $S^\Gamma$ and $\chi_\Gamma$ are all evaluated from the same reconstructed interface, Eq.~\eqref{eq:moving_conservative_advection} maintains consistency between field transport and interface motion.

Although Eq.~\eqref{eq:moving_conservative_advection} provides the natural conservative discretization, its direct application to momentum becomes numerically delicate in severely cut cells. When $c\ll1$, division by the small liquid volume amplifies the local convective update, making the scheme susceptible to the classical small-cell instability of Cartesian cut-cell methods. For $\bm u_l$, we therefore adopt the stabilization proposed in Refs.~\citep{ghigo:hal-03948786,sverdrup2019embedded}, in which the conservative operator is locally replaced by a weighted neighborhood average,
\begin{equation}
    \widetilde{\mathcal A}_{\mathrm{adv}}^{\,n+\frac12}(\bm u_l)
    =
    \frac{
        \displaystyle
        \sum_{k\in\mathcal N}
        \left(c_k^{n+\frac12}\right)^2
        \mathcal A_{\mathrm{adv},k}^{\,n+\frac12}(\bm u_l)
    }{
        \displaystyle
        \sum_{k\in\mathcal N}
        \left(c_k^{n+\frac12}\right)^2
    },
    \label{eq:nonconservative_average}
\end{equation}
where $\mathcal N$ denotes the neighboring-cell stencil. The momentum-advection operator is then renewed by
\begin{equation}
    \mathcal A_{\mathrm{adv}}^{n+\frac12}(\bm u_l)
    =
    \eta\,
    \mathcal A_{\mathrm{adv}}^{\,n+\frac12}(\bm u_l)
    +
    (1-\eta)\,
    \widetilde{\mathcal A}_{\mathrm{adv}}^{\,n+\frac12}(\bm u_l),
    \label{eq:momentum_hybrid}
\end{equation}
where $\eta$ is determined from the local cut-cell configuration. Away from severely cut cells, $\eta\rightarrow1$ and the conservative discretization is recovered, whereas the smoothed contribution is activated locally as $c$ becomes small, thereby limiting excessive momentum updates.

This stabilization is appropriate for momentum but is not used for $T_l$ or $Y_l$, for which strict local conservation is substantially more important. The reason lies in how the local discretization error enters the phase-change closure. For momentum, the neighborhood averaging in Eq.~\eqref{eq:momentum_hybrid} introduces only a localized departure from strict conservation while providing a stable velocity field. For the scalar fields, however, a local conservation defect directly perturbs the thermal or solutal boundary layer adjacent to $\Gamma_\Delta$. The resulting error propagates into $(\left.\partial T_l/\partial n\right)|_\Gamma$, $Y_{l,\Gamma}$ and $(\left.\partial Y_l/\partial n\right)|_\Gamma$, and therefore directly contaminates the Stefan and solute-rejection balances that determine $U_m$.

This sensitivity is particularly relevant to the present binary phase-change problem, especially at relatively large Prandtl number and Schimit number, for which the thermal and solutal boundary layers may be considerably thinner than the hydrodynamic one. As demonstrated in \ref{sec:ncon}, applying the same conservative/non-conservative blending to the scalar equations produces measurable local heat- and solute-conservation errors and consequently appreciable errors in the interfacial fluxes, even when the corresponding bulk fields remain qualitatively reasonable. Therefore, $T_l$ and $Y_l$ retain the fully conservative moving-embedded-interface operator $\mathcal A_{\mathrm{adv}}^{n+\frac12}$ of Eq.~\eqref{eq:moving_conservative_advection}.
For these scalar fields, the embedded-boundary convective flux
$F_C^\Gamma=(\bm u_{l,\Gamma}\cdot\bm n)\chi_\Gamma$
requires the corresponding sharp interfacial state of $\chi_\Gamma$. For $T_l$, this state is $T_\Gamma$, determined from the interfacial equilibrium relation~\eqref{eq:gibbs_thomson}; for $Y_l$, the unknown $Y_{l,\Gamma}$ is obtained from the Robin reconstruction~\eqref{eq:robin_value}. Thus, the variation of the cut-cell volume, the scalar amount swept by the moving interface and the interfacial convective flux are all evaluated consistently from the same instantaneous geometry and interfacial state. The conservative advection step then gives
\begin{equation}
    \frac{\chi^{\rm adv}-\chi^n}{\Delta t}
    =
    -\mathcal A_{\mathrm{adv}}^{\,n+\frac12}(\chi),
    \label{eq:scalar_advection_update}
\end{equation}
where $\chi^{\rm adv}$ provides the input state for the subsequent TGA diffusion update in Eqs.~\eqref{eq:TGA_first} and \eqref{eq:TGA_second}.

The different treatments of momentum and scalar advection are therefore deliberate. Both originate from the same moving-control-volume balance of Eq.~\eqref{eq:moving_conservative_advection}, but momentum permits a localized relaxation of conservation to suppress the small-cell instability, whereas $T_l$ and $Y_l$ retain strict local conservation because their near-interface errors enter directly into the thermosolutal phase-change closure.
Finally, the conservative liquid-scalar formulation alone does not ensure a geometry-consistent transport of $T_s$. Since $T_s$ is advected by the rigid-body velocity $\bm U_s$, its numerical support must remain locked to the translating and rotating solid domain without leakage across $\Gamma_\Delta$. This additional constraint is addressed by Technique~II-C.

\subsubsection{Technique II-C: Geometry-consistent transport of the solid temperature}
\label{sec:solid_temperature}

The conservative formulation in \S~\ref{sec:conservative_advection} is sufficient for the liquid scalars $T_l$ and $Y_l$, which are transported within the liquid domain by $\bm u_l$. The solid temperature $T_s$ is subject to an additional geometrical constraint: its numerical support must translate and rotate together with the solid body. A conservative update of the scalar amount alone does not guarantee this property near a moving sharp boundary.

Direct Eulerian advection of the cell-centered $T_s$ field may smear its support across $\Gamma_\Delta$ as the solid translates and rotates, thereby producing temperature values in cells that no longer belong to the solid. To preserve the consistency between the transported temperature field and the instantaneous solid geometry, we instead advect $T_s$ as a conservative tracer attached to the reconstructed solid volume.
Let $c$ denote the liquid volume fraction reconstructed from the instantaneous level-set geometry and $c_s=1-c$ the corresponding solid fraction. We introduce the solid-supported temperature tracer
\begin{equation}
    E_s=c_sT_s ,
    \qquad
    \frac{\partial E_s}{\partial t}
    +
    \nabla\cdot(\bm U_sE_s)
    =
    0,
    \label{eq:solid_geometric_advection}
\end{equation}
for the rigid-body advection step, while thermal diffusion is treated separately using the sharp operator developed in \S~\ref{sec:sharp_diffusion}. Importantly, $c_s$ is a geometric measure reconstructed from the level-set interface rather than an independently evolved interface variable.

\begin{figure}
	\centering
	\includegraphics[width=0.6\textwidth]{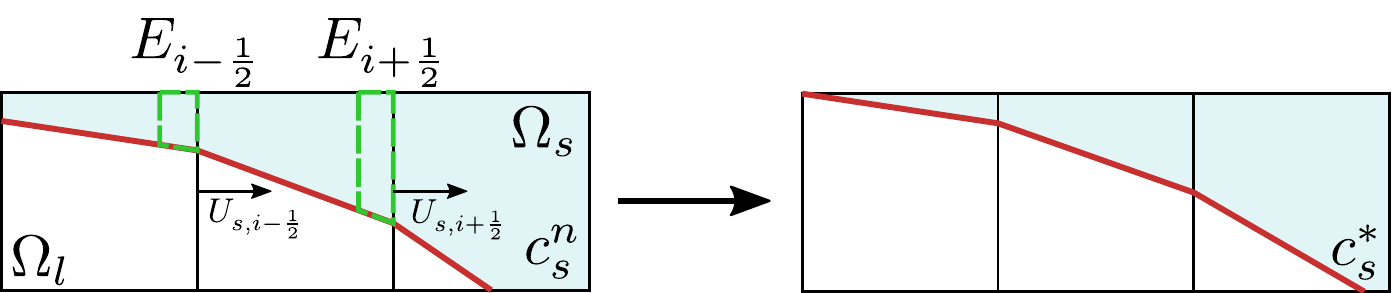}
	\caption{Schematic of the geometry-consistent transport of the solid temperature tracer $E_{s} = c_sT_s$ using the reconstructed solid volume.}	
    \label{fig:ts-transport}
\end{figure}

The advection of $E_s$ follows the directionally split geometric-VOF formulation used by Zhao \textit{et al}.~\citep{zhao2022boiling}. For illustration, in two dimensions the successive $x$- and $y$-sweeps read
\begin{subequations}
\label{eq:solid_split_advection}
\begin{align}
    \frac{
        E_{s,i,j}^{\ast}-E_{s,i,j}^{n}
    }{\Delta t}
    +
    \frac{
        \mathcal{F}_{i+\frac12,j}^{\,n}
        -
        \mathcal{F}_{i-\frac12,j}^{\,n}
    }{\Delta x}
    &=
    E_{s,c}
    \frac{
        U_{s,x,i+\frac12,j}^{\,n}
        -
        U_{s,x,i-\frac12,j}^{\,n}
    }{\Delta x},
    \label{eq:solid_split_x}
    \\
    \frac{
        E_{s,i,j}^{\ast\ast}-E_{s,i,j}^{\ast}
    }{\Delta t}
    +
    \frac{
        \mathcal{G}_{i,j+\frac12}^{\,\ast}
        -
        \mathcal{G}_{i,j-\frac12}^{\,\ast}
    }{\Delta y}
    &=
    E_{s,c}
    \frac{
        U_{s,y,i,j+\frac12}^{\,n}
        -
        U_{s,y,i,j-\frac12}^{\,n}
    }{\Delta y}.
    \label{eq:solid_split_y}
\end{align}
\end{subequations}
Here, $\mathcal{F}$ and $\mathcal{G}$ denote the conservative tracer fluxes associated with the geometrically swept solid volumes in the $x$- and $y$-directions, respectively. The corresponding face values of $T_s$ are reconstructed using the same geometric-tracer procedure, ensuring that the transport of $E_s$ remains consistent with that of the solid volume. The treatment for the $x$-directional sweep is illustrated schematically in Fig.~\ref{fig:ts-transport}. The term involving $E_{s,c}$ is the directional-splitting correction required by the geometric tracer formulation, as proposed by Weymouth and Yue \citep{weymouth2010conservative}. An additional $z$-sweep gives the three-dimensional extension.

After the directional sweeps, the solid temperature is recovered as
$T_s=E_s/c_s$ wherever $c_s>0$. Since the fluxes of $E_s$ are constructed from the same geometrically swept solid volumes that determine the evolution of $c_s$, the temperature field and the solid geometry remain consistent at the discrete level. As the body translates or rotates across the Cartesian mesh, the solid-temperature tracer is transferred together with the swept solid volume. The cut-cell volume variation is therefore incorporated intrinsically into the geometric transport, without requiring the additional swept-volume correction used for the liquid scalars in Eq.~\eqref{eq:moving_conservative_advection}.
It should be stressed that $c_s$ is not an independently advected VOF representation of the solid--liquid interface. The interface is transported exclusively by the Level-Set field $\phi$ in Technique~I; $c_s$ and the associated geometric fluxes are reconstructed from the instantaneous Level-Set geometry and serve only as conservative geometric measures for transporting $E_s$. This division of roles combines the long-time geometric fidelity of the Level-Set representation with a geometry-consistent conservative transport of $T_s$.

After advection, solid-phase thermal diffusion is advanced using the sharp cut-cell operator of Technique~II-A, with the liquid fraction $c$ replaced by $c_s=1-c$. The prescribed $T_\Gamma$ and the corresponding solid-side normal gradient are therefore evaluated on the same instantaneous interface $\Gamma_\Delta$. Technique~II-C thus ensures not only conservative transport of the solid temperature, but also that its numerical support remains locked to the translating and rotating solid geometry.

\textit{\textbf{Completion of the transport update.}}
After Techniques~II-A$\sim$II-C have been applied, the temperature and concentration fields are advanced over the current Picard sweep, yielding $T_l^{n+1,r}$, $T_s^{n+1,r}$ and $Y_l^{n+1,r}$ on the reconstructed sharp geometry $\Gamma_\Delta^{n+1,r}$. The momentum equation requires an additional projection step. The advection and viscous-diffusion treatments provide a provisional velocity $\bm u_l^{\star\star}$, which does not in general satisfy the discrete incompressibility constraint. Following the approximate projection method used in \textit{Basilisk}~\citep{popinet2009accurate} and in our previous sharp-interface formulations~\citep{zhao2022boiling,xue2023three}, the pressure is obtained from
\begin{equation}
    \nabla^2 p^{n+1,r}
    =
    \frac{\rho_l}{\Delta t}
    \nabla\cdot\bm u_l^{f,\star\star},
    \label{eq:pressure_projection}
\end{equation}
with a homogeneous Neumann condition for $p^{n+1,r}$ on $\Gamma_\Delta^{n+1,r}$, while the provisional velocity satisfies the prescribed moving-boundary condition. The cell- and face-centered velocities are then corrected as
\begin{equation}
    \bm u_l^{n+1,r}
    =
    \bm u_l^{\star\star}
    -
    \frac{\Delta t}{\rho_l}\nabla p^{n+1,r},
    \qquad
    \bm u_l^{f,n+1,r}
    =
    \bm u_l^{f,\star\star}
    -
    \frac{\Delta t}{\rho_l}\nabla^f p^{n+1,r},
    \label{eq:velocity_projection}
\end{equation}
so that the corrected face velocity satisfies the discrete divergence-free constraint. Technique~II therefore advances the Eulerian fields $(\bm u_l,p,T_l,T_s,Y_l)$ consistently on the instantaneous sharp geometry within each Picard iteration. The complete sequence is summarized in Algorithm~\ref{alg:transport_update}.

\begin{algorithm}
\caption{Technique II: Eulerian-field update on the moving sharp geometry.}
\label{alg:transport_update}
\begin{algorithmic}[1]

\Require
$\Gamma_\Delta^{n+1,r}$ and the associated cut-cell geometry;
interfacial states $(T_\Gamma,Y_{l,\Gamma},\bm u_{l,\Gamma})$;
rigid-body velocity $\bm U_s$;
and the restored fields at $t^n$.

\State
Construct the sharp embedded-boundary operators for
$\bm u_l$, $T_l$, $T_s$ and $Y_l$.
\Comment{Technique II-A}

\State
Advance momentum advection using the stabilized
moving-cut-cell operator.

\State
Advance $T_l$ and $Y_l$ using the fully conservative
moving-cut-cell operator.
\Comment{Technique II-B}

\State
Apply the sharp TGA diffusion update to
$\bm u_l$, $T_l$ and $Y_l$.

\State
Advect $E_s=(1-c)T_s$ geometrically, and apply the sharp solid-side diffusion update.
\Comment{Technique II-C}

\State
Obtain the provisional velocity $\bm u_l^{\star\star}$.

\State
Solve Eq.~\eqref{eq:pressure_projection} for $p^{n+1,r}$
and project $\bm u_l^{\star\star}$ using
Eq.~\eqref{eq:velocity_projection}.

\State
Return
$(\bm u_l,p,T_l,T_s,Y_l)^{n+1,r}$.

\end{algorithmic}
\end{algorithm}

\subsection{Technique III: Sharp hydrodynamic loads and strongly coupled rigid-body update}
\label{sec:rigid_body_update}

Techniques~I and II determine the instantaneous sharp geometry and the corresponding Eulerian fields for a given rigid-body state. Technique~III closes the mechanical feedback from the fluid to the freely moving solid. The hydrodynamic force and torque update $(\bm v_s,\bm\omega_s)$, which in turn modify both the interfacial fluid velocity $\bm u_{l,\Gamma}$ and the geometrical velocity $\bm V_\Gamma$, thereby feeding back into the interface geometry, transport fields and phase-change rate. The loads are therefore evaluated on the same reconstructed interface and incorporated into the fixed-time nonlinear iteration of Algorithm~\ref{alg:overall}. 

\textit{\textbf{Sharp evaluation of hydrodynamic loads.}}
Once $(\bm u_l^{n+1, r},p^{n+1, r})$ has been obtained from Technique~II, the hydrodynamic force and torque defined in \S~\ref{sec:governing_equations} are evaluated directly on $\Gamma_\Delta$. Since the reconstructed interface consists of linear segments in 2D or planar facets in 3D, the surface integrals are discretized as
\begin{equation}
    \bm F_H
    \simeq
    \sum_{m\in\Gamma_\Delta}
    \left[
        p_m\bm I-\bm\tau_m
    \right]\cdot\bm n_m\,A_m^\Gamma,
    \qquad
    \bm T_H
    \simeq
    \sum_{m\in\Gamma_\Delta}
    \bm r_m\times
    \left\{
        \left[
            p_m\bm I-\bm\tau_m
        \right]\cdot\bm n_m
    \right\}A_m^\Gamma ,
    \label{eq:discrete_hydrodynamic_load}
\end{equation}
where $\bm I$ is the identity tensor, $\bm r_m=\bm x_m^\Gamma-\bm X_c$, while $\bm x_m^\Gamma$, $\bm n_m$ and $A_m^\Gamma$ denote the centroid, liquid-to-solid normal and measure of the $m-$th embedded segment or facet, respectively. The viscous stress
$\bm\tau_m=\mu_l[\nabla\bm u_l+(\nabla\bm u_l)^T]_m$
is evaluated from velocity gradients reconstructed using the neighboring liquid-phase solution through a local least-squares approximation~\citep{fang2021spatiotemporal}; the pressure $p_m$ is reconstructed at the same interface location. Hence, both pressure and viscous contributions are evaluated directly on the geometrically resolved particle surface.

This direct surface evaluation is particularly important for freely rotating non-spherical bodies, for which errors in the local traction are amplified in the torque through the moment arm $\bm r_m$ and may lead to appreciable errors in the predicted orientation. Moreover, as melting continuously changes the particle shape, the surface used for load integration must remain identical to that on which the hydrodynamic boundary condition is imposed. In the present formulation, both are based on the same instantaneous $\Gamma_\Delta$, ensuring geometrical consistency between the resolved flow and the rigid-body response.

\textit{\textbf{Rigid-body update.}}
The mass properties entering the Newton--Euler equations are evaluated from the same instantaneous solid geometry reconstructed from $\Gamma_\Delta$. In each Cartesian cell, the reconstructed solid fraction $(1-c)$ defines a solid volume $V^s=(1-c)\Delta^{\mathcal D}$ and its centroid $\bm x^s$, where $\mathcal D$ denotes the spatial dimension. The total solid volume and centroid are then obtained as $V_s=\sum V^s$ and $\bm X_c=\sum V^s\bm x^s/V_s$, respectively. The corresponding mass and inertia tensor are evaluated as
\begin{equation}
    M_s=\rho_s V_s,
    \qquad
    \bm I_s
    \simeq
    \rho_s
    \sum
    V^s
    \left[
        |\bm x^s-\bm X_c|^2\bm I
        -
        (\bm x^s-\bm X_c)(\bm x^s-\bm X_c)
    \right],
    \label{eq:discrete_mass_inertia}
\end{equation}
where the summation extends over all Cartesian cells containing solid. During each mechanical update, the mass $M_s$ and inertia tensor $\bm I_s$ are frozen at the values associated with the current reconstructed geometry, and are recomputed when the phase-change geometry is updated in the next Picard sweep. Since the present step explicitly defines the Picard mapping, the iteration index $r$ is restored below.

Using a first-order time discretization, the translational velocity is updated as
\begin{equation}
    \bm v_s^{\,n+1,r+1}
    =
    \bm v_s^n
    +
    \frac{\Delta t}{M_s^{\,n+1,r}}
    \left[
        \bm F_H^{\,n+1,r}
        +
        (\rho_s-\rho_l)V_s^{\,n+1,r}\bm g
    \right],
    \label{eq:translation_update}
\end{equation}
while the angular velocity is obtained from
\begin{equation}
    \bm\omega_s^{\,n+1,r+1}
    =
    \bm\omega_s^n
    +
    \Delta t
    \left(\bm I_s^{\,n+1,r}\right)^{-1}
    \left[
        \bm T_H^{\,n+1,r}
        -
        \bm\omega_s^{\,n+1,r}
        \times
        \left(
            \bm I_s^{\,n+1,r}
            \bm\omega_s^{\,n+1,r}
        \right)
    \right].
    \label{eq:rotation_update}
\end{equation}
Here, the nonlinear gyroscopic term is evaluated using the current Picard estimate. Freezing $M_s$ and $\bm I_s$ within one mechanical update avoids introducing an additional variable-mass closure into the present formulation, while their evolution due to phase change is retained through recomputation from the updated solid geometry at successive Picard iterations. As we stated previously, it is also possible to consider the time-varying $M_s$ and $\bm I_s$ in the numerical computations while Suzuki \textit{et al}. \citep{suzuki2026equations} derived more rigorous expressions for Eqs.~\eqref{eq:translation_update}\eqref{eq:rotation_update}.

The updated velocities $(\bm v_s^{\,n+1,r+1},\bm\omega_s^{\,n+1,r+1})$ define the next estimate of the rigid-body velocity $\bm U_s$ through Eq.~\eqref{eq:rigid_body_velocity}. This updated $\bm U_s$ enters both the liquid-side boundary velocity $\bm u_{l,\Gamma}$ in Eq.~\eqref{eq:velocity_interface} and the geometrical interface velocity $\bm V_\Gamma$ in Eq.~\eqref{eq:interface_velocity}, thereby providing the mechanical feedback required for the next Picard sweep.

\textit{\textbf{Strong closure at a fixed physical time level.}}
The mechanical update completes the nonlinear feedback loop,
\[
\Gamma_\Delta
\rightarrow
(\bm u_l,p,T_l,T_s,Y_l)
\rightarrow
(\bm F_H,\bm T_H)
\rightarrow
(\bm v_s,\bm\omega_s)
\rightarrow
(\bm u_{l,\Gamma},\bm V_\Gamma)
\rightarrow
\Gamma_\Delta ,
\]
within which the thermosolutal closure determines $Y_{l,\Gamma}$, $T_\Gamma$ and $U_m$. A single traversal of this sequence would generally leave the flow, phase-change rate and rigid-body motion associated with different interface states. The complete sequence is therefore iterated, as summarized in Algorithm~\ref{alg:overall}, until the coupled variables in $\mathcal Q$ converge at the target time $t^{n+1}$. At every Picard sweep, the transient fields $(\bm u_l,p,T_l,T_s,Y_l,\phi)^{n+1, r}$ are restored to their converged states at $t^n$ and advanced again over the same interval $\Delta t$; only the coupled estimates entering the nonlinear closure are updated between sweeps. Hence, the iteration enforces multiphysics consistency without introducing repeated physical-time advancement.

Upon convergence, the interface geometry, Eulerian fields and rigid-body state are accepted simultaneously at $t^{n+1}$. Technique~III therefore closes the coupling among sharp-interface transport, thermosolutal phase change, hydrodynamic loading and free rigid-body motion within a common physical time step.

\section{Numerical tests}
\label{sec:numerical_tests}

The numerical tests are organized to assess the individual components of the method progressively before examining the fully coupled problem. \S~\ref{sec:motion_without_m} first considers rigid-body motion without phase change, thereby isolating the long-time transport of the moving geometry and the hydrodynamic force/torque--rigid-body coupling developed in Techniques~I and III. \S\S~\ref{sec:melting_sphere_trans} and \ref{sec:binary_m} then prescribe the particle motion and successively introduce pure and binary melting. In this way, the moving-boundary scalar transport and sharp interfacial closure of Technique~II can be assessed without the additional feedback from hydrodynamic loads to the particle motion. Finally, \S~\ref{sec:coupled} restores the complete coupling and considers freely moving particles undergoing pure and binary melting. The test sequence therefore increases progressively from rigid-body motion, to moving phase change, and ultimately to the fully coupled thermosolutal--hydrodynamic problem.

Unless otherwise stated, all computations employ the adaptive mesh refinement capability provided by \textit{Basilisk}. Refinement is concentrated around the solid--liquid interface and in regions of strong velocity, thermal and solutal gradients. At least seven layers of cells at the finest resolution $\Delta_{\min}$ are maintained around $\Gamma_\Delta$, so that the near-interface viscous, thermal and solutal boundary layers are adequately resolved.

\subsection{Rigid-body motion without phase change}
\label{sec:motion_without_m}

We first suppress phase change and assess the accuracy of the moving-body solver in isolation. These tests examine the transport of the rigid geometry, the sharp evaluation of the hydrodynamic loads and the resulting translational and rotational dynamics, i.e. the principal ingredients of Techniques~I and III.

The relevant dimensionless parameters are the Reynolds number $Re_p = \rho_lUD/\mu_l$, the density ratio $\rho_r=\rho_s/\rho_l$ and, for non-spherical particles, the aspect ratio $\alpha = a/b$. Here $U$ and $D$ denote the characteristic velocity and particle size, respectively. For a circle or sphere, $D$ is the particle diameter, whereas for the elliptical and oblate-ellipsoidal particles considered in \S\S~\ref{sec:ellipse}, \ref{sec:ellipsoid_rotate} and \ref{sec:ellipsoid_fall}, $D=a$, with $a$ and $b$ denoting the major and minor axis lengths.
When the particle velocity is prescribed, $U=|\bm v_s|$. For gravity-driven motion, the terminal velocity is not known \textit{a priori}; unless otherwise stated, we therefore use the gravitational scale $U_g=\sqrt{(\rho_r-1)|\bm g|D}$ and characterize the problem by the Galileo number $Ga=\rho_lU_gD/\mu_l$.

\subsubsection{Sedimentation of an elliptical cylinder between two walls}
\label{sec:ellipse}

\begin{figure}[htbp]
	\centering
	\includegraphics[width=0.8\textwidth]{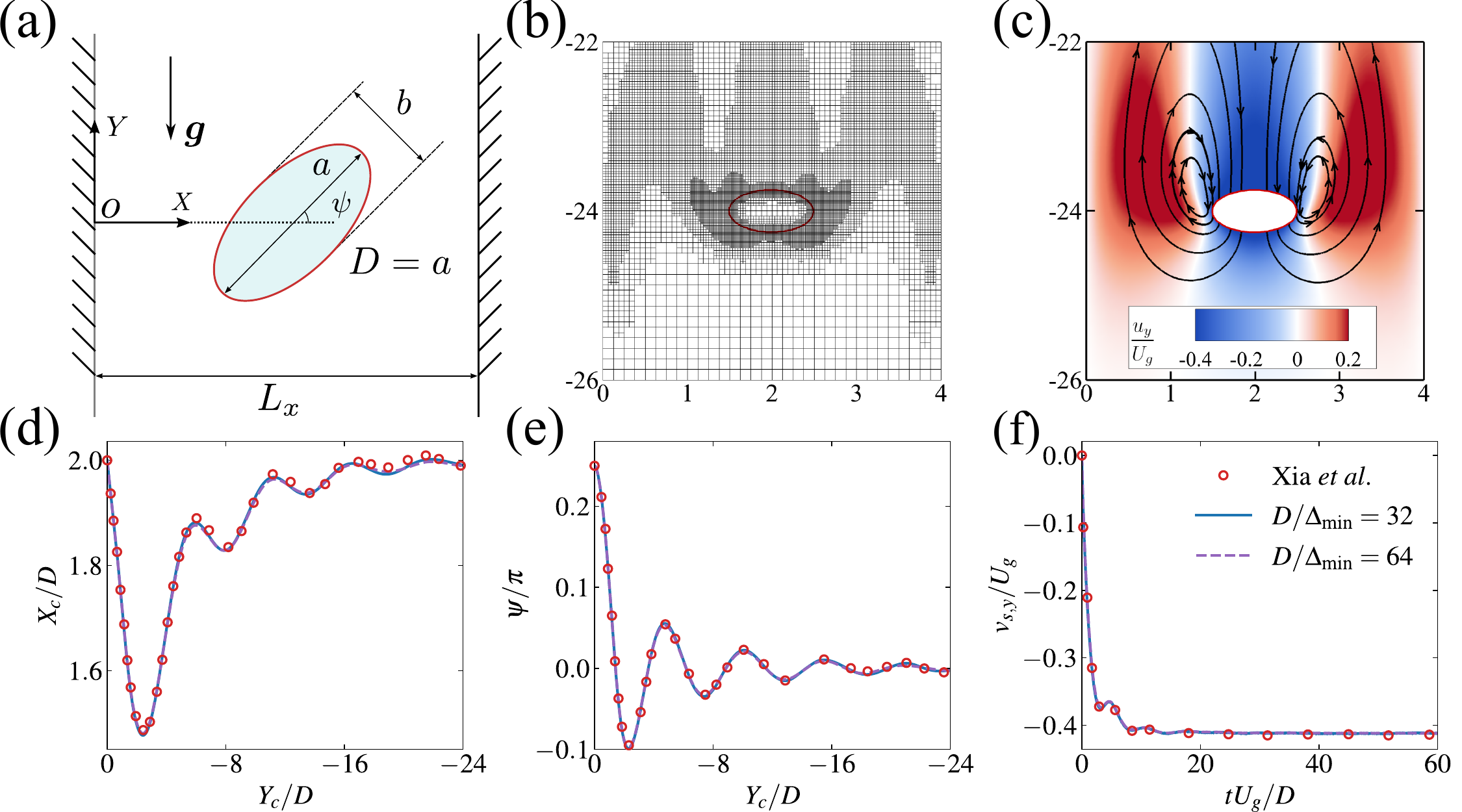}
	\caption{Sedimentation of an elliptical cylinder between two parallel walls. 
	(a) Problem configuration. 
	(b) Adaptive mesh and (c) vertical-velocity contours with streamlines at $Y_c=-24D$ ($tU_g/D=59.9$, $D/\Delta_{\min}=64$). 
	(d) Centroid trajectory, (e) inclination angle and (f) vertical settling velocity for two grid resolutions, compared with the results of Xia \textit{et al.}~\cite{xiaFlowPatternsSedimentation2009}.}
	\label{fig:ellipse}
\end{figure}

This first test examines the coupled translation and rotation of a freely moving non-spherical rigid body, and therefore directly assesses the long-time interface transport and rigid-body coupling introduced in Techniques~I and III. We consider the sedimentation of a two-dimensional elliptical cylinder between two vertical parallel walls, following Xia \textit{et al.}~\cite{xiaFlowPatternsSedimentation2009}; the configuration is shown in Fig.~\ref{fig:ellipse}(a). The aspect ratio is $\alpha = 2$, the density ratio is $\rho_r=1.1$, the wall spacing is $L_x/D=4$, and $Ga=31.32$. The computational domain has dimensions $L_x\times L_y=4D\times64D$, extending $10D$ above and $54D$ below the initial particle centroid. No-slip conditions are imposed on the two lateral walls, while zero-normal-gradient conditions are applied at the upper and lower boundaries. Initially, the particle is released from rest with $\bm X_{c,0}=(2D,0)$ and inclination angle $\psi_0=\pi/4$. Two finest-grid resolutions, $D/\Delta_{\min}=32$ and $64$, are considered.

As the particle settles, it simultaneously migrates laterally and rotates toward the stable broad-side orientation $\psi=0$. At long times, both the lateral and angular velocities vanish, while the vertical velocity approaches a constant terminal value corresponding to $Re_p\simeq12.7$, consistent with the reference solution~\cite{xiaFlowPatternsSedimentation2009}. Figures~\ref{fig:ellipse}(b,c) show the adaptive mesh and the corresponding laboratory-frame flow field at $Y_c=-24D$. The finest cells remain concentrated around the moving particle and its wake, while the nearly symmetric velocity and streamline patterns indicate that the settling state is close to steady.
Figures~\ref{fig:ellipse}(d--f) provide the quantitative validation. The centroid trajectory, inclination angle and vertical settling velocity obtained with $D/\Delta_{\min}=32$ and $64$ are nearly indistinguishable, showing that the solution is insensitive to further grid refinement at these resolutions. More importantly, all three quantities agree closely with the lattice-Boltzmann results of Xia \textit{et al.}~\cite{xiaFlowPatternsSedimentation2009}. The agreement in both translational and rotational dynamics confirms that the present sharp-interface formulation accurately evaluates the hydrodynamic force and torque while preserving the particle geometry over a long settling distance.

The same configuration is revisited in \ref{sec:appA} to compare Level-Set and VOF interface transport. Although VOF preserves the enclosed volume more accurately, cumulative shape distortion develops during the long-distance rigid-body motion and eventually affects the hydrodynamic response. The Level-Set representation exhibits a larger global volume error, of order $\mathcal O(10^{-2})$, but maintains the particle shape substantially better. This comparison motivates the use of Level Set as the primary geometry representation in Technique~I.

\subsubsection{Rotation of an oblate ellipsoid in Couette flow}
\label{sec:ellipsoid_rotate}

\begin{figure}[htbp]
	\centering
	\includegraphics[width=0.8\textwidth]{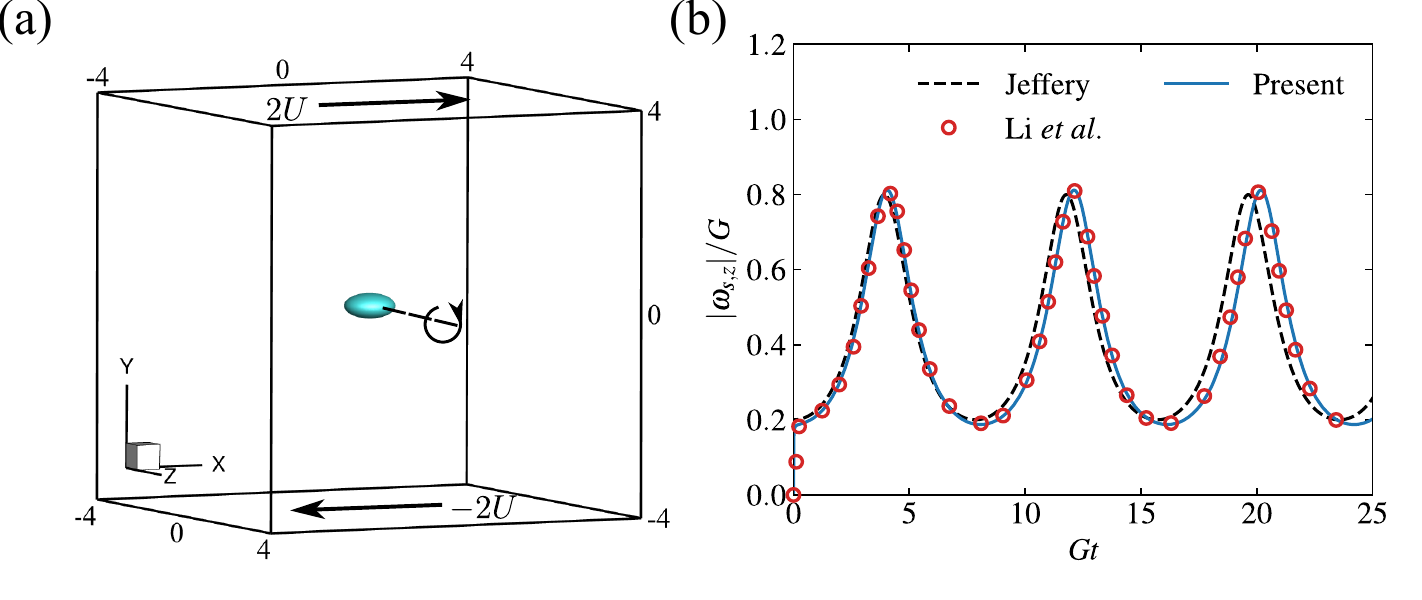}
	\caption{Rotation of an oblate ellipsoid in Couette flow. 
	(a) Problem configuration. 
	(b) Angular velocity of the ellipsoid compared with Jeffery's creeping-flow solution~\cite{jeffery} and the numerical results of Li \textit{et al.}~\cite{liFully3DSimulation2020}.}
	\label{fig:jeffery}
\end{figure}

We next isolate the rotational dynamics of a three-dimensional non-spherical particle in a prescribed shear flow. This test primarily assesses the sharp evaluation of the hydrodynamic torque and the subsequent angular-velocity update in Technique~III, together with the three-dimensional transport of the rotating geometry in Technique~I. As sketched in Fig.~\ref{fig:jeffery}(a), a neutrally buoyant oblate ellipsoid with aspect ratio $\mathcal \alpha = 2$ is placed at the center of a cubic domain of size $L_x=L_y=L_z=8D$. The upper ($y = 4D$) and lower ($y = -4D$) walls move in the $x$ direction with velocities $+2U$ and $-2U$, respectively, producing a uniform shear rate $G=0.5U/D$ along the $y-$direction, while periodic conditions are imposed in the $x$ and $z$ directions. The minor axis of the ellipsoid is initially aligned with the $y$ direction, and thus the particle rotates about the $z$ axis in the shear flow. The finest mesh resolution is $D/\Delta_{\min}=64$.

In the creeping-flow limit, the rotational motion of an ellipsoid in simple shear is described analytically by Jeffery~\cite{jeffery}. With the angle convention adopted in Fig.~\ref{fig:jeffery}(a), the orientation $\psi$ and angular velocity $\omega_{s,z}$ satisfy
\begin{equation}
	\psi = \tan^{-1} \left( \alpha \tan\left[ \frac{\alpha G t}{\alpha^2 + 1} \right] \right),
	\qquad
	\omega_{s,z} = \dot{\psi} = \frac{G}{\alpha^2 + 1}
	\left( \alpha^2 \cos^2 \psi + \sin^2 \psi \right).
	\label{eq:jeffery}
\end{equation}

Figure~\ref{fig:jeffery}(b) compares the computed angular velocity $\omega_{s,z}$ at $Re=0.5$ with Jeffery's analytical solution and the immersed-boundary results of Li \textit{et al.}~\cite{liFully3DSimulation2020}, obtained at the same finite Reynolds number. The present result closely follows the reference numerical solution throughout the rotation cycle. A small systematic departure from Jeffery's prediction \citep{jeffery} is observed, as expected because the latter applies to an unbounded creeping shear flow, whereas the numerical simulations are performed at finite $Re$ in a confined domain. The agreement with the finite-$Re$ benchmark of Li \textit{et al.}~\cite{liFully3DSimulation2020} confirms the accuracy of the three-dimensional torque evaluation and rigid-body rotational update.

\subsubsection{Sedimentation of an oblate ellipsoid}
\label{sec:ellipsoid_fall}

\begin{figure}[htbp]
	\centering
	\includegraphics[width=0.8\textwidth]{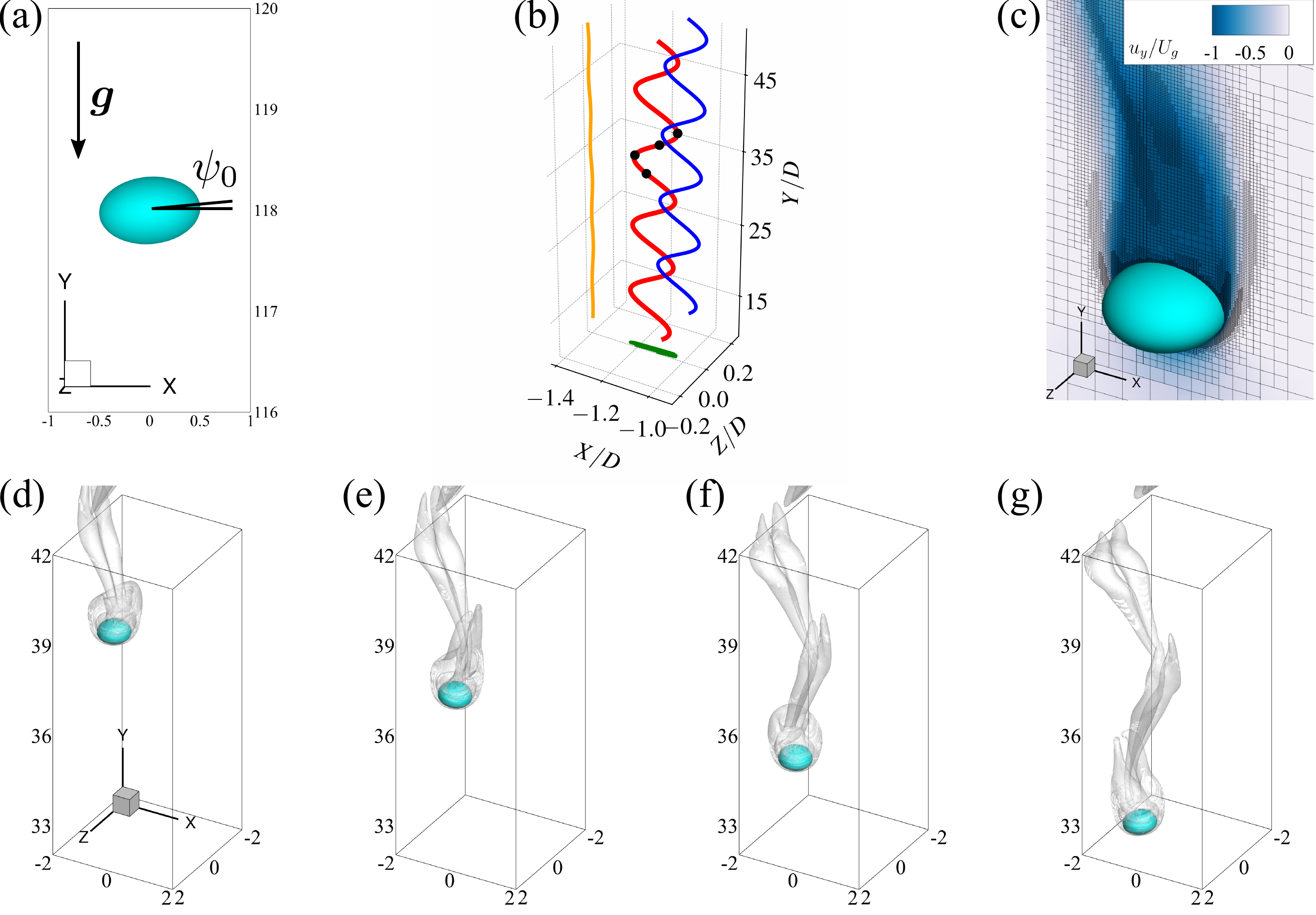}
	\caption{Sedimentation of an oblate ellipsoid in the periodic regime ($D/\Delta_{\min}=128$).
	(a) Initial configuration.
	(b) Particle trajectory and its projections onto the coordinate planes.
	(c) Adaptive mesh and vertical-velocity contours at $tU_g/D=77$.
	(d--g) Particle configurations and vortical structures characterized by $\lambda_2 = -0.05$ at $tU_g/D=77$, $79$, $81$, and $83$.}
	\label{fig:ellipsoid}
\end{figure}

\begin{figure}[htbp]
	\centering
	\includegraphics[width=0.8\textwidth]{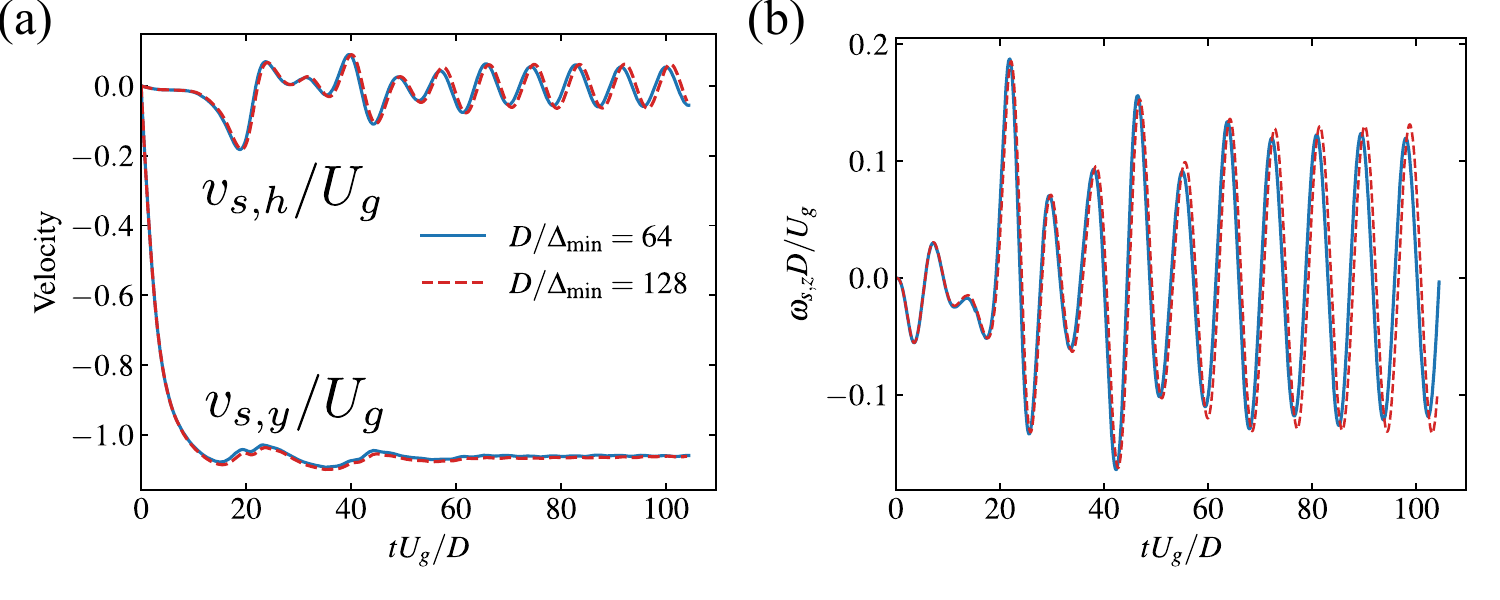}
	\caption{Grid convergence of the horizontal, vertical and angular velocities during the periodic sedimentation of the oblate ellipsoid.}
	\label{fig:ellipsoid_vel}
\end{figure}

As a more stringent test of the freely moving rigid-body solver, we next consider the three-dimensional sedimentation of an oblate ellipsoid in the periodic regime reported by Moriche \textit{et al.}~\cite{moricheSingleOblateSpheroid2021} and Jiang \textit{et al.}~\cite{jiangFlowreconstructionBasedApproach2024}. In contrast to the preceding cases, the particle undergoes long-distance translation, finite rotation and sustained periodic oscillations, providing a combined assessment of the interface transport in Technique~I and the force/torque--rigid-body coupling in Technique~III. The particle has aspect ratio $\alpha = 1.5$ and density ratio $\rho_r=2.14$, with $Ga=253.89$. As shown in Fig.~\ref{fig:ellipsoid}(a), the computational domain is a cube of size $L_x=L_y=L_z=128D$, with gravity acting in the negative $y$ direction. The bottom boundary is no-slip, while the slip conditions are imposed at the remaining boundaries. The particle is released from rest, and the particle centroid is initially located at $\bm X_{c,0}=(0,118D,0)$, i.e. $10D$ below the upper boundary, while its equatorial plane is inclined by $\psi_0=5^\circ$ about the $x$--$z$ plane to onset the oscillatory motion. The grid resolution is refined to $D/\Delta_\mathrm{min} = 128$ in the following discussions.

After the initial transient, the ellipsoid develops a periodic lateral oscillation while settling vertically on average. Figure~\ref{fig:ellipsoid}(b) shows the three-dimensional trajectory and its projections onto the coordinate planes. The motion remains essentially confined to the $x$--$y$ plane, so that the particle dynamics may be characterized by the vertical velocity $v_{s,y}$, the horizontal velocity $v_{s,h}$ and the angular velocity $\omega_{s,z}$. The four marked positions in Fig.~\ref{fig:ellipsoid}(b) correspond to the instantaneous states shown in Figs.~\ref{fig:ellipsoid}(d--g). These snapshots illustrate the periodic variation of the particle orientation together with the accompanying wake structures, characterized by $\lambda_2 = -0.05$. Figure~\ref{fig:ellipsoid}(c) further shows that the adaptive mesh remains concentrated around the particle and its wake, as colored by the vertical velocity, during the long-distance motion.

The grid sensitivity of the rigid-body dynamics is also examined in Fig.~\ref{fig:ellipsoid_vel}, with different resolutions of $D/\Delta_\mathrm{min} = 64$ and $128$, respectively. The time histories of $v_{s,h}$, $v_{s,y}$ and $\omega_{s,z}$ obtained on the different finest-grid resolutions converge toward nearly identical periodic signals, indicating that the oscillation frequency, mean settling velocity and rotational response are adequately resolved. Quantitative comparisons are summarized in Table~\ref{table:ellipsoid}. Here $St=fD/U_g$ is the Stroul number characterizing the oscillatory motion, where $f$ is the oscillation frequency; the overbar denotes the period-averaged value and the prime denotes the oscillation amplitude. The present predictions of $St$, $\bar v_{s,y}$, $v'_{s,h}$ and $\omega'_{s,z}$ are close to the body-fitted and immersed-boundary-type results reported in Refs.~\cite{moricheSingleOblateSpheroid2021,jiangFlowreconstructionBasedApproach2024}. This agreement provides a demanding validation of the present three-dimensional force/torque evaluation and rigid-body update for sustained non-axisymmetric motion.

\begin{table}[t]
	\centering
	\caption{Quantitative comparison for the periodic sedimentation of an oblate ellipsoid.}
	\begin{tabular}{c c c c c}
		\toprule
		Numerical Method & $St$ & $\bar{v}_{s,y}/U_g$ & $v'_{s,h}/U_g$ & $\omega'_{s,z}D/U_g$\\
		\midrule
		body-fitted grid \cite{moricheSingleOblateSpheroid2021}
		& 0.1172 & -1.028 & 0.1145 & 0.2409\\
		forcing points IBM \cite{moricheSingleOblateSpheroid2021}
		& 0.1125 & -1.012 & 0.1134 & 0.2086\\
		force integration IBM \cite{jiangFlowreconstructionBasedApproach2024}
		& 0.116 & -1.04 & -- & --\\
		stress integration IBM \cite{jiangFlowreconstructionBasedApproach2024}
		& 0.115 & -1.05 & -- & --\\
		Present EBM ($D/\Delta_{\min}=128$)
		& 0.1175 & -1.064 & 0.1260 & 0.2600\\
		\bottomrule
	\end{tabular}
	\label{table:ellipsoid}
\end{table}

The same case is reconsidered in~\ref{sec:appA} to assess the influence of interface transport over long distances. In the VOF calculation, cumulative distortion of the ellipsoidal geometry modifies the hydrodynamic response and leads to a spurious inclination of the mean settling path. By contrast, the Level-Set representation preserves the ellipsoidal shape more effectively and maintains the expected predominantly vertical mean trajectory. This comparison further supports the use of Level Set as the primary geometry representation in Technique~I.

\subsection{Pure melting of an ice sphere under prescribed translation}
\label{sec:melting_sphere_trans}

This test introduces phase change while prescribing the particle motion, thereby isolating the coupling between rigid-body translation, sharp thermal transport and interface recession without the additional feedback from hydrodynamic loads. The relevant dimensionless parameters are the initial Reynolds number $Re_0=\rho_l U D_0/\mu_l$, the Prandtl number $Pr=\mu_l/(\rho_l\alpha_l)$ and the Stefan number $Ste=c_{p,l}(T_\infty-T_m)/L$, where $D_0$ is the initial particle diameter. Since the particle shrinks continuously during melting, $D_0$ is used throughout as the reference length.

\begin{figure}[t]
	\centering
	\includegraphics[width=0.8\textwidth]{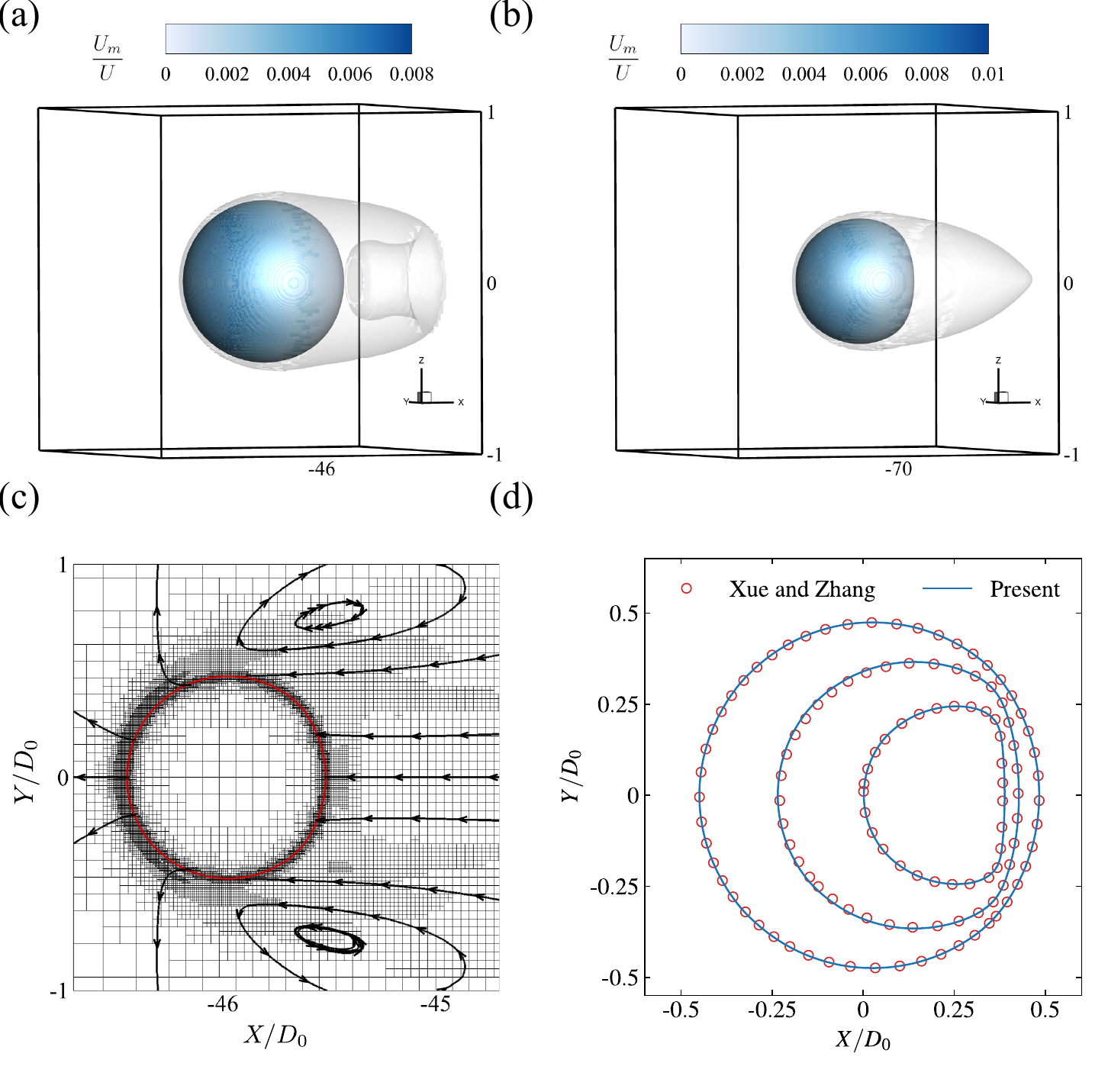}
	\caption{Pure melting of an ice sphere under prescribed translation.
	(a,b) Three-dimensional interface colored by the local melting rate $U_m$, together with the isosurface $(T_l - T_m)/(T_\infty-T_m) = 0.5$, at $tU/D_0=6$ and $30$, respectively.
	(c) Adaptive mesh and streamlines at $tU/D_0=6$, with the red line denoting the interface of the melting sphere.
	(d) Mid-plane interface profiles compared with the stationary-sphere results of Xue and Zhang~\cite{xue2026wake}, corresponding to the instants of $tU/D_0=6$, $30$, and $52.2$, respectively.}
	\label{fig:pure_melting3D}
\end{figure}

Xue and Zhang~\cite{xue2026wake} considered the melting dynamics of a fixed ice sphere exposed to a uniform flow. Here the complementary configuration is adopted: the surrounding liquid is initially quiescent, while the sphere translates at the prescribed velocity $\bm v_s=(-U,0,0)$. In the particle frame, the two problems are physically equivalent under a Galilean transformation and therefore provide a direct benchmark for the present moving-boundary formulation. Numerically, however, the present configuration is more demanding because the solid--liquid interface simultaneously translates across the fixed Cartesian mesh and recedes due to melting. Consequently, the cut-cell volumes vary continuously in time, and the momentum and scalar advection must remain conservative and geometrically consistent with the moving interface. This makes the comparison a direct test of the moving-boundary transport developed in Techniques~I and II.

The computational domain has dimensions $L_x=L_y=L_z=160D_0$, extending from $x=-150D_0$ to $10D_0$ in the direction of translation. Before melting is activated, the sphere is translated from the origin to $X_c=-40D_0$ at the prescribed velocity $\bm v_s=(-U,0,0)$, allowing the surrounding flow to develop; the time origin is then reset. At the left boundary $x=-150D_0$, $\bm u_l=\bm 0$ and $T_l=T_\infty>T_m$ are prescribed, while an outflow condition is imposed at $x=10D_0$. The lateral boundaries $y=\pm80D_0$ and $z=\pm80D_0$ are free-slip and adiabatic. The solid is maintained isothermal at $T_s=T_m$, buoyancy is neglected, and the governing parameters are $(\rho_r,Re_0,Pr,Ste)=(1,100,7,0.25)$. The finest mesh resolution is $D_0/\Delta_{\min}=204$.

Figures~\ref{fig:pure_melting3D}(a,b) show the instantaneous interface and thermal field at $tU/D_0=6$ and $30$, respectively. The sphere shrinks while remaining nearly axisymmetric, whereas the local melting rate varies appreciably over the surface owing to the non-uniform thermal boundary layer. As the particle size decreases, both the wake and the associated thermal disturbance weaken. Figure~\ref{fig:pure_melting3D}(c) shows the adaptive mesh and streamlines at $tU/D_0=6$. The finest cells remain concentrated around the moving interface and in the near wake, while the streamline pattern retains the expected near-axisymmetry.
A more stringent quantitative comparison is provided in Fig.~\ref{fig:pure_melting3D}(d), where the instantaneous mid-plane interface shapes are compared with the fixed-sphere results of Xue and Zhang~\cite{xue2026wake}. After translating the present profiles into the particle-centered frame, the interfaces at $tU/D_0=6$, $30$, and $52.2$ closely reproduce the corresponding reference shapes, the last instant corresponding to $V_s/V_{s,0}\simeq0.1$. This agreement confirms that the present formulation consistently combines prescribed rigid translation, sharp thermal transport and Stefan-driven interface recession, despite the continuous variation of the cut-cell geometry as the melting sphere crosses the fixed Cartesian mesh.

\subsection{Binary melting of a 2D circular cylinder under prescribed translation}
\label{sec:binary_m}

We next extend the prescribed-translation test to binary melting, thereby introducing the full thermosolutal coupling while still excluding the feedback from hydrodynamic loads to the particle motion. In addition to the parameters defined above, the binary system is characterized by the Lewis number $Le=\alpha_l/D_l$ and the dimensionless liquidus slope $m_L^*=m_LY_\infty/(T_\infty-T_{s,0})$, where $Y_\infty$ is the far-field solute concentration. The parameters used below are summarized in Table~\ref{table:binary_param}.

\begin{table}[t]
	\centering
	\caption{Parameters for the prescribed-translation binary-melting test.}
	\label{table:binary_param}
	\begin{tabular}{l c c}
		\toprule
		Parameter & Symbol & Value \\
		\midrule
		Density ratio & $\rho_r$ & 1 \\
		Initial Reynolds number & $Re_0$ & 40 \\
		Prandtl number & $Pr$ & 7 \\
		Lewis number & $Le$ & 5 \\
		Stefan number & $Ste$ & 0.13 \\
		Dimensionless liquidus slope & $m_L^*$ & -0.33 \\
		Thermal-diffusivity ratio & $\alpha_l/\alpha_s$ & 0.13 \\
		Thermal-conductivity ratio & $k_l/k_s$ & 0.13 \\
		Dimensionless pure-water melting temperature
		& $T_m^*=(T_m-T_{s,0})/(T_\infty-T_{s,0})$ & 0.83 \\
		\bottomrule
	\end{tabular}
\end{table}

\begin{figure}[t]
	\centering
	\includegraphics[width=0.8\textwidth]{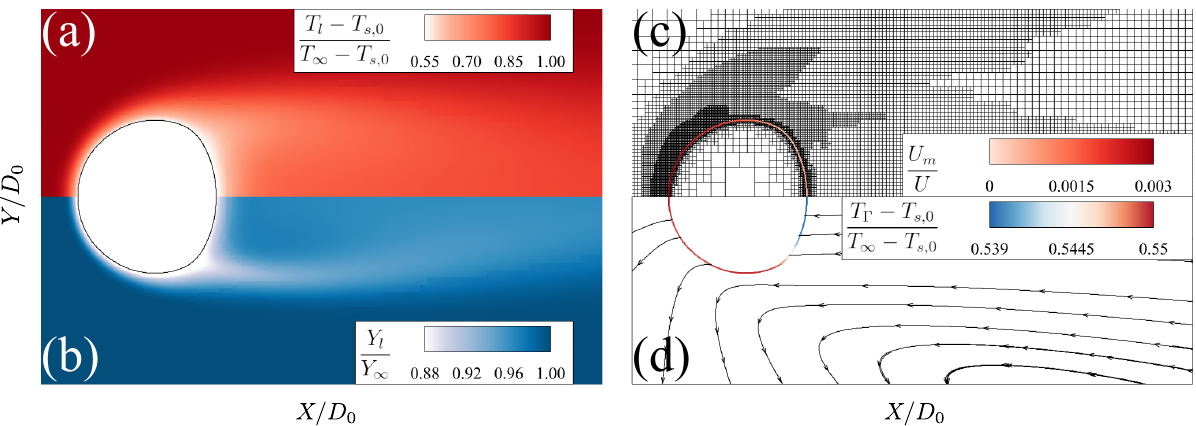}
	\caption{Binary melting of a translating ice cylinder at $tU/D_0=75$.
	(a) Temperature and (b) concentration fields.
	(c) Adaptive mesh and local melting rate $U_m$ at the interface.
	(d) Streamlines and interfacial temperature $T_\Gamma$.}
	\label{fig:binary2D_field}
\end{figure}

\begin{figure}[t]
	\centering
	\includegraphics[width=0.5\textwidth]{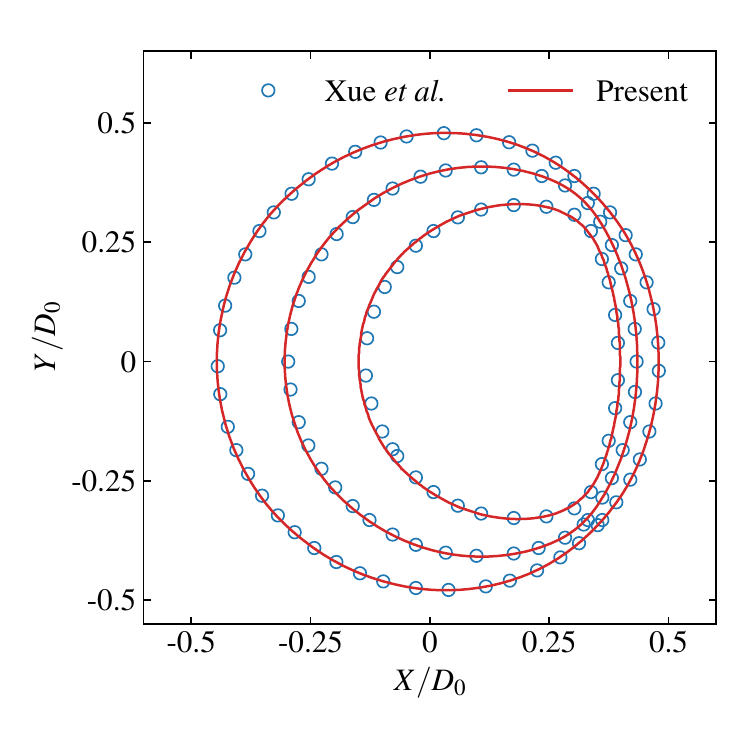}
	\caption{Interface profiles at $tU/D_0=25$, $75$, and $125$ for the binary melting of a translating circular cylinder, compared with the stationary-cylinder reference computed using the method of Xue \textit{et al.}~\cite{xue2023three}.}
	\label{fig:binary_2D_inter}
\end{figure}

To construct a reference solution, we perform a companion calculation with the sharp and conservative binary phase-change method of Xue \textit{et al.}~\cite{xue2023three}, in which the 2D ice cylinder is held fixed and subjected to a uniform incoming flow of velocity $U$. The present calculation adopts the Galilean-equivalent configuration: the surrounding saline water is initially quiescent, while the ice cylinder translates at the prescribed velocity $\bm v_s=(-U,0)$. The stationary reference is therefore not a previously reported test case, but is recomputed here using our earlier fixed-body formulation under otherwise identical physical and numerical parameters. Any discrepancy between the two configurations may consequently be attributed primarily to the treatment of the moving and melting boundary.

The computational domain still has dimensions $L_x=L_y=160D_0$, extending from $x=-150D_0$ to $10D_0$. At the left boundary, $\bm u_l=\bm 0$, $T_l=T_\infty$ and $Y_l=Y_\infty$ are prescribed, while an outflow condition is imposed at $x=10D_0$. The lateral boundaries $y=\pm80D_0$ are free-slip and adiabatic, with zero normal solute flux. The cylinder is initially centered at the origin with $T_s=T_{s,0}<T_m$ and begins to translate and melt at $t=0$. The finest mesh resolution is $D_0/\Delta_{\min}=204$.

Compared with the pure-substance test of \S~\ref{sec:melting_sphere_trans}, the present case additionally requires the simultaneous solution of the solute field and the binary interfacial closure. In particular, the local quantities $Y_{l,\Gamma}$, $T_\Gamma$ and $U_m$ are mutually coupled through the liquidus relation, solute balance and Stefan condition while the interface continuously crosses the fixed Cartesian mesh. Figure~\ref{fig:binary2D_field}(a) shows the resulting fields at $tU/D_0=75$, the thermal and solutal boundary layers remain sharply resolved around the moving interface, with the latter being thinner because $Le>1$. Their non-uniform distributions generate corresponding variations of $T_\Gamma$ and $U_m$ along the ice surface, as shown in the lower part of Fig.~\ref{fig:binary2D_field}(b), which also displays the adaptive mesh concentrated around the interfacial boundary layers and the near wake.

Compared with the pure-substance test of \S~\ref{sec:melting_sphere_trans}, the present case additionally requires the solution of the solute field and the binary interfacial closure. In particular, $Y_{l,\Gamma}$, $T_\Gamma$ and $U_m$ are mutually coupled through the liquidus relation, solute balance and Stefan condition while the interface continuously crosses the fixed Cartesian mesh. Figure~\ref{fig:binary2D_field}(a, b) shows the temperature and concentration fields at $tU/D_0=75$. Both thermal and solutal boundary layers remain sharply resolved around the moving interface, with the latter being thinner owing to $Le>1$. The adaptive mesh in Fig.~\ref{fig:binary2D_field}(c) remains concentrated around the interfacial boundary layers and the near wake, while the resulting non-uniform interfacial heat and solute transport leads to spatial variations of $T_\Gamma$ and $U_m$, as shown in the lower part of Fig.~\ref{fig:binary2D_field}(d).

Figure~\ref{fig:binary_2D_inter} provides the quantitative validation. After translating the present profiles into the particle-centered frame, the interface shapes at $tU/D_0=25$, $75$, and $125$ closely reproduce those obtained from the stationary-cylinder reference calculation. Since both calculations employ the same sharp binary phase-change treatment while differing in whether the solid boundary crosses the Eulerian mesh, this comparison specifically isolates the additional numerical error associated with moving cut-cell transport. The agreement therefore validates the conservative thermal and solutal transport, geometry-consistent solid-temperature treatment, and sharp binary interfacial closure of Technique~II for a translating phase-change boundary.

\subsection{Fully coupled melting with free particle motion}
\label{sec:coupled}

\subsubsection{Sedimentation of a melting 2D circular particle}
\label{sec:melting_circle_g}

\begin{figure}[t]
	\centering
	\includegraphics[width=0.6\textwidth]{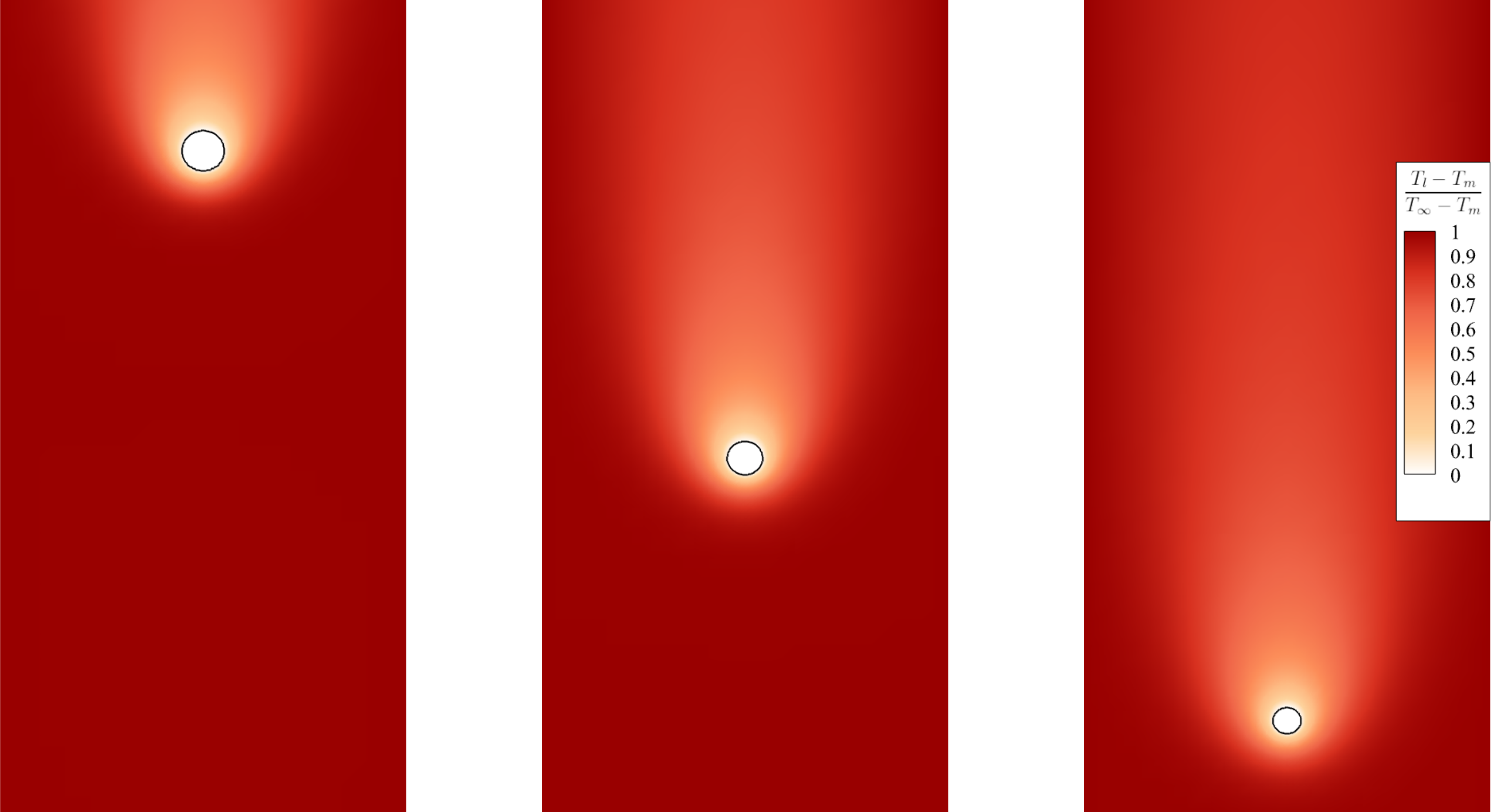}
	\caption{Sedimentation and pure melting of a circular particle. Temperature field and solid interface at $tU_g/D_0=100$, $110$, and $120$.}
	\label{fig:pure_melting2D_field}
\end{figure}

\begin{figure}[htbp]
	\centering
	\includegraphics[width=0.8\textwidth]{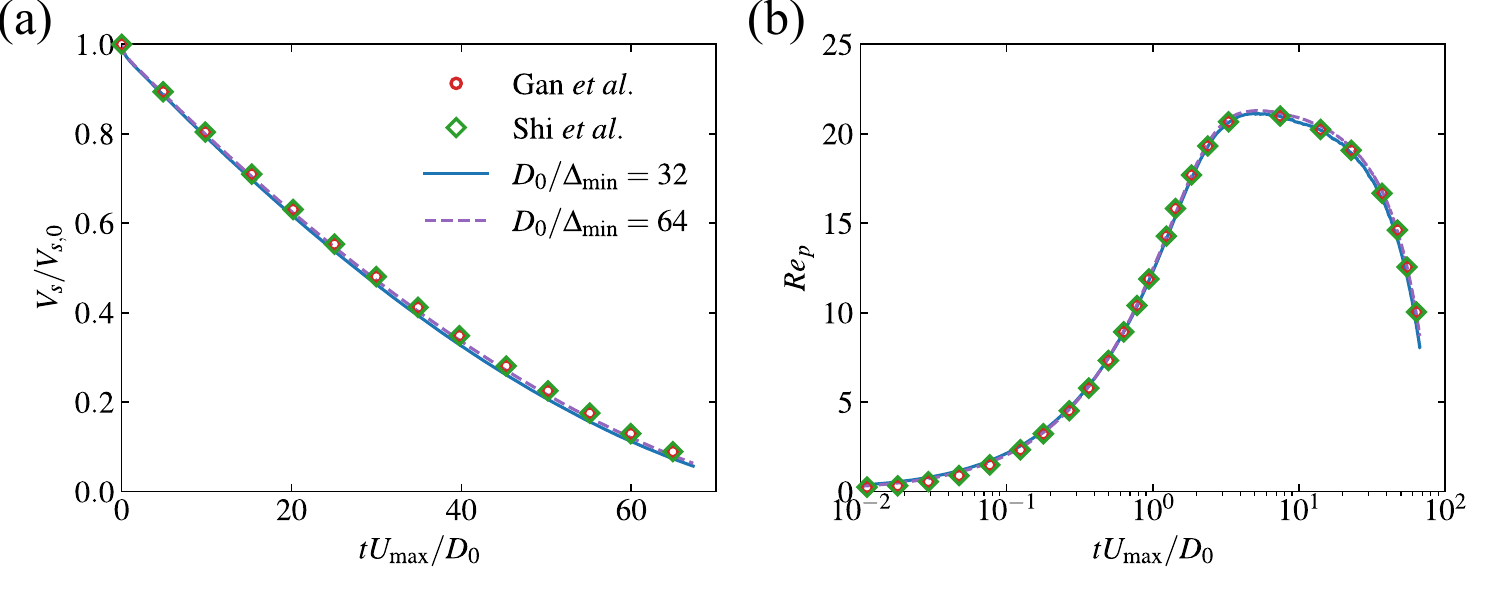}
	\caption{Grid convergence and quantitative validation for the sedimentation of a melting circular particle.
	(a) Remaining solid volume and (b) particle Reynolds number, compared with Refs.~\cite{gan2003simulation,shiNumericalSimulationMelting2024}.}
	\label{fig:pure_melting2D}
\end{figure}

We now activate the complete coupling between phase change and free object motion. Unlike the prescribed-translation tests considered above, the particle velocity is determined by the instantaneous hydrodynamic load, while melting continuously modifies the particle geometry and hence its subsequent motion. This first fully coupled test considers pure melting only, allowing the combined performance of Techniques~I--III and the nonlinear coupling of Algorithm~\ref{alg:overall} to be assessed before introducing solutal effects.

Following Gan \textit{et al.}~\cite{gan2003simulation} and Shi \textit{et al.}~\cite{shiNumericalSimulationMelting2024}, a 2D circular particle is released in a vertical channel of size $L_x\times L_y=4D_0\times64D_0$. For direct comparison with these references, the gravitational velocity scale is defined as $U_g=\sqrt{0.5\pi(\rho_r-1)|\bm g|D_0}$, with $Ga=\rho_lU_gD_0/\mu_l$. The side and bottom walls are no-slip and maintained at $T_l=T_\infty$, whereas the top boundary is free-slip and adiabatic. Initially, the liquid is at rest and uniformly at $T_\infty$, and the particle is released from rest on the channel centerline, $10D_0$ below the upper boundary. The solid is maintained isothermal at $T_s=T_m$, and buoyancy is included through $Gr=\rho_l^2D_0^3\beta_T(T_\infty-T_m)|\bm g|/\mu_l^2$. The governing parameters are $\rho_r=1.00232$, $Ga=38.7$, $Pr=0.7$, $Gr=100$, and $Ste=0.025125$. The instantaneous particle Reynolds number is defined as $Re_p=\rho_l|v_{s,y}|D_0/\mu_l=Ga|v_{s,y}|/U_g$.

Figure~\ref{fig:pure_melting2D_field} shows the coupled evolution of the particle and temperature field during the late stage of sedimentation. The particle continuously shrinks while maintaining an approximately circular shape and a predominantly vertical trajectory. At the same parameters but without melting, the particle approaches a steady sedimentation state with $Re_p\simeq21$~\cite{shiNumericalSimulationMelting2024}. With melting activated, however, the decrease in particle size and mass continuously modifies the hydrodynamic balance, so that the settling velocity evolves together with the interface.
The quantitative comparison is shown in Fig.~\ref{fig:pure_melting2D}. Results obtained with $D_0/\Delta_{\min}=32$ and $64$ are nearly indistinguishable for both the remaining solid volume (panel (a)) and $Re_p$ (panel (b)), indicating adequate grid convergence. Furthermore, both panels indicate that the present predictions also closely follow the ALE finite-element results of Gan \textit{et al.}~\cite{gan2003simulation} and the immersed-boundary results of Shi \textit{et al.}~\cite{shiNumericalSimulationMelting2024}. The particle first accelerates under gravity, reaches a maximum settling speed of approximately $0.55U_g$, and subsequently decelerates as melting reduces its size and gravitational driving. The simultaneous agreement in melting history and particle velocity confirms that the present framework consistently couples Stefan-driven interface recession, hydrodynamic loading and freely evolving rigid-body motion.

\subsubsection{Wake-induced rotation of a melting ice sphere}
\label{sec:melting_sphere_rotate}

\begin{figure}
	\centering
	\includegraphics[width=0.8\textwidth]{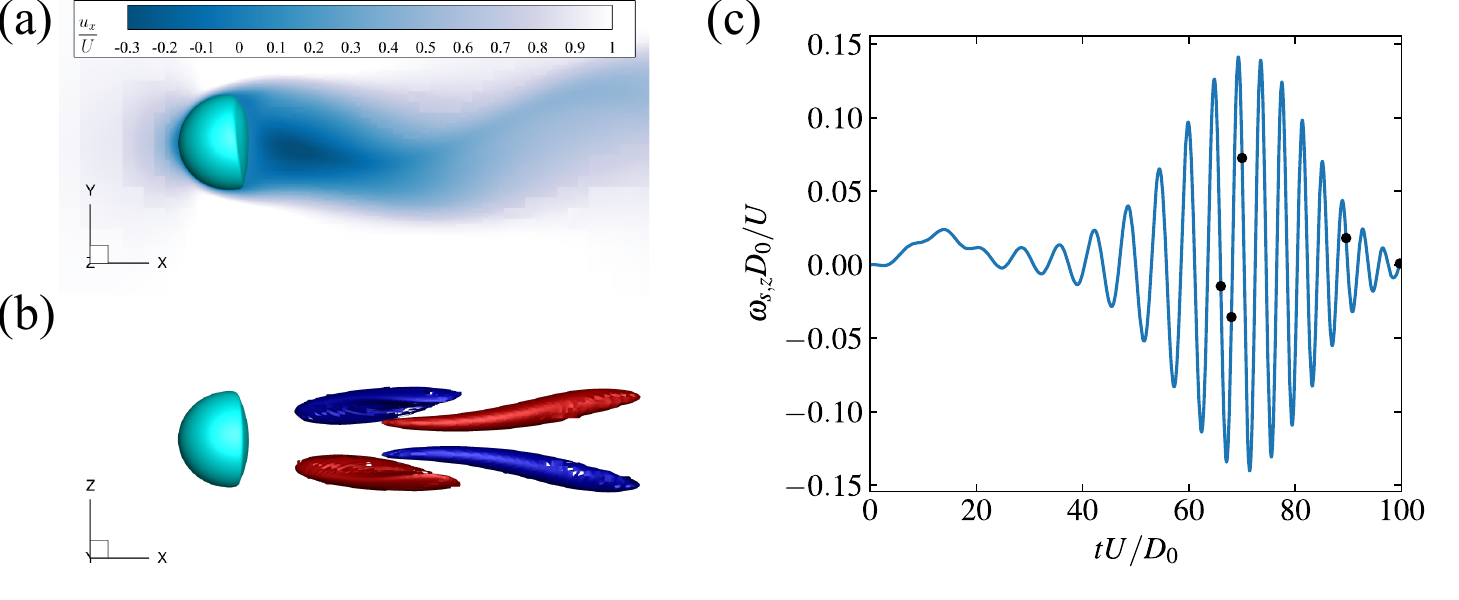}
	\caption{Wake-induced rotational response of a melting ice sphere at $Re_0=300$.
	(a) Streamwise velocity and (b) streamwise-vorticity isosurfaces of $\omega_x = \pm 2$ in the presence of coupled melting and rotation at $tU/D_0=66$.
	(c) Angular velocity $\omega_{s,z}$; markers indicate the instants shown in Fig.~\ref{fig:ice_sphere_field}.}
	\label{fig:ice_sphere_angular}
\end{figure}

\begin{figure}
	\centering
	\includegraphics[width=\textwidth]{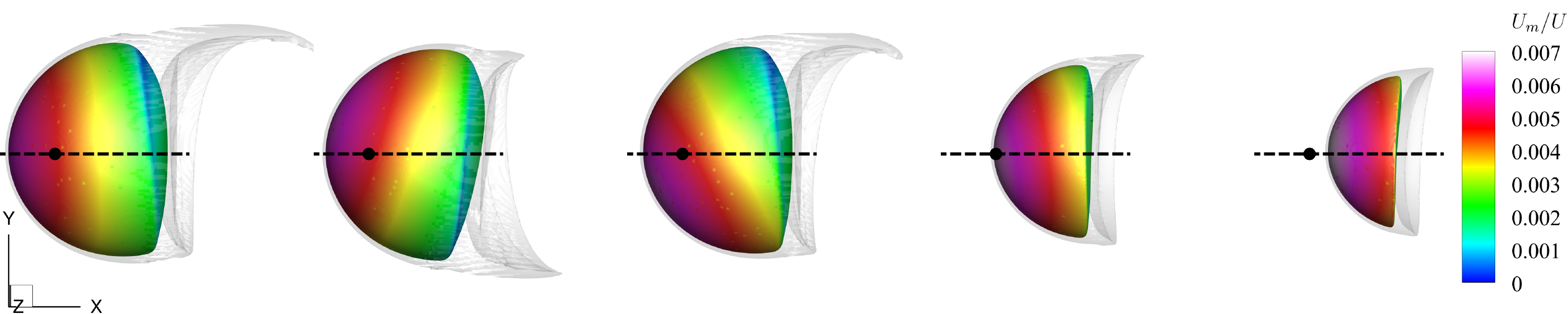}
\caption{Coupled rotation and melting of an initially spherical particle at $tU/D_0=66$, $68$, $70$, $90$, and $100$.
The particle surface is colored by the local melting rate $U_m$, while the gray isosurface denotes $(T_l-T_m)/(T_\infty-T_m)=0.5$.
The last two instants correspond to $V_s/V_{s,0}=0.1$ and $0.05$, respectively.
The solid circles mark the initial centroid position before melting, and the dashed lines indicate the initial horizontal orientation.}
	\label{fig:ice_sphere_field}
\end{figure}

The preceding test demonstrates the coupling between translational motion and phase change, whereas the present case further examines the rotational response induced by asymmetric melting. Xue and Zhang~\cite{xue2026wake} showed that, at sufficiently large $Re_0$, a melting sphere with constrained orientation develops an inclined rear surface accompanied by a non-zero hydrodynamic torque. Releasing the rotational degree of freedom therefore provides a direct test of the feedback between wake asymmetry, particle rotation and non-uniform melting, before both translation and rotation are fully released in \S~\ref{sec:ellipsoid_melt}. Such flow--melting feedback and the resulting rotational motion of phase-changing bodies have also been highlighted as an important aspect of freely moving melting problems in the recent review of Du \textit{et al.}~\citep{du2024physics}.

We consider $Re_0=300$, $Pr=7$, $Ste=0.25$, $Gr=0$, and $\rho_r=1$, corresponding to the periodic planar-symmetric wake regime identified by Xue and Zhang~\cite{xue2026wake}. A small transverse perturbation is introduced to select the $x$--$y$ symmetry plane, after which an asymmetric but planar-symmetric periodic wake is allowed to develop with melting suppressed and all particle degrees of freedom constrained. At $tU/D_0=0$, melting is activated and rotation about the $z-$ axis is released, while particle translation remains constrained.

Figures~\ref{fig:ice_sphere_angular}(a,b) show the streamwise velocity field and the vorticity isosurfaces $\omega_x = \pm 2$, respectively, at $tU/D_0=66$, when melting and rotation are fully coupled. The asymmetric wake produces a non-zero hydrodynamic torque about the $z$ axis once the rotational degree of freedom is released. The resulting angular response is quantified in Fig.~\ref{fig:ice_sphere_angular}(c). The oscillation amplitude of $\omega_{s,z}$ initially grows, then progressively decreases and eventually vanishes as melting reduces the particle size and weakens the associated wake unsteadiness.
The corresponding snapshots in Fig.~\ref{fig:ice_sphere_field} illustrate the evolution of the particle orientation and melting morphology. The dashed line indicates the horizontal reference axis and the dot marks the center of rotation. The particle surface is colored by the local melting rate $U_m$, while the gray isosurface represents $(T_l-T_m)/(T_\infty-T_m)=0.5$. As the particle shrinks, its effective Reynolds number decreases, leading to a progressively weaker wake and hence a reduction of the oscillatory hydrodynamic torque.

At the late stages, corresponding to $V_s/V_{s,0}=0.1$ and $0.05$, the inclined rear surface observed when rotation is constrained~\cite{xue2026wake} evolves toward a surface nearly normal to the incoming flow. This comparison demonstrates that releasing the rotational degree of freedom can qualitatively modify the melting morphology, highlighting the necessity of coupling rigid-body motion and phase change in the present framework.

\subsubsection{Sedimentation of an 3D oblate ellipsoid with pure and binary melting}
\label{sec:ellipsoid_melt}

\begin{figure}[t]
	\centering
	\includegraphics[width=0.8\textwidth]{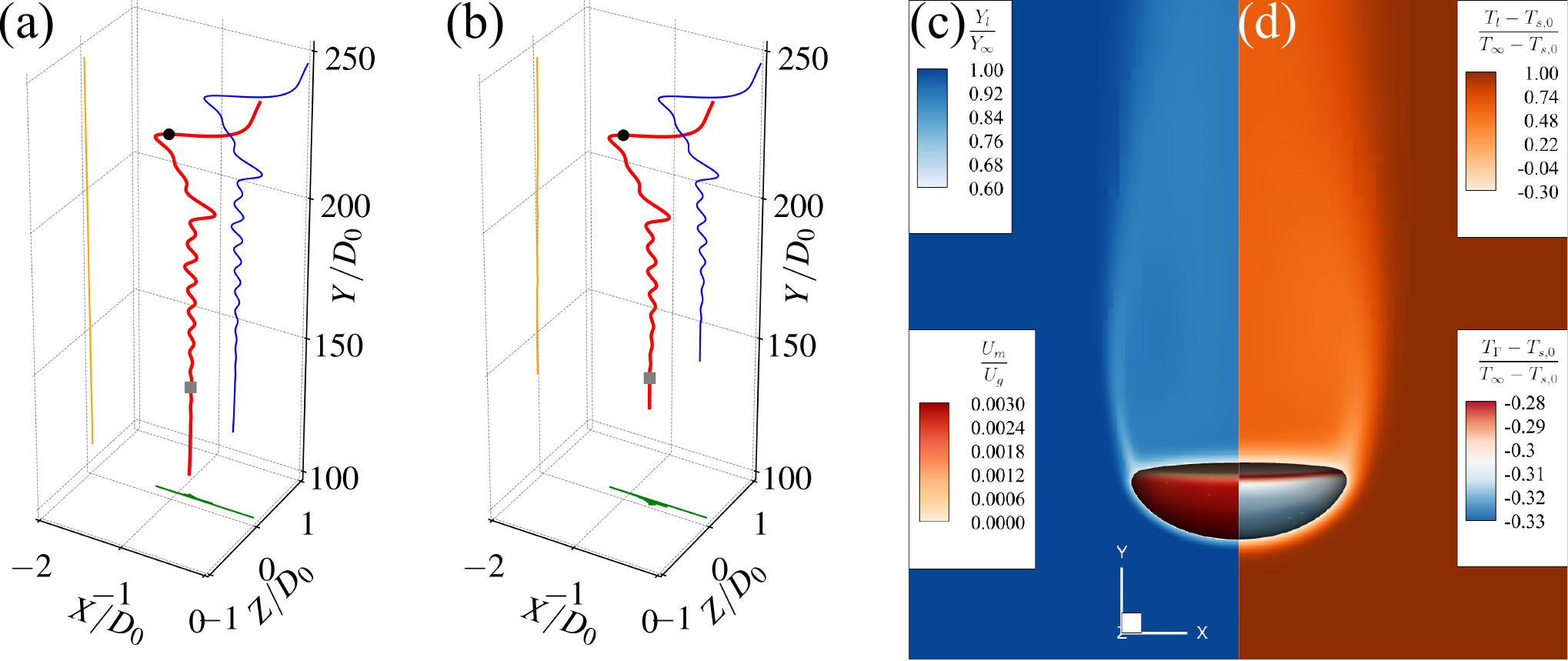}
	\caption{Sedimentation of a melting oblate ellipsoid.
	Three-dimensional trajectories for (a) pure and (b) binary melting.
	The circle marks the onset of melting at $tU_g/D_0=20$, and the square marks $tU_g/D_0=119$.
	(c) Solute concentration and surface melting rate, and
	(d) temperature field and interfacial temperature $T_\Gamma$, for the binary case at $V_s/V_{s,0}=0.1$.}
	\label{fig:ellipsoid_melt_tra}
\end{figure}

\begin{figure}[t]
	\centering
	\includegraphics[width=0.8\textwidth]{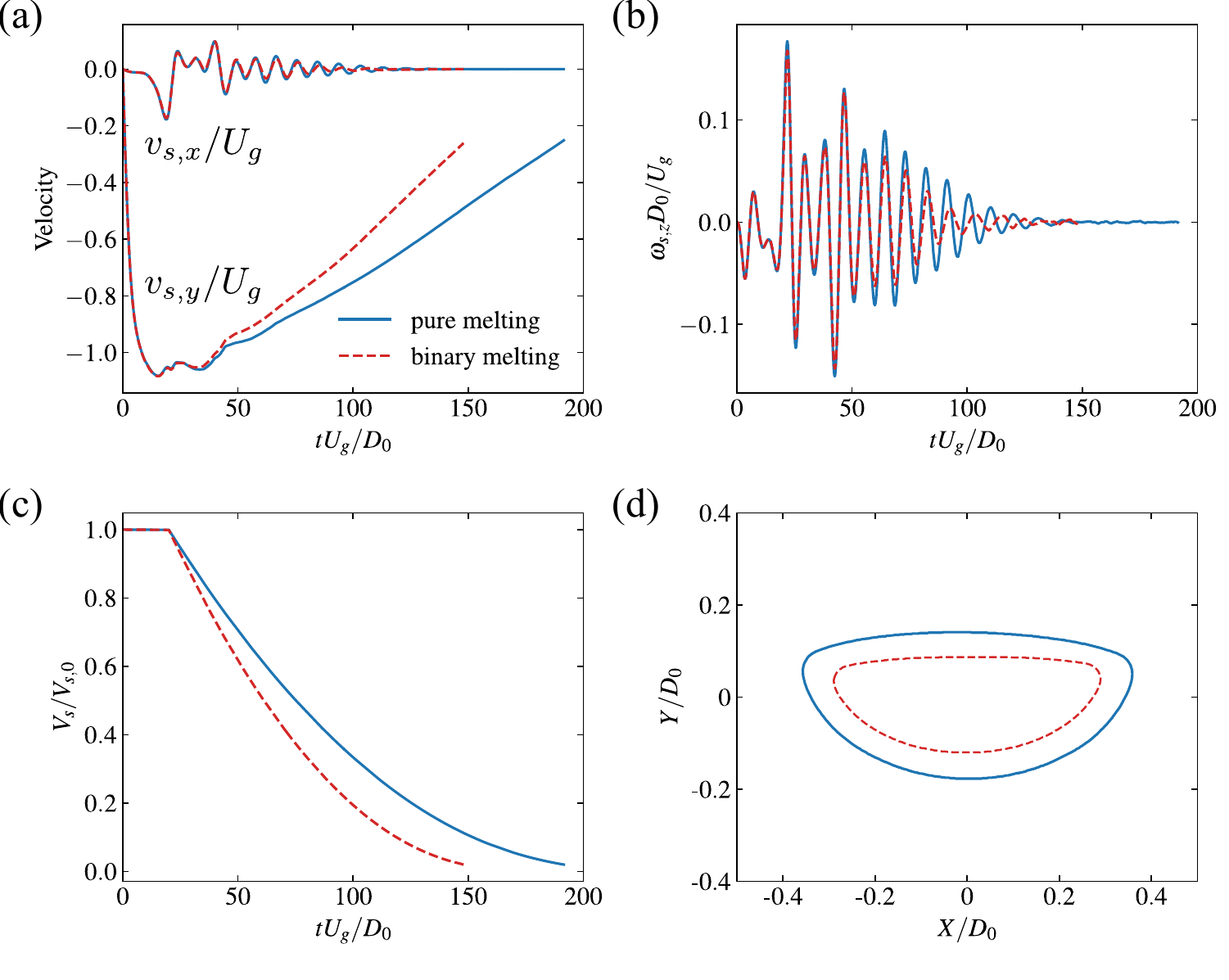}
	\caption{Evolution of the melting oblate ellipsoid.
	(a) Transverse velocity, (b) settling velocity, and (c) remaining solid volume for the pure- and binary-melting cases.
	(d) Particle profiles at the same time moments of $tU_g/D_0=119$, shifted to a common centroid.}
	\label{fig:ellipsoid_melt_plot}
\end{figure}

As the final and most strongly coupled test, we consider the sedimentation of an oblate ellipsoid undergoing either pure or binary melting. This configuration combines all numerical ingredients introduced above: long-distance transport of a three-dimensional non-spherical interface, freely evolving translation and rotation, sharp thermal and solutal transport, phase-change deformation, and the nonlinear feedback between hydrodynamic loading and the instantaneous particle geometry. The hydrodynamic parameters are identical to those of \S~\ref{sec:ellipsoid_fall} for non-melting ellipsoid, namely $\alpha = 1.5$, $\rho_r=2.14$, and $Ga=253.89$, with $U_g=\sqrt{(\rho_r-1)|\bm g|D_0}$.

The particle is initially located on the central $x$--$z$ plane, $10D_0$ below the upper boundary, with an initial inclination $\psi_0=5^\circ$. To minimize confinement effects during the combined settling and melting process, the computational domain is enlarged to $L_x=L_y=L_z=256D_0$, with the finest resolution $D_0/\Delta_{\min}=256$ to capture the thin solutal boundary owing to $Le = 5$. The bottom and lateral boundaries are maintained at $T_l=T_\infty$ and, for the binary case, $Y_l=Y_\infty$, whereas zero normal temperature and concentration gradients are imposed at the upper boundary. The pure-melting particle is maintained at $T_s=T_m$. For binary melting, the solid is initially at $T_{s,0}=T_m$, while the subsequent interfacial temperature is determined locally from the liquidus relation and therefore varies with $Y_{l,\Gamma}$. The remaining phase-change parameters are given in Table~\ref{table:ellipsoid_param}.

\begin{table}[t]
	\centering
	\caption{Phase-change parameters for the melting oblate-ellipsoid cases.}
	\label{table:ellipsoid_param}
	\begin{tabular}{l c c c}
		\toprule
		Parameter & Symbol & Pure & Binary \\
		\midrule
		Prandtl number & $Pr$ & 7 & 7 \\
		Lewis number & $Le$ & -- & 5 \\
		Stefan number & $Ste$ & 0.25 & 0.25 \\
		Dimensionless liquidus slope & $m_L^*$ & -- & -0.45 \\
		Thermal-diffusivity ratio & $\alpha_l/\alpha_s$ & 0.13 & 0.13 \\
		Thermal-conductivity ratio & $k_l/k_s$ & 0.13 & 0.13 \\
		Dimensionless pure-water melting temperature
		& $T_m^*=(T_m-T_{s,0})/(T_\infty-T_{s,0})$ & 0 & 0 \\
		\bottomrule
	\end{tabular}
\end{table}

To separate the initial hydrodynamic transient from the subsequent phase-change dynamics, melting is activated only after the particle has sedimented to $tU_g/D_0=20$. Figures~\ref{fig:ellipsoid_melt_tra}(a,b) compare the subsequent trajectories for pure and binary melting. In both cases, the oblate ellipsoid initially retains the lateral oscillations characteristic of the non-melting regime, but these oscillations progressively weaken as melting reduces the particle size and, consequently, the effective particle Reynolds number. The motion therefore evolves toward a predominantly vertical settling path. This transition occurs over a shorter settling distance in the binary case because the particle melts more rapidly in the solutal environment.
The origin of this difference is illustrated more directly in Figs.~\ref{fig:ellipsoid_melt_tra}(c,d). In the binary liquid, solute accumulates near the moving interface and lowers the local equilibrium temperature through the liquidus relation, thereby enhancing the melting rate relative to the pure-substance case. The resulting difference in geometry is quantified in Fig.~\ref{fig:ellipsoid_melt_plot}(d), where particle profiles at the same time instants are shifted to a common centroid for comparison. The binary-melting profiles exhibit a systematically larger reduction in volume, providing direct geometrical evidence of the enhanced melting rate.

Figures~\ref{fig:ellipsoid_melt_plot}(a--c) further quantify the coupled dynamics. The decay of the transverse-velocity oscillations accompanies the progressive reduction in particle volume, while the settling velocity continuously adjusts to the evolving hydrodynamic and gravitational balance. Because binary melting reduces $V_s$ more rapidly, the lateral oscillations are suppressed earlier than in the pure-melting case, consistently with the trajectories in Fig.~\ref{fig:ellipsoid_melt_tra}. These results demonstrate that the present framework can consistently couple three-dimensional shape evolution, hydrodynamic force and torque, rigid-body motion, and sharp thermosolutal phase change within a strongly evolving moving-boundary problem.

\subsection{Thermoconvection-driven motion and melting of an ice disc}
\label{sec:ice_disc}

As a final application of the present framework, we consider the spontaneous motion of a cold ice disc floating at the water surface, driven by thermoconvection induced by the disc--water temperature difference. This problem provides a particularly convincing setting because the particle motion is not imposed externally but emerges from a symmetry-breaking instability of the surrounding thermal flow. Li and Zhu~\cite{li2025hydrodynamic} showed that a cold, non-melting disc initially generates a downward plume associated with the density anomaly of water near its temperature of maximum density (TMD). Once this plume loses its rotational and mirror symmetries, a net hydrodynamic force and torque develop and drive the disc to translate and rotate spontaneously. Their calculations deliberately excluded melting in order to isolate this instability mechanism. Here we first reproduce the similar thermoconvective fluid--structure interaction and subsequently introduce phase change, allowing the feedback among plume instability, rigid-body motion and evolving particle geometry to be examined within a single framework. The spontaneous motion of melting ice discs has also been reported experimentally~\citep{dorbolo2016rotation}.

The density anomaly of water is represented by the generalized equation of state
$\rho_l(T_l)=\rho_{l,0}[1-\beta|T_l-T_{\mathrm{TMD}}|^{1.895}]$,
where $\rho_{l,0}$ is the maximum reference density, $T_{\mathrm{TMD}}\simeq4^\circ\mathrm{C}$, and $\beta$ is the generalized thermal-expansion coefficient. The corresponding density variation is incorporated into the momentum equation through the Boussinesq approximation. The disc has an initial diameter $D_0$ and thickness $0.1D_0$, and the ambient water is initially quiescent at $T_\infty>T_{\mathrm{TMD}}$, while the ice is maintained initially at $T_m<T_{\mathrm{TMD}}$.

\begin{figure}
	\centering
	\includegraphics[width=\textwidth]{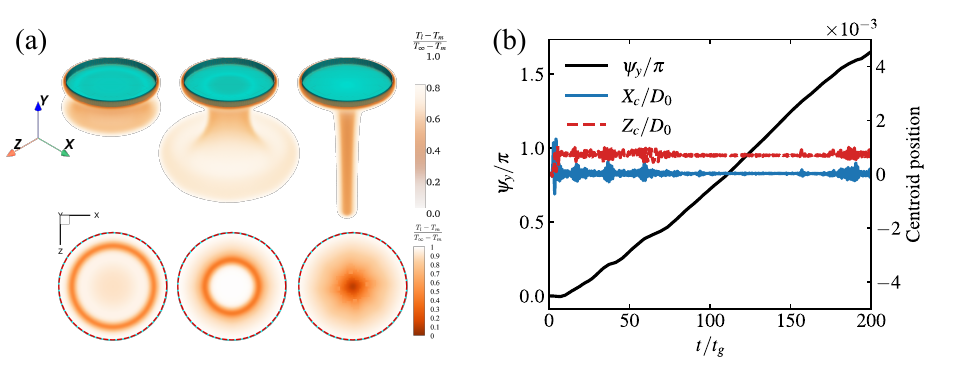}
	\caption{Thermoconvection-driven motion of the non-melting ice disc in Case A, with a domain height of $20D_0$.
	(a) Instantaneous configurations at $tU_g/D_0=5$, $10$, and $100$. The upper panels show the dimensionless temperature field $\widetilde{T}$, while the lower panels show the corresponding bottom views on a horizontal plane located $0.1D_0$ beneath the disc. The red dashed circle marks the initial disc position.
	(b) Temporal evolution of the rotation angle $\psi_y$ and centroid displacements $X_c$ and $Z_c$.}
	\label{fig:move_large}
\end{figure}

The present liquid--solid formulation does not explicitly resolve the air phase or the free surface. We therefore adopt a reduced floating-disc configuration in which the upper surface of the disc coincides with the top boundary of the liquid domain. Vertical motion is constrained, while the disc is free to translate in the horizontal $x$--$z$ plane with velocities $(u_{s,x},u_{s,z})$ and to rotate about the vertical $y$ axis with angular velocity $\omega_{s,y}$, as sketched in Fig.~\ref{fig:move_large}(a). In this way, the upper surface is excluded from both hydrodynamic and thermal interactions with the liquid, while the immersed part of the disc remains fully coupled to the surrounding flow. This approximation avoids introducing an additional gas-phase/free-surface solver and is consistent with the fact that only a small fraction of a floating ice body is exposed to air~\citep{bellincioni2025melting}.

Using $D_0$ and $T_\infty-T_m$ as the characteristic length and temperature differences, respectively, we define
$U_g=\sqrt{g\beta(T_\infty-T_m)^{1.895}D_0}$ and $t_g=D_0/U_g$. The dimensionless temperature of maximum density, Rayleigh number and Prandtl number are
\begin{equation}
	\mathcal{T}
	=
	\frac{T_{\mathrm{TMD}}-T_m}{T_\infty-T_m}
	=0.16,
	\qquad
	Ra
	=
	\frac{\rho_{l,0}g\beta(T_\infty-T_m)^{1.895}D_0^3}
	{\mu_l\alpha_l}
	=10^6,
	\qquad
	Pr
	=
	\frac{\mu_l}{\rho_{l,0}\alpha_l}
	=6.67 .
\end{equation}
With $\widetilde T=(T_l-T_m)/(T_\infty-T_m)$, the dimensionless buoyancy term becomes
$|\widetilde T-\mathcal T|^{1.895}\bm e_y$, while the viscous and thermal-diffusion coefficients are $\sqrt{Pr/Ra}$ and $1/\sqrt{RaPr}$, respectively.

The reference configuration employs a cubic domain of size $L_x=L_y=L_z=20D_0$, with gravity acting in the negative $y$ direction. For all simulations, the finest mesh resolution is set to $D_0/\Delta_{\min}=102$ in the vicinity of the melting interface. The top boundary is impermeable and free-slip, allowing the disc to move freely along it, whereas no-slip conditions are imposed on the remaining walls. All outer boundaries are thermally insulated. Starting from this common configuration, three cases are considered successively to separate the roles of thermoconvective instability, geometric confinement and phase change in determining the disc motion.

\paragraph{Case A: Reference non-melting disc.}
We first suppress phase change and establish the thermoconvective fluid--structure interaction that serves as the reference for the following cases. The disc is initially at rest, and its subsequent motion is driven solely by the buoyancy-induced flow associated with the density anomaly of water. Cooling beneath the disc produces a downward-moving layer of dense fluid, whose development determines the hydrodynamic force and torque acting on the freely moving disc.

Figure~\ref{fig:move_large}(a) shows the evolution of dimensionless temperature $\widetilde{T}$ at $tU_g/D_0=5$, $10$, and $100$. The temperature field initially forms an annular curtain beneath the disc and progressively converges into a centralized descending column. Both the three-dimensional distributions and the corresponding bottom views indicate that the large-scale thermal structure remains nearly axisymmetric throughout this stage. Consistently, the bottom view reveals that the disc centroid stays close to its initial position, indicating an almost complete cancellation of the horizontal force components.

The rigid-body response is quantified in Fig.~\ref{fig:move_large}(b). Although the horizontal displacements $X_c$ and $Z_c$ remain only of $\mathcal{O}(10^{-3}D_0)$, the rotation angle $\psi_y=\int\omega_{s,y}\,\mathrm{d}t$ increases continuously. Thus, the weak azimuthal asymmetry that remains in the thermoconvective flow is sufficient to generate a finite hydrodynamic torque, while producing almost no net horizontal force. Case A therefore provides a clean reference state characterized by appreciable rotation but negligible translation, against which the effects of geometric confinement and phase change are examined below.

\begin{figure}
	\centering
	\includegraphics[width=\textwidth]{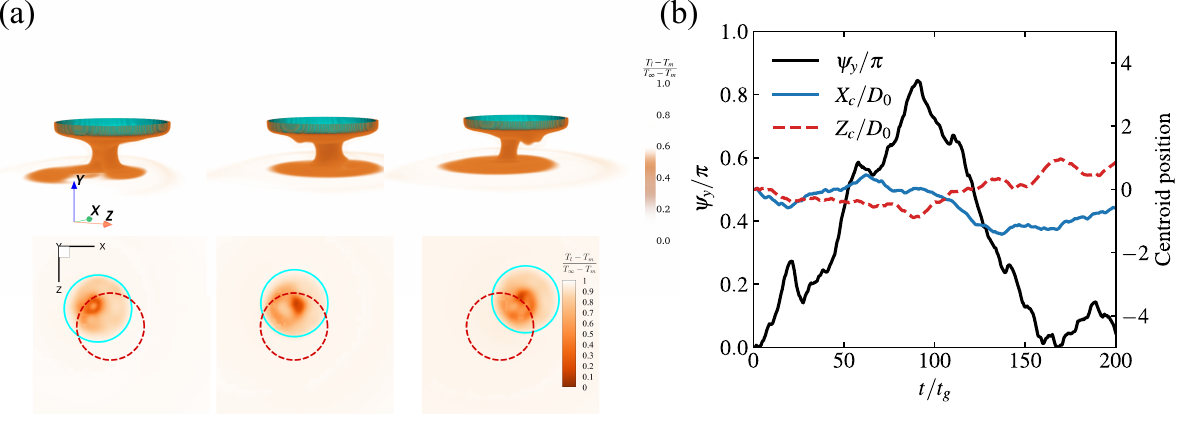}
	\caption{Effect of vertical confinement on the thermoconvection-driven motion of the non-melting ice disc in Case B, with the domain height reduced to $L_y=0.5D_0$.
	(a) Instantaneous configurations at $tU_g/D_0=30$, $45$, and $60$.
	(b) Temporal evolution of the rotation angle $\psi_y$ and centroid displacements $X_c$ and $Z_c$. Other descriptions are the same as in Fig.~\ref{fig:move_large}.}
	\label{fig:move_small}
\end{figure}

\paragraph{Case B: Vertically confined non-melting disc.}
Case A shows that, in the reference deep domain, the large-scale thermoconvective flow remains nearly axisymmetric: a finite hydrodynamic torque develops from weak azimuthal asymmetry, whereas the net horizontal force remains very small. We next examine how this force--torque balance is altered when the downward development of the cold flow is strongly confined. The vertical extent of the domain is reduced to $L_y=0.5D_0$, while $L_x=L_z=20D_0$ and all other physical and numerical parameters are kept unchanged. In this shallow configuration, the descending cold fluid encounters the bottom wall shortly after leaving the disc and is forced to spread laterally, so that the redistribution of the thermoconvective flow occurs much closer to the particle.

The resulting evolution is shown in Fig.~\ref{fig:move_small}(a). In contrast to Case A, the descending thermal structure cannot develop into a long, nearly axisymmetric column. Its interaction with the bottom wall produces a pronounced lateral redistribution and a strongly non-axisymmetric temperature field beneath the disc, as is particularly evident from the bottom views of $\widetilde{T}$. The disc consequently undergoes not only rotation about the $y$ axis but also appreciable translation in the $x$--$z$ plane, with its centroid progressively departing from the initial position marked by the red dashed circle. Thus, vertical confinement qualitatively changes the hydrodynamic response: the thermoconvective flow now generates a substantial net horizontal force in addition to the torque.

Figure~\ref{fig:move_small}(b) quantifies this change. Both $X_c$ and $Z_c$ grow to $\mathcal{O}(10^{-1}D_0)$, approximately two orders of magnitude larger than in Case A, while $\psi_y$ continues to evolve, confirming the coexistence of appreciable translational and rotational motion. The rotational response also becomes non-monotonic once significant lateral displacement develops. As the disc moves away from the domain center, the distances to the lateral boundaries become unequal and the associated return flow becomes increasingly asymmetric, thereby modifying the hydrodynamic torque acting on the disc.

The comparison between Cases A and B therefore shows that the disc response depends strongly on how the thermoconvective flow develops relative to the particle. In the deep domain, the descending cold flow remains nearly axisymmetric over a long vertical distance, so that the horizontal force components almost cancel and the motion is dominated by rotation. Under strong vertical confinement, the cold flow is redirected horizontally close to the disc, producing a much stronger near-particle asymmetry and, consequently, substantial horizontal translation in addition to rotation. This confinement-induced change provides a second reference state for assessing how phase change modifies the coupled disc--flow dynamics in Case C.

\begin{figure}
	\centering
	\includegraphics[width=\textwidth]{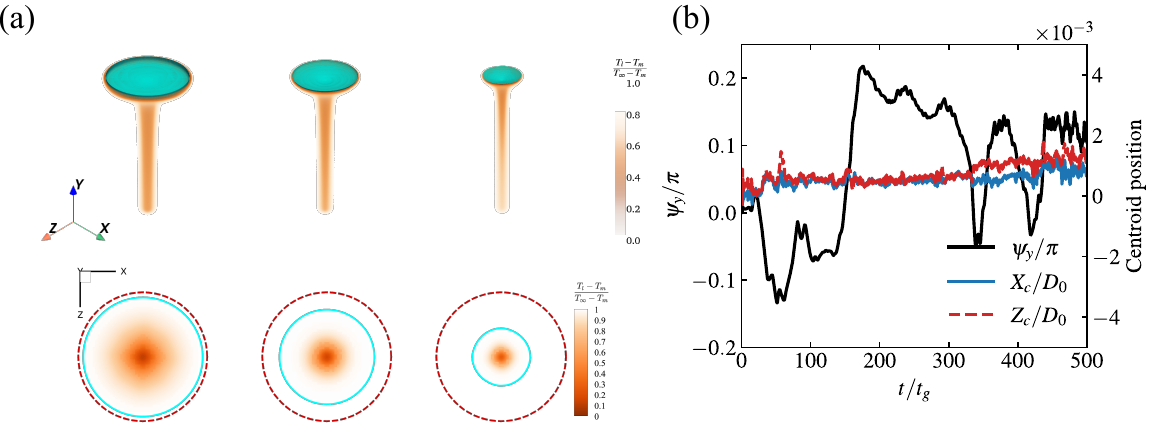}
	\caption{Thermoconvection-driven motion of the melting ice disc in Case C.
	(a) Instantaneous configurations at $tU_g/D_0=100$, $300$, and $500$.
	(b) Temporal evolution of the rotation angle $\psi_y$ and centroid displacements $X_c$ and $Z_c$. Other descriptions are the same as in Fig.~\ref{fig:move_large}.}
	\label{fig:move_melt}
\end{figure}

\paragraph{Case C: Melting disc.}
Finally, we return to the reference configuration of Case A and activate phase change, thereby closing the feedback among thermoconvection, rigid-body motion and the evolving disc geometry. The computational domain and boundary conditions are unchanged, while the Stefan number is set to $Ste=10^{-3}$. This small value allows the disc to melt gradually, so that the influence of phase change on the thermoconvective motion can be followed over a sufficiently long time interval.

The resulting evolution is shown in Fig.~\ref{fig:move_melt}(a). Compared with the non-melting reference case, the thermal structure beneath the disc remains more axisymmetric over the same stage of evolution, while the lower surface of the disc gradually develops an approximately axisymmetric conical shape due to the wake-enhanced melting concentrated beneath its center. As the disc shrinks, the characteristic length scale of the buoyancy-driven flow decreases, which weakens the thermoconvective motion and delays the growth of the azimuthal asymmetry responsible for spontaneous rigid-body motion. The corresponding bottom views confirm that the disc centroid remains very close to its initial position throughout the simulation. Thus, melting introduces a stabilizing feedback: thermoconvection drives phase change, whereas the resulting reduction of the disc size weakens the same thermoconvective forcing that would otherwise promote disc motion.

The rigid-body response is quantified in Fig.~\ref{fig:move_melt}(b). The disc still rotates about the $y$ axis, but the accumulated rotation is markedly smaller than in Case A, while $X_c$ and $Z_c$ remain of only $\mathcal{O}(10^{-3}D_0)$ throughout the calculation. Hence, phase change does not merely alter the body geometry passively; it also modifies the thermoconvective forcing and thereby feeds back on both the translational and rotational dynamics.

Taken together, Cases A--C show that the present framework can capture the coupled effects of thermoconvection, geometric confinement, free rigid-body motion and phase change within a single sharp-interface formulation. In particular, Case C indicates that the evolving particle geometry may feed back on the thermoconvective instability and thereby modify the resulting motion. This coupled melting--motion problem appears worthy of a more systematic physical investigation, which will be pursued in future work using the present method.

\section{Summary and perspectives}
\label{sec:conclusion}

We have developed a sharp-interface method for the simulation of freely moving solid particles undergoing pure or binary melting. The method couples rigid-body translation and rotation, incompressible flow, heat and solute transport, and phase change within a fixed Cartesian framework. In contrast to our previous formulation for phase-changing bodies with prescribed motion, the present method closes the two-way coupling between the evolving solid geometry, hydrodynamic force and torque, rigid-body dynamics, and interfacial melting.

Three numerical ingredients are central to the formulation. First, the solid--liquid interface is represented and transported by a Level-Set method, which preserves the particle geometry during long-distance translation and rotation. Second, heat and solute are solved sharply in the corresponding liquid and solid sub-domains, separated by the moving and melting embedded interface: the liquid scalars employ a fully conservative moving-control-volume discretization, while the solid temperature is transported through the geometrical approach, so that its support remains consistent with the moving solid domain. Third, the hydrodynamic force and torque, rigid-body motion, interfacial heat and mass transfer, and phase-change velocity are coupled within a common nonlinear iteration at each physical time step. These treatments address respectively the difficulties of long-time geometry preservation, local scalar conservation on moving cut cells, and strong coupling between phase change and particle motion.

The individual components were assessed progressively before considering fully coupled problems. Rigid-body translation and rotation without melting reproduced established two- and three-dimensional benchmarks, while prescribed-motion melting tests verified the moving-boundary thermal and binary transport against corresponding fixed-body calculations. Fully coupled simulations further reproduced reference results for a sedimenting melting particle and demonstrated wake-induced rotation of a melting sphere when the rotational degree of freedom was released. The sedimentation of a three-dimensional oblate ellipsoid with pure and binary melting then showed that the method can simultaneously resolve free translation and rotation, non-uniform interface recession, and thermosolutal coupling. As a final application, the thermoconvection-driven motion of a floating-like ice disc was considered to examine the interaction among flow instability, geometric confinement, rigid-body motion, and melting. The results showed that phase change can feed back on the thermoconvective instability through the evolving particle geometry and thereby modify the resulting translational and rotational response. Additional tests in the appendices clarified the numerical choices: the Level-Set formulation provides better long-distance shape preservation than the present VOF implementation, strict local conservation is required for $T_l$ and $Y_l$, and geometry-consistent transport is necessary to keep $T_s$ attached to the moving solid domain.

Several extensions remain worthwhile. A more complete mechanical formulation should account explicitly for momentum and angular-momentum transfer associated with phase change when the particle mass and inertia vary appreciably within a time step. From the numerical side, combining the long-time geometrical fidelity obtained here with the superior global conservation of geometrical VOF methods would be particularly attractive. These developments would further extend the present framework toward strongly coupled phase-change problems involving larger density changes, more complex material systems and multiple interacting solid bodies.

\section{Acknowledgement}

The authors gratefully acknowledge the support of the the NSFC under grants numbers W2511004, 12588201, 12472256, and that of the National Key R$\&$D Program of China under grants number 2023YFA1011000.

\appendix
\section{VOF versus Level Set for long-distance particle transport}
\label{sec:appA}

\begin{figure}
	\centering
	\includegraphics[width=0.9\textwidth]{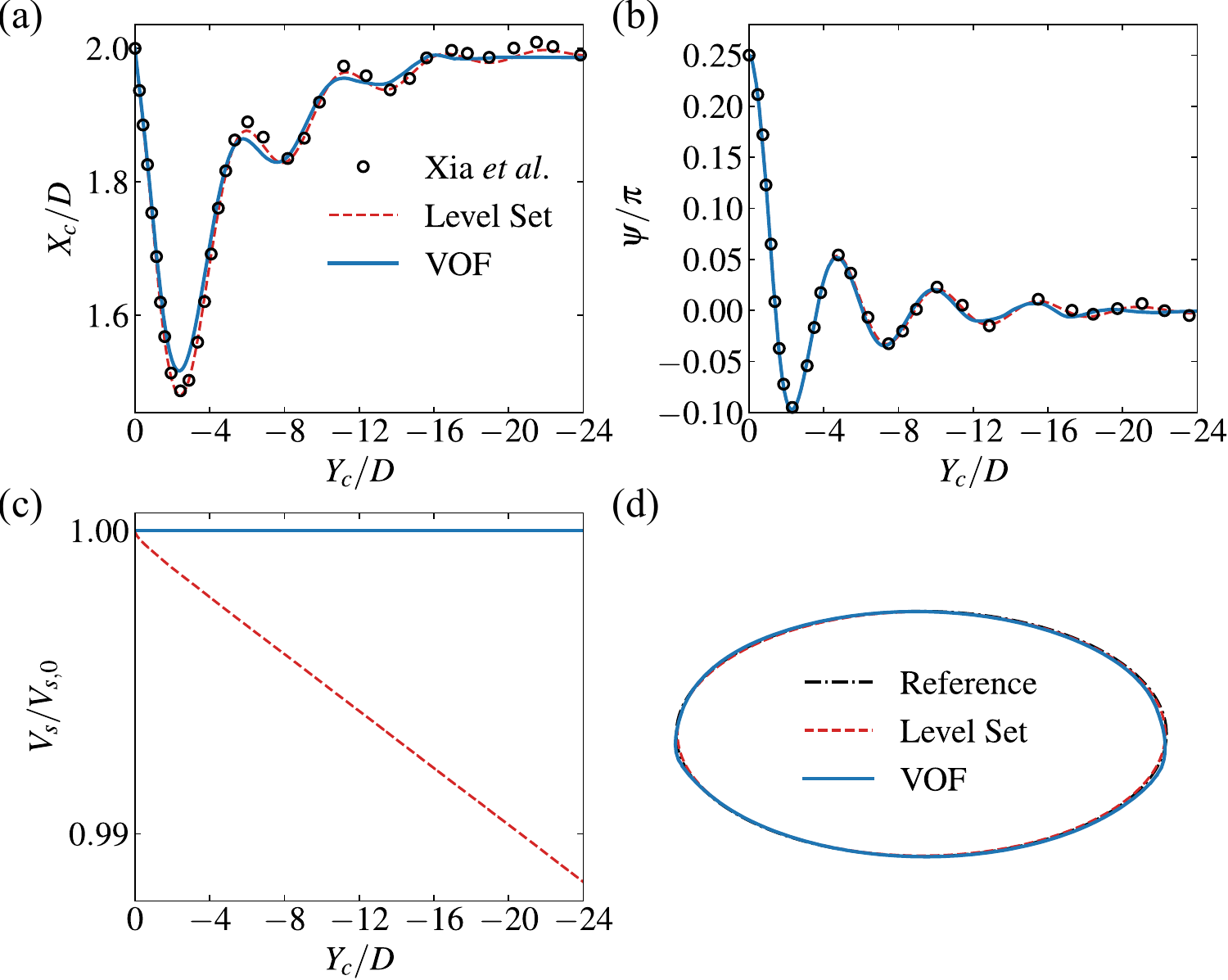}
	\caption{Long-distance sedimentation of an elliptical cylinder using VOF, compared with the Level-Set results of \S~\ref{sec:ellipse}.
	(a) Centroid trajectory.
	(b) Inclination angle.
	(c) Particle-volume evolution.
	(d) Interface at $Y_c=-24D$, compared with the exact ellipse and the Level-Set result.}
	\label{fig:vof-ellipse}
\end{figure}

The choice of the Level-Set method as the primary geometrical representation in the present framework is motivated mainly by the requirement of preserving the particle shape during long-distance translation and rotation. This appendix examines this issue more directly by repeating the rigid-body sedimentation tests of \S\S~\ref{sec:ellipse} and \ref{sec:ellipsoid_fall} with a VOF representation of the solid--liquid interface. The comparison is not intended to establish a general superiority of one interface method over the other. In particular, geometrical VOF methods possess the important advantage of excellent global volume conservation. The objective here is instead to quantify a specific difficulty encountered in the present application, namely the accumulation of geometrical errors when a nominally rigid interface is repeatedly reconstructed and advected over distances much larger than the particle size.

In the VOF formulation, the liquid volume fraction $c$ is advected according to
\begin{equation}
	\frac{\partial c}{\partial t}
	+\bm V_\Gamma^{\mathrm{ext}}\cdot\nabla c=0,
	\label{eq:ad_vof}
\end{equation}
where $\bm V_\Gamma^{\mathrm{ext}}$ still denotes the narrow-band extension of the geometrical interface velocity $\bm V_\Gamma$. Equation~\eqref{eq:ad_vof} is discretized using the conservative geometrical directional-splitting scheme of Weymouth and Yue~\cite{weymouth2010conservative}, consistent with the VOF transport employed for the moving solid-supported quantities described in \S~\ref{sec:solid_temperature}. After each advection step, the interface is reconstructed by the piecewise-linear interface construction (PLIC) procedure~\cite{scardovelli_direct_1999,popinet2009accurate}. Because the interface segment in each cut cell is reconstructed locally, neighboring PLIC segments are not constrained to form a globally smooth surface. The resulting small geometrical inconsistencies are usually acceptable for conventional multiphase-flow calculations and do not compromise the excellent volume conservation of the VOF method. For repeated rigid-body transport, however, reconstruction and advection errors may accumulate and progressively distort a shape that should remain unchanged. This behavior is already evident in classical rigid-interface advection tests such as the Zalesak slotted-disk problem~\cite{aulisaGeometricalAreapreservingVolumeofFluid2003}. In the present solid-particle problem, the absence of a physical restoring mechanism such as surface tension makes such geometrical errors particularly persistent.

Figure~\ref{fig:vof-ellipse} first considers the two-dimensional elliptical cylinder of \S~\ref{sec:ellipse}, using the same physical parameters and spatial resolution. The VOF and Level-Set calculations predict comparable overall settling dynamics, although the VOF result approaches the long-time settling state somewhat earlier, as seen from the trajectory and inclination in Figs.~\ref{fig:vof-ellipse}(a,b). The origin of this difference becomes clearer from the geometrical comparison in Fig.~\ref{fig:vof-ellipse}(d). After the particle has travelled to $Y_c=-24D$, the VOF interface has developed visible departures from the original elliptical geometry, including spurious protrusions near the equatorial region, whereas the Level-Set interface remains substantially closer to the prescribed ellipse. These local shape errors alter the instantaneous distribution of pressure and viscous stresses and can therefore modify the subsequent translational and rotational response. The trade-off is evident in Fig.~\ref{fig:vof-ellipse}(c): the VOF method retains the particle volume almost exactly, whereas the Level-Set calculation exhibits a relative volume error of order $\mathcal O(10^{-2})$. Thus, for this test, VOF provides superior global conservation while Level Set provides superior long-time geometrical fidelity.

\begin{figure}
	\centering
	\includegraphics[width=0.9\textwidth]{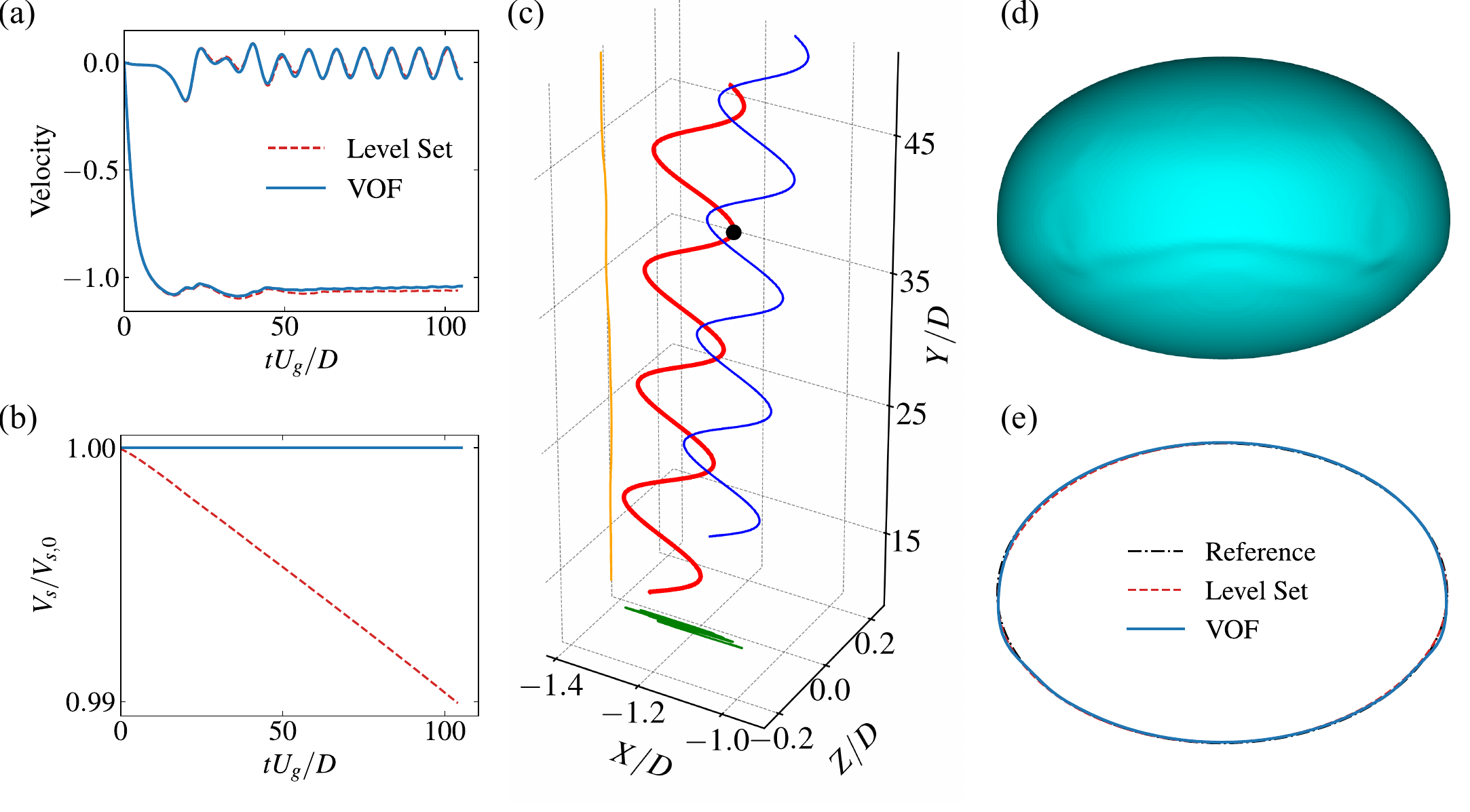}
	\caption{Long-distance sedimentation of an oblate ellipsoid using VOF, compared with the Level-Set results of \S~\ref{sec:ellipsoid_fall}.
	(a) Horizontal and vertical particle velocities.
	(b) Particle-volume evolution.
	(c) Three-dimensional trajectory in the periodic regime.
	(d) Particle geometry at $tU_g/D=77$.
	(e) Corresponding $x$--$y$ cross-section compared with the reference ellipse.}
	\label{fig:vof-ellipsoid}
\end{figure}

The same tendency becomes more consequential in the three-dimensional oblate-ellipsoid problem of \S~\ref{sec:ellipsoid_fall}. As shown in Fig.~\ref{fig:vof-ellipsoid}(a), the VOF and Level-Set calculations initially predict similar horizontal and vertical velocity histories. Nevertheless, their long-time trajectories eventually differ. The Level-Set calculation retains the predominantly vertical mean settling path reported in \S~\ref{sec:ellipsoid_fall}, whereas the VOF result develops a noticeable mean lateral drift, as shown in Fig.~\ref{fig:vof-ellipsoid}(c). Figures~\ref{fig:vof-ellipsoid}(d,e), corresponding to the instant of $tU_g/D=77$, reveal that this drift is accompanied by an appreciable distortion of the particle geometry: the initially smooth oblate shape develops asymmetric protrusions and depressions after repeated VOF advection. Such numerically generated asymmetry produces an imbalance in the hydrodynamic loading and is consistent with the inclined mean trajectory. Again, Fig.~\ref{fig:vof-ellipsoid}(b) shows the opposite behavior in terms of global conservation: the VOF particle volume is preserved substantially better, whereas the Level-Set representation incurs a relative volume error of order $\mathcal O(10^{-2})$.

These two tests clarify the numerical compromise underlying the present choice of interface representation. For the long-distance transport of freely moving solid particles, preserving the instantaneous geometry is essential because even small accumulated shape errors feed directly into the hydrodynamic force, torque and hence the subsequent particle trajectory. The present PLIC-based VOF implementation provides excellent volume conservation but gradually accumulates geometrical distortion under repeated rigid-body transport. The Level-Set method sacrifices part of this global conservation accuracy, but preserves the particle geometry much more faithfully and is therefore adopted here as the primary interface representation.

It should also be noted that the distortion observed above is less pronounced once melting is activated, because phase-change-induced interface recession tends to continuously modify and partially smooth the advected geometry. This effect, however, should not be regarded as a numerical remedy: the geometrical fidelity of the interface representation should remain independent of whether a particular physical process happens to regularize the shape. Given the superior conservation property of VOF, an important direction for further development is therefore to construct a conservative interface-transport strategy that retains the long-time shape fidelity required for freely translating and rotating solids without sacrificing the intrinsic volume-conservation advantage of geometrical VOF methods.

\section{Effect of local scalar conservation on interfacial heat transfer}
\label{sec:ncon}

Technique~II-B distinguishes deliberately between the advection of momentum and that of the liquid scalars. All variables are first formulated from the same conservative moving-control-volume balance, which accounts consistently for both the convective fluxes and the variation of the liquid volume as $\Gamma_\Delta$ sweeps across a Cartesian cell. For momentum, however, the strictly conservative update becomes vulnerable to the small-cell instability when the liquid fraction is very small. The velocity equation therefore employs the stabilized operator
\[
   \mathcal A_{\mathrm{adv}}^{n+\frac12}(\bm u_l)
    =
    \eta\,
    \mathcal A_{\mathrm{adv}}^{\,n+\frac12}(\bm u_l)
    +
    (1-\eta)\,
    \widetilde{\mathcal A}_{\mathrm{adv}}^{\,n+\frac12}(\bm u_l),
\]
where $\widetilde{\mathcal A}_{\mathrm{adv}}^{\,n+\frac12}(\bm u_l)$ is the locally smoothed operator defined in Technique~II-B. In contrast, the liquid temperature and concentration retain the fully conservative formulation,
$\mathcal A_{\mathrm{adv}}^{n+\frac12}(\chi)$ for $\chi\in\{T_l,Y_l\}$. The purpose of this appendix is to demonstrate why the momentum stabilization cannot be transferred directly to the scalar equations.

\begin{figure}[htbp]
	\centering
	\includegraphics[width=0.5\textwidth]{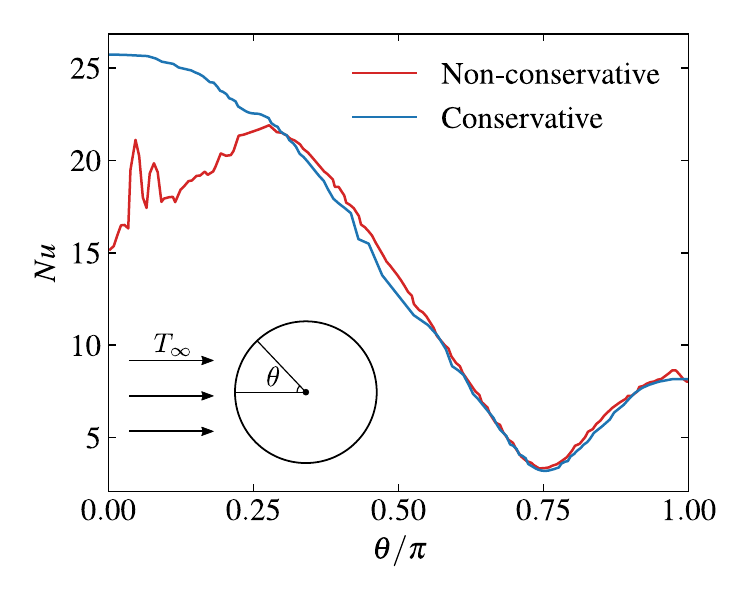}
	\caption{Effect of the scalar-advection treatment on the local Nusselt number for the translating ice sphere of \S~\ref{sec:melting_sphere_trans} at $tU/D_0=0$. The fully conservative scalar operator of Technique~II-B is compared with the conservative/non-conservative blending used for momentum. The inset defines the polar angle $\theta$, measured from the front stagnation point.}
	\label{fig:nonconservation}
\end{figure}

To isolate this effect, we repeat the prescribed-translation ice-sphere problem of \S~\ref{sec:melting_sphere_trans}, keeping the geometry, flow solver, thermal boundary conditions, mesh and all physical parameters unchanged. The only modification is that the temperature advection is replaced by the same blended treatment used for momentum,
\[
   \mathcal A_{\mathrm{adv}}^{n+\frac12}(T_l)
    =
    \eta\,
    \mathcal A_{\mathrm{adv}}^{\,n+\frac12}(T_l)
    +
    (1-\eta)\,
    \widetilde{\mathcal A}_{\mathrm{adv}}^{\,n+\frac12}(T_l),
\]
with the same locally determined blending coefficient $\eta$. The comparison is made at $tU/D_0=0$, immediately before melting is activated. At this instant the particle geometry is identical in the two calculations, so that any difference in the interfacial heat transfer originates directly from the scalar-advection discretization rather than from a subsequent difference in interface evolution.
A particularly sensitive measure is the local Nusselt number,
\begin{equation}
	Nu(\theta)
	=
	-\frac{D_0}{T_\infty-T_m}
	\left.
	\frac{\partial T_l}{\partial n}
	\right|_\Gamma ,
	\label{eq:local_Nu}
\end{equation}
where $\theta$ is measured along the particle surface from the front stagnation point. For the present forced-convection problem, the thermal boundary layer is thinnest near the front stagnation region and the corresponding interfacial temperature gradient, and hence $Nu$, reaches its largest value there.

Figure~\ref{fig:nonconservation} shows that the fully conservative formulation preserves this characteristic distribution and resolves the strong front-stagnation heat transfer. In contrast, once the momentum-type blending is applied to $T_l$, the local Nusselt-number distribution is visibly distorted, with the largest discrepancy occurring in the region of strongest interfacial heat transfer. The important point is that this error is highly localized: the bulk temperature field may still appear qualitatively reasonable, while the one-sided gradient evaluated directly at $\Gamma_\Delta$ has already lost the accuracy required by the phase-change problem.

This distinction explains the different treatments adopted in Technique~II-B. A localized relaxation of momentum conservation can suppress the small-cell instability while retaining an accurate hydrodynamic solution. For $T_l$, however, the same local smoothing directly alters $\partial_nT_l|_\Gamma$, which enters the Stefan condition and therefore immediately contaminates the predicted melting velocity $U_m$. The requirement is even more restrictive for binary melting, because the generally thinner solutal boundary layer makes $Y_{l,\Gamma}$ and $\partial_nY_l|_\Gamma$ equally sensitive to local conservation errors. Consequently, the liquid scalars must retain the fully conservative moving-cut-cell operator even though the momentum equation employs the stabilized conservative/non-conservative blending. This comparison provides the numerical justification for that central design choice in Technique~II-B.

\section{Need for geometry-consistent transport of the solid temperature}
\label{sec:bcg_Ts}

Technique~II-C is designed to satisfy an additional requirement that does not arise for the liquid scalars: the numerical support of the solid temperature $T_s$ must remain locked to the instantaneous solid domain as the particle translates and rotates across the fixed Cartesian mesh. A conventional Eulerian advection scheme, such as the BCG scheme~\citep{bell1989second} for transporting the momentum, may transport $T_s$ with acceptable accuracy in the bulk, but its numerical diffusion inevitably spreads the scalar across the sharp solid--liquid boundary. To avoid this inconsistency, Technique~II-C does not advect $T_s$ directly. Instead, the conservative quantity $E_s=c_sT_s$, with $c_s=1-c$ the solid volume fraction reconstructed from the Level-Set geometry, is transported using the same geometrically swept solid volumes as $c_s$, after which $T_s=E_s/c_s$ is recovered wherever $c_s>0$. The following test isolates the necessity of this geometrical treatment by comparing it with a direct BCG advection of $T_s$.

\begin{figure}[htbp]
	\centering
	\includegraphics[width=0.8\textwidth]{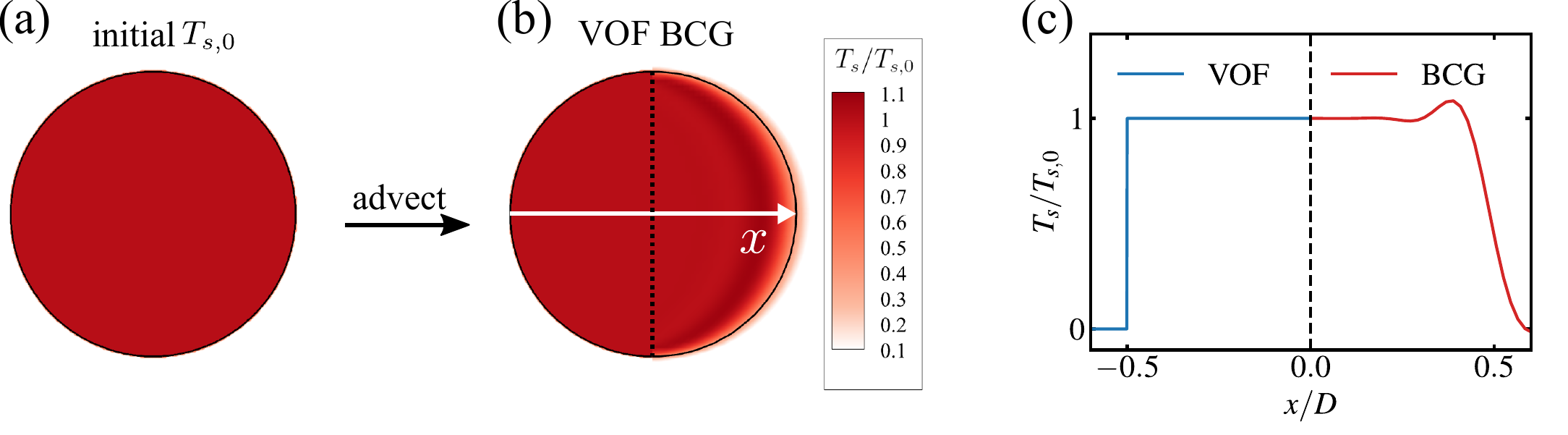}
	\caption{Transport of the solid temperature during rigid translation of a circular particle over a distance of $2D$.
	(a) Initial and final temperature fields obtained with the geometry-consistent tracer transport of Technique~II-C and with direct BCG advection of $T_s$.
	(b) Corresponding temperature profiles along the particle centerline.}
	\label{fig:geo_advect}
\end{figure}

We consider a 2D circular particle with an initially uniform solid temperature $T_{s,0}$ and prescribe a constant leftward rigid-body velocity $\bm U_s=(-U,0)$. Phase change and thermal diffusion are both suppressed, so that the exact solution after any displacement is simply the initial temperature field translated together with the particle, with $T_s=T_{s,0}$ throughout the solid and no solid-temperature support outside $\Omega_s$. The particle is translated over a distance of $2D$ using a finest resolution $D/\Delta_{\min}=51$ and $\mathrm{CFL}=U\Delta t/\Delta_{\min}=0.5$. The two calculations differ only in the treatment of solid-temperature advection: one uses the geometric tracer transport of Technique~II-C, whereas the other applies the second-order BCG reconstruction directly to $T_s$.

Figure~\ref{fig:geo_advect}(a) shows that the geometry-consistent formulation transports the temperature field together with the particle without altering its support: the temperature field remains confined to the reconstructed solid region and retains the uniform value $T_{s,0}$ using a geometrical advection scheme, while it spreads out of the solid region using the BCG advection scheme. The corresponding centerline profile in Fig.~\ref{fig:geo_advect}(b) therefore preserves the sharp transition at the particle boundary. In contrast, direct BCG advection produces a smeared temperature field across the moving interface, with non-zero values extending into cells that belong to the liquid phase. Because $T_s$ is physically defined only in $\Omega_s$, this leakage represents a loss of consistency between the scalar field and the instantaneous solid geometry rather than a physical thermal boundary layer.

This comparison clarifies why the treatment of $T_s$ in Technique~II-C differs fundamentally from that of $T_l$ and $Y_l$ in Technique~II-B. For the solid temperature, scalar conservation alone is insufficient: the transported temperature content must remain attached to the same solid volume that is displaced by the rigid-body motion. Advecting $E_s=c_sT_s$ with the geometrically swept solid fluxes enforces this requirement by construction, whereas direct Eulerian BCG advection cannot prevent numerical leakage across $\Gamma_\Delta$. The geometric tracer formulation is therefore essential for maintaining both conservation and support consistency of $T_s$ on a translating or rotating solid domain.

\bibliographystyle{elsarticle-num}
\bibliography{ref}

@article{popinet2015quadtree,
  title={A quadtree-adaptive multigrid solver for the Serre--Green--Naghdi equations},
  author={Popinet, St{\'e}phane},
  journal={Journal of Computational Physics},
  volume={302},
  pages={336--358},
  year={2015},
  publisher={Elsevier}
}

@article{schwartz2006cartesian,
  title={A Cartesian grid embedded boundary method for the heat equation and Poisson’s equation in three dimensions},
  author={Schwartz, Peter and Barad, Michael and Colella, Phillip and Ligocki, Terry},
  journal={Journal of Computational Physics},
  volume={211},
  number={2},
  pages={531--550},
  year={2006},
  publisher={Elsevier}
}

@article{meng2020phase,
  title={Phase-field-lattice Boltzmann simulation of dendrite motion using an immersed boundary method},
  author={Meng, Shaoxing and Zhang, Ang and Guo, Zhipeng and Wang, Qigui},
  journal={Computational Materials Science},
  volume={184},
  pages={109784},
  year={2020},
  publisher={Elsevier}
}

@article{echebarria2004quantitative,
  author    = {Echebarria, B. and Folch, R. and Karma, A. and Plapp, M.},
  title     = {Quantitative phase-field model of alloy solidification},
  journal   = {Physical Review E},
  volume    = {70},
  number    = {6},
  pages     = {061604},
  year      = {2004}
}

@article{beckermann1999modeling,
  author    = {Beckermann, C. and Diepers, H.-J. and Steinbach, I. and Karma, A. and Tong, X.},
  title     = {Modeling melt convection in phase-field simulations of solidification},
  journal   = {Journal of Computational Physics},
  volume    = {154},
  number    = {2},
  pages     = {468--496},
  year      = {1999}
}

@article{karma1998quantitative,
  author    = {Karma, A. and Rappel, W.-J.},
  title     = {Quantitative phase-field modeling of dendritic growth in two and three dimensions},
  journal   = {Physical Review E},
  volume    = {57},
  number    = {4},
  pages     = {4323--4349},
  year      = {1998}
}

@article{limare2023hybrid,
  author  = {Limare, A. and Popinet, S. and Josserand, C. and Xue, Z. and Ghigo, A.},
  title   = {A hybrid level-set / embedded boundary method applied to solidification--melt problems},
  journal = {Journal of Computational Physics},
  volume  = {474},
  pages   = {111829},
  year    = {2023}
}

@article{zhao2022boiling,
  author  = {Zhao, S. and Zhang, J. and Ni, M. J.},
  title   = {Boiling and evaporation model for liquid--gas flows: A sharp and conservative method based on the geometrical {VOF} approach},
  journal = {Journal of Computational Physics},
  volume  = {452},
  pages   = {110908},
  year    = {2022}
}

@article{xue2023three,
  title={Three-dimensional sharp and conservative VOF method for the simulation of binary solidification},
  author={Xue, Zhong-Han and Zhao, Shuo and Ni, Ming-Jiu and Zhang, Jie},
  journal={Journal of Computational Physics},
  volume={491},
  pages={112380},
  year={2023},
  publisher={Elsevier}
}

@article{Appolaire1998FreeGrowthSettling,
  author  = {Appolaire, B. and Albert, V. and Combeau, H. and Lesoult, G.},
  title   = {Free growth of equiaxed crystals settling in undercooled NH4Cl-H2O melts},
  journal = {Acta Materialia},
  year    = {1998},
  volume  = {46},
  number  = {15},
  pages   = {5851--5862},
  publisher = {Elsevier}
}

@article{Badillo2007DendriteTipKinetics,
  author  = {Badillo, A. and Ceynar, D. and Beckermann, C.},
  title   = {Growth of equiaxed dendritic crystals settling in an undercooled melt, Part 1: Tip kinetics},
  journal = {Journal of Crystal Growth},
  year    = {2007},
  volume  = {309},
  number  = {2},
  pages   = {197--215},
  publisher = {Elsevier}
}

@article{Rojas2015PhaseFieldLBM,
  author  = {Rojas, R. and Takaki, T. and Ohno, M.},
  title   = {A phase-field-lattice Boltzmann method for modeling motion and growth of a dendrite for binary alloy solidification in the presence of melt convection},
  journal = {Journal of Computational Physics},
  year    = {2015},
  volume  = {298},
  pages   = {29--40},
  publisher = {Elsevier}
}

@article{Takaki2018PhaseFieldLBMGrainGrowth,
  author  = {Takaki, T. and Sato, R. and Rojas, R. and Ohno, M. and Shibuta, Y.},
  title   = {Phase-field lattice Boltzmann simulations of multiple dendrite growth with motion, collision, and coalescence and subsequent grain growth},
  journal = {Computational Materials Science},
  year    = {2018},
  volume  = {147},
  pages   = {124--131},
  publisher = {Elsevier}
}

@article{Sakane2019PhaseFieldLBM,
  author  = {Sakane, S. and Takaki, T. and Ohno, M. and Shibuta, Y.},
  title   = {Simulation method based on phase-field lattice Boltzmann model for long-distance sedimentation of single equiaxed dendrite},
  journal = {Computational Materials Science},
  year    = {2019},
  volume  = {164},
  pages   = {39--45},
  publisher = {Elsevier},
}

@article{fang2021spatiotemporal,
  title={Spatiotemporal evolutions of forces and vortices of flow past ellipsoidal bubbles: Direct numerical simulation based on a Cartesian grid scheme},
  author={Fang, Zhiheng and Zhang, Jie and Xiong, Qingang and Xu, Fei and Ni, Mingjiu},
  journal={Physics of Fluids},
  volume={33},
  number={1},
  year={2021},
  publisher={AIP Publishing}
}

@article{popinet2009accurate,
  title={An accurate adaptive solver for surface-tension-driven interfacial flows},
  author={Popinet, St{\'e}phane},
  journal={Journal of Computational Physics},
  volume={228},
  number={16},
  pages={5838--5866},
  year={2009},
  publisher={Elsevier}
}

@article{bell1989second,
  title={A second-order projection method for the incompressible Navier-Stokes equations},
  author={Bell, John B and Colella, Phillip and Glaz, Harland M},
  journal={Journal of computational physics},
  volume={85},
  number={2},
  pages={257--283},
  year={1989},
  publisher={Elsevier}
}

@article{weymouth2010conservative,
  title={Conservative volume-of-fluid method for free-surface simulations on cartesian-grids},
  author={Weymouth, Gabriel D and Yue, Dick K-P},
  journal={Journal of Computational Physics},
  volume={229},
  number={8},
  pages={2853--2865},
  year={2010},
  publisher={Elsevier}
}

@article{mao2024anisotropic,
  title={An anisotropic lattice Boltzmann-phase field model for dendrite growth and movement in rapid solidification of binary alloys},
  author={Mao, Shilin and Cao, Yuting and Chen, Wei and Sun, Dongke},
  journal={npj Computational Materials},
  volume={10},
  number={1},
  pages={63},
  year={2024},
  publisher={Nature Publishing Group UK London}
}

@article{gan2003simulation,
  title={Simulation of the sedimentation of melting solid particles},
  author={Gan, Hui and Feng, James J and Hu, Howard H},
  journal={International journal of multiphase flow},
  volume={29},
  number={5},
  pages={751--769},
  year={2003},
  publisher={Elsevier}
}

@article{cenedese2023icebergs,
  title={Icebergs melting},
  author={Cenedese, Claudia and Straneo, Fiamma},
  journal={Annual Review of Fluid Mechanics},
  volume={55},
  number={1},
  pages={377--402},
  year={2023},
  publisher={Annual Reviews}
}

@Article{bochkov2021sharp,
  Title                    = {Sharp-Interface Simulations of Multicomponent Alloy Solidification},
  Author                   = {Bochkov, Daniil and Pollock, Tresa and Gibou, Frederic},
  Journal                  = {arXiv preprint arXiv:2112.08650},
  Year                     = {2021}
}

@Article{boettinger_phase-field_2002,
  Title                    = {Phase-{Field} {Simulation} of {Solidification}},
  Author                   = {Boettinger, W. J. and Warren, J. A. and Beckermann, C. and Karma, A.},
  Journal                  = {Annual Review of Materials Research},
  Year                     = {2002},
  Month                    = aug,
  Number                   = {1},
  Pages                    = {163--194},
  Volume                   = {32},
  ISSN                     = {1531-7331, 1545-4118},
  Language                 = {en},
}

@Book{davis2001theory,
  Title                    = {Theory of solidification},
  Author                   = {Davis, Stephen H},
  Publisher                = {Cambridge University Press},
  Year                     = {2001}
}

@Article{hester_improved_2020,
  Title                    = {Improved phase-field models of melting and dissolution in multi-component flows},
  Author                   = {Hester, Eric W. and Couston, Louis-Alexandre and Favier, Benjamin and Burns, Keaton J. and Vasil, Geoffrey M.},
  Journal                  = {Proceedings of the Royal Society A: Mathematical, Physical and Engineering Sciences},
  Year                     = {2020},
  Month                    = oct,
  Number                   = {2242},
  Pages                    = {20200508},
  Volume                   = {476},
  ISSN                     = {1364-5021, 1471-2946},
  Language                 = {en},
}

@Article{johansen1998cartesian,
  Title                    = {A Cartesian grid embedded boundary method for Poisson's equation on irregular domains},
  Author                   = {Johansen, Hans and Colella, Phillip},
  Journal                  = {Journal of Computational Physics},
  Year                     = {1998},
  Number                   = {1},
  Pages                    = {60--85},
  Volume                   = {147},
  Publisher                = {Elsevier}
}

@Article{karagadde_coupled_2012,
  Title                    = {A coupled {VOF}–{IBM}–enthalpy approach for modeling motion and growth of equiaxed dendrites in a solidifying melt},
  Author                   = {Karagadde, Shyamprasad and Bhattacharya, Anirban and Tomar, Gaurav and Dutta, Pradip},
  Journal                  = {Journal of Computational Physics},
  Year                     = {2012},
  Month                    = may,
  Number                   = {10},
  Pages                    = {3987--4000},
  Volume                   = {231},
  ISSN                     = {00219991},
  Language                 = {en},
}

@Article{karma_phase-field_2001,
  Title                    = {Phase-{Field} {Formulation} for {Quantitative} {Modeling} of {Alloy} {Solidification}},
  Author                   = {Karma, Alain},
  Journal                  = {Physical Review Letters},
  Year                     = {2001},
  Month                    = aug,
  Number                   = {11},
  Pages                    = {115701},
  Volume                   = {87},
  ISSN                     = {0031-9007, 1079-7114},
  Language                 = {en},
  Urldate                  = {2021-11-11}
}

@Article{ramirez_phase-field_2004,
  Title                    = {Phase-field modeling of binary alloy solidification with coupled heat and solute diffusion},
  Author                   = {Ramirez, J. C. and Beckermann, C. and Karma, A. and Diepers, H.-J.},
  Journal                  = {Physical Review E},
  Year                     = {2004},
  Month                    = may,
  Number                   = {5},
  Pages                    = {051607},
  Volume                   = {69},
  ISSN                     = {1539-3755, 1550-2376},
  Language                 = {en},
  Urldate                  = {2021-11-11}
}

@Article{scardovelli_direct_1999,
  Title                    = {{DIRECT} {NUMERICAL} {SIMULATION} {OF} {FREE}-{SURFACE} {AND} {INTERFACIAL} {FLOW}},
  Author                   = {Scardovelli, Ruben and Zaleski, Stéphane},
  Journal                  = {Annual Review of Fluid Mechanics},
  Year                     = {1999},
  Month                    = jan,
  Number                   = {1},
  Pages                    = {567--603},
  Volume                   = {31},
  ISSN                     = {0066-4189, 1545-4479},
  Language                 = {en},
  Urldate                  = {2021-07-10}
}

@Article{theillard_sharp_2015,
  Title                    = {A {Sharp} {Computational} {Method} for the {Simulation} of the {Solidification} of {Binary} {Alloys}},
  Author                   = {Theillard, Maxime and Gibou, Frédéric and Pollock, Tresa},
  Journal                  = {Journal of Scientific Computing},
  Year                     = {2015},
  Month                    = may,
  Number                   = {2},
  Pages                    = {330--354},
  Volume                   = {63},
  ISSN                     = {0885-7474, 1573-7691},
  Language                 = {en},
  Urldate                  = {2021-07-21}
}

@Article{udaykumar_computation_1999,
  Title                    = {Computation of {Solid}–{Liquid} {Phase} {Fronts} in the {Sharp} {Interface} {Limit} on {Fixed} {Grids}},
  Author                   = {Udaykumar, H.S and Mittal, R and Shyy, Wei},
  Journal                  = {Journal of Computational Physics},
  Year                     = {1999},
  Month                    = aug,
  Number                   = {2},
  Pages                    = {535--574},
  Volume                   = {153},
  ISSN                     = {00219991},
  Language                 = {en},
  Urldate                  = {2021-09-23}
}

@Article{yang2005sharp,
  Title                    = {Sharp interface Cartesian grid method III: solidification of pure materials and binary solutions},
  Author                   = {Yang, Yi and Udaykumar, HS},
  Journal                  = {Journal of Computational Physics},
  Year                     = {2005},
  Number                   = {1},
  Pages                    = {55--74},
  Volume                   = {210},
  Publisher                = {Elsevier}
}

@article{sverdrup2019embedded,
  title={An embedded boundary approach for efficient simulations of viscoplastic fluids in three dimensions},
  author={Sverdrup, Knut and Almgren, Ann and Nikiforakis, Nikolaos},
  journal={Physics of Fluids},
  volume={31},
  number={9},
  year={2019},
  publisher={AIP Publishing}
}

@article{min2007second,
  title={A second order accurate level set method on non-graded adaptive cartesian grids},
  author={Min, Chohong and Gibou, Fr{\'e}d{\'e}ric},
  journal={Journal of Computational Physics},
  volume={225},
  number={1},
  pages={300--321},
  year={2007},
  publisher={Elsevier}
}

@article{sussman1994level,
  title={A level set approach for computing solutions to incompressible two-phase flow},
  author={Sussman, Mark and Smereka, Peter and Osher, Stanley},
  journal={Journal of Computational physics},
  volume={114},
  number={1},
  pages={146--159},
  year={1994},
  publisher={Elsevier}
}

@article{suzuki2026equations,
  title={On the equations of motion for a solid phase moving in a liquid phase while melting and solidification},
  author={Suzuki, Kosuke and Aoyagi, Yuto and Yoshino, Masato},
  journal={Journal of Fluid Mechanics},
  volume={1036},
  pages={A25},
  year={2026},
  publisher={Cambridge University Press}
}

@article{xiaFlowPatternsSedimentation2009,
	title = {Flow Patterns in the Sedimentation of an Elliptical Particle},
	author = {Xia, Zhenhua and Connington, Kevin W. and Rapaka, Saikiran and Yue, Pengtao and Feng, James J. and Chen, Shiyi},
	year = {2009},
	month = apr,
	journal = {Journal of Fluid Mechanics},
	volume = {625},
	pages = {249--272},
	issn = {0022-1120, 1469-7645}
}

@article{jiangFlowreconstructionBasedApproach2024,
	title = {A Flow-Reconstruction Based Approach for the Computation of Hydrodynamic Stresses on Immersed Body Surface},
	author = {Jiang, Xinyu and Huang, Weixi and Xu, Chunxiao and Zhao, Lihao},
	year = {2024},
	month = jul,
	journal = {Journal of Computational Physics},
	volume = {508},
	pages = {113025},
	publisher = {Elsevier BV},
	issn = {0021-9991}
}

@article{moricheSingleOblateSpheroid2021,
	title = {A Single Oblate Spheroid Settling in Unbounded Ambient Fluid: {{A}} Benchmark for Simulations in Steady and Unsteady Wake Regimes},
	shorttitle = {A Single Oblate Spheroid Settling in Unbounded Ambient Fluid},
	author = {Moriche, Manuel and Uhlmann, Markus and Du{\v s}ek, Jan},
	year = {2021},
	month = mar,
	journal = {International Journal of Multiphase Flow},
	volume = {136},
	pages = {103519},
	issn = {03019322}
}

@unpublished{ghigo:hal-03948786,
  TITLE = {{A conservative finite volume cut-cell method on an adaptive Cartesian tree grid for moving rigid bodies in incompressible flows}},
  AUTHOR = {Ghigo, Arthur R. and Popinet, St{\'e}phane and Wachs, Anthony},
  NOTE = {working paper or preprint},
  YEAR = {2021},
  MONTH = Aug,
  HAL_ID = {hal-03948786},
  HAL_VERSION = {v1},
}

@article{xue2026wake,
  title={Wake transitions and melting dynamics of a translating sphere in warm liquid},
  author={Xue, Zhonghan and Zhang, Jie},
  journal={Journal of Fluid Mechanics},
  volume={1029},
  pages={A22},
  year={2026},
  publisher={Cambridge University Press}
}

@article{shiNumericalSimulationMelting2024,
	title = {Numerical {{Simulation}} of the {{Melting}} of {{Solid Particles}} in {{Thermal Convection}} with a {{Modified Immersed Boundary Method}}},
	author = {Shi, Yang and Shao, Xueming and Xu, Jian and Yu, Zhaosheng},
	year = {2024},
	month = nov,
	journal = {Processes},
	volume = {12},
	number = {11},
	pages = {2533},
	issn = {2227-9717}
}

@article{twizellSecondorderL0stableMethods1996,
	title = {Second-Order,{{L}} 0-Stable Methods for the Heat Equation with Time-Dependent Boundary Conditions},
	author = {Twizell, E. H. and Gumel, A. B. and Arigu, M. A.},
	year = {1996},
	month = dec,
	journal = {Advances in Computational Mathematics},
	volume = {6},
	number = {1},
	pages = {333--352},
	issn = {1019-7168, 1572-9044}
}

@article{zhongFronttrackingImmersedboundaryFramework2025,
	title = {A Front-Tracking Immersed-Boundary Framework for Simulating {{Lagrangian}} Melting Problems},
	author = {Zhong, Kevin and Howland, Christopher J. and Lohse, Detlef and Verzicco, Roberto},
	year = {2025},
	month = mar,
	journal = {Journal of Computational Physics},
	volume = {525},
	pages = {113762},
	issn = {00219991}
}

@article{mccorquodaleCartesianGridEmbedded2001,
	title = {A {{Cartesian Grid Embedded Boundary Method}} for the {{Heat Equation}} on {{Irregular Domains}}},
	author = {McCorquodale, Peter and Colella, Phillip and Johansen, Hans},
	year = {2001},
	month = nov,
	journal = {Journal of Computational Physics},
	volume = {173},
	number = {2},
	pages = {620--635},
	issn = {00219991}
}

@article{johnsonShapeEvolutionCapsize2025,
	title = {Shape Evolution and Capsize Dynamics of Melting Ice},
	author = {Johnson, Bobae and Weady, Scott and Zhang, Zihan and Kim, Alison and Ristroph, Leif},
	year = {2025},
	month = sep,
	journal = {Physical Review Fluids},
	volume = {10},
	number = {9},
	pages = {093801},
	issn = {2469-990X}
}

@misc{xu2025EulerianLSIBM,
	author       = {Xu, Chang and Jiang, Xinyu and Satheeshchandran, Kiran and Brandt, Luca},
	title = {An {Eulerian} {Level Set}--{Immersed Boundary} Method for Fluid-Solid Interactions with Phase Change},
	year         = {2025},
	publisher    = {SSRN},
	note         = {SSRN preprint}
}

@article{jeffery,
	author = {Jeffery, George Barker},
	title = {The motion of ellipsoidal particles immersed in a viscous fluid},
	journal = {Proceedings of the Royal Society of London. Series A, Containing Papers of a Mathematical and Physical Character},
	volume = {102},
	number = {715},
	pages = {161-179},
	year = {1922},
	month = {11},
	issn = {0950-1207}
}

@article{liFully3DSimulation2020,
	title = {A Fully {{3D}} Simulation of Fluid-Structure Interaction with Dynamic Wetting and Contact Angle Hysteresis},
	author = {Li, Hai-Long and Liu, Hao-Ran and Ding, Hang},
	year = {2020},
	month = nov,
	journal = {Journal of Computational Physics},
	volume = {420},
	pages = {109709},
	issn = {00219991}
}

@misc{Popinet2026Basilisk,
	author = {St{\'e}phane Popinet},
	title  = {Basilisk: Free Software for the Solution of Partial Differential Equations},
	year   = {2026}
}

@article{aulisaGeometricalAreapreservingVolumeofFluid2003,
	title = {A Geometrical Area-Preserving {{Volume-of-Fluid}} Advection Method},
	author = {Aulisa, Eugenio and Manservisi, Sandro and Scardovelli, Ruben and Zaleski, Stephane},
	year = {2003},
	month = nov,
	journal = {Journal of Computational Physics},
	volume = {192},
	number = {1},
	pages = {355--364},
	issn = {00219991}
}

@article{du2024physics,
  title={The physics of freezing and melting in the presence of flows},
  author={Du, Yihong and Calzavarini, Enrico and Sun, Chao},
  journal={Nature Reviews Physics},
  volume={6},
  number={11},
  pages={676--690},
  year={2024},
  publisher={Nature Publishing Group UK London}
}

@article{guo2025effects,
  title={The effects of double diffusive convection on the basal melting of solid ice in seawater},
  author={Guo, Rongfu and Yang, Yantao},
  journal={Journal of Fluid Mechanics},
  volume={1013},
  pages={A24},
  year={2025},
  publisher={Cambridge University Press}
}

@article{yang2023morphology,
  title={Morphology evolution of a melting solid layer above its melt heated from below},
  author={Yang, Rui and Howland, Christopher J and Liu, Hao-Ran and Verzicco, Roberto and Lohse, Detlef},
  journal={Journal of Fluid Mechanics},
  volume={956},
  pages={A23},
  year={2023},
  publisher={Cambridge University Press}
}

@article{peng1999pde,
  title={A PDE-based fast local level set method},
  author={Peng, Danping and Merriman, Barry and Osher, Stanley and Zhao, Hongkai and Kang, Myungjoo},
  journal={Journal of computational physics},
  volume={155},
  number={2},
  pages={410--438},
  year={1999},
  publisher={Elsevier}
}

@article{li2025hydrodynamic,
  title={Hydrodynamic instability induces spontaneous motion of floating ice discs},
  author={Li, Min and Zhu, Lailai},
  journal={arXiv preprint arXiv:2511.13184},
  year={2025}
}

@article{dorbolo2016rotation,
  title={Rotation of melting ice disks due to melt fluid flow},
  author={Dorbolo, St{\'e}phane and Adami, N and Dubois, C and Caps, H and Vandewalle, N and Darbois-Texier, B},
  journal={Physical Review E},
  volume={93},
  number={3},
  pages={033112},
  year={2016},
  publisher={APS}
}

@article{bellincioni2025melting,
  title={Melting of floating ice cylinders in fresh and saline environments},
  author={Bellincioni, Edoardo and Lohse, Detlef and Huisman, Sander G},
  journal={Journal of Fluid Mechanics},
  volume={1019},
  pages={A29},
  year={2025},
  publisher={Cambridge University Press}
}

\end{document}